\documentclass[twocolumn, twocolappendix]{aastex63}

\usepackage{graphicx}
\usepackage{subfigure}
\usepackage{longtable}
\usepackage{wrapfig}
\usepackage[figuresleft]{rotating}\usepackage{epstopdf}
\usepackage{amsmath}
\usepackage{multirow}
\usepackage[mathscr]{eucal}
\usepackage{hyperref}

\newcommand{\kms}{\mbox{${\rm km~s^{-1}}$}}
\newcommand{\e}{\mbox{$^{-1}$} }

\newcommand{\kmsMpc}{km s$^{-1}$ Mpc$^{-1}$}
\newcommand{\hi}{H{\sc\,i} }
\newcommand{\dg}{$^{\circ}$}

\newcommand{\hnut}{$H_0$}

\graphicspath{{./}{}}

\received{July 10, 2020}
\revised{August 27, 2020}
\accepted{September 1, 2020}
\published{October 23, 2020}

\submitjournal{ApJ}

\shorttitle{Cosmicflows-4: The Tully-Fisher Distances}
\shortauthors{Kourkchi et al.}

\begin{document}

\title{Cosmicflows-4: The Catalog of $\sim$10000 Tully-Fisher Distances}


\author[0000-0002-5514-3354]{Ehsan Kourkchi}
\email{ehsan@ifa.hawaii.edu}
\affil{Institute for Astronomy, University of Hawaii, 2680 Woodlawn Drive, Honolulu, HI 96822, USA}

\author[0000-0002-9291-1981]{R. Brent Tully}
\email{tully@ifa.hawaii.edu}
\affil{Institute for Astronomy, University of Hawaii, 2680 Woodlawn Drive, Honolulu, HI 96822, USA}

\author[0000-0002-8281-8388]{Sarah Eftekharzadeh}
\affil{Department of Physics and Astronomy, University of Utah, 115 S. 1400 E., Salt Lake City, UT 84112, USA}

\author{Jordan Llop}
\affil{Department of Physics \& Astronomy, University of Hawaii, 2505 Correa Road, Honolulu, HI 96822, USA}

\author[0000-0003-0509-1776]{H\'el\`ene M. Courtois}
\affil{University of Lyon, UCB Lyon 1, IUF, CNRS/IN2P3,  IP2I Lyon, UMR5822, F-69622 Villeurbanne. France}

\author{Daniel Guinet}
\affil{University of Lyon, UCB Lyon 1, IUF, CNRS/IN2P3,  IP2I Lyon, UMR5822, F-69622 Villeurbanne. France}

\author[0000-0001-9005-2792]{Alexandra Dupuy}
\affil{University of Lyon, UCB Lyon 1, IUF, CNRS/IN2P3,  IP2I Lyon, UMR5822, F-69622 Villeurbanne. France}

\author[0000-0002-0466-1119]{James D. Neill}
\affil{California Institute of Technology, 1200 East California Boulevard, MC 278-17, Pasadena, CA 91125, USA}

\author[0000-0002-1143-5515]{Mark Seibert}
\affil{The Observatories of the Carnegie Institute of Washington, 813 Santa Barbara Street, Pasadena, CA 91101, USA}

\author{Michael Andrews}
\affil{Department of Physics \& Astronomy, University of Hawaii, 2505 Correa Road, Honolulu, HI 96822, USA}

\author{Juana Chuang}
\affil{Department of Physics \& Astronomy, University of Hawaii, 2505 Correa Road, Honolulu, HI 96822, USA}

\author{Arash Danesh}
\affil{Department of Physics, University of Zanjan, University Blvd, Zanjan, 45371-38791, Iran}

\author{Randy Gonzalez}
\affil{Department of Physics \& Astronomy, University of Hawaii, 2505 Correa Road, Honolulu, HI 96822, USA}

\author{Alexandria Holthaus}
\affil{Department of Physics \& Astronomy, University of Hawaii, 2505 Correa Road, Honolulu, HI 96822, USA}

\author{Amber Mokelke}
\affil{Department of Physics \& Astronomy, University of Hawaii, 2505 Correa Road, Honolulu, HI 96822, USA}

\author{Devin Schoen}
\affil{Department of Physics \& Astronomy, University of Hawaii, 2505 Correa Road, Honolulu, HI 96822, USA}

\author{Chase Urasaki}
\affil{Department of Physics \& Astronomy, University of Hawaii, 2505 Correa Road, Honolulu, HI 96822, USA}

\begin{abstract}

We present the distances of 9792 spiral galaxies lying within 15,000 \kms\ using the relation between luminosity and rotation rate of spiral galaxies. The sample is dominantly, but not exclusively, drawn from galaxies detected in the course of the ALFALFA \hi survey with the Arecibo Telescope. Relations between \hi line widths and luminosity are calibrated at SDSS {\it u, g, r, i, z} bands and WISE {\it W}1 and {\it W}2 bands. By exploiting secondary parameters, particularly color indices, we address discrepancies between measured distances at different wave bands with unprecedented detail.  
We provide a catalog that includes reduced kinematic, photometric, and inclination parameters. We also describe a machine learning algorithm, based on the random forest technique that predicts the dust attenuation in spirals lacking infrared photometry.
We determine a Hubble Constant value of $H_0 = 75.1\pm0.2$(stat.), with potential systematics up to $\pm$ 3 \kmsMpc.
 
\end{abstract}

\keywords{
\href{http://astrothesaurus.org/uat/590}{Galaxy distances (590)}; \href{http://astrothesaurus.org/uat/1560}{Spiral galaxies (1560)}; \href{http://astrothesaurus.org/uat/611}{Galaxy photometry (611)};
\href{http://astrothesaurus.org/uat/758}{Hubble constant (758)}; \href{http://astrothesaurus.org/uat/690}{H I line emission (690)}; \href{http://astrothesaurus.org/uat/902}{Large-scale structure of the universe (902)}; \href{http://astrothesaurus.org/uat/780}{Inclination (780)}; \href{http://astrothesaurus.org/uat/1464}{Sky surveys (1464)}; \href{http://astrothesaurus.org/uat/205}{Catalogs (205)}; \href{http://astrothesaurus.org/uat/395}{Distance measure (395)}; \href{http://astrothesaurus.org/uat/1935}{Random Forests (1935)};
}

\section{Introduction} \label{sec:intro}

The {\it Cosmicflows} program is an ongoing project to map the structure of the universe from departures in the motions of galaxies from the mean cosmic expansion.  Galaxies are test particles experiencing "peculiar velocities" in the line of sight due to the distribution of (mostly dark) matter: $V_{pec}=V_r-H_0 d$, where $V_r$ is the radial velocity, $d$ is the radial distance of a galaxy and $H_0$ is the Hubble constant.

Errors on individual measurements are substantial, but meaningful signals can be discerned because of coherence in the motions of adjacent systems. A robust modeling of the complexities of large-scale structure requires dense coverage of space with many thousands of accurate distance measurements.

Over the course of successive releases, {\it Cosmicflows} has expanded in distance and density coverage.  With {\it Cosmicflows-3}, the most important incrementation involved the inclusion of the Fundamental Plane measures from the Six Degree Field Redshift Survey \citep{2012MNRAS.427..245M, 2014MNRAS.445.2677S}.  This component is restricted to $\delta\le0$; hence, whereas {\it Cosmicflows-2} was relatively deficient in the south celestial hemisphere, {\it Cosmicflows-3} is heavily weighted toward coverage of the south \citep{2013AJ....146...86T, 2016AJ....152...50T}.

The next release, {\it Cosmicflows-4}, will largely redress the hemispheric imbalance.  This paper presents the most important new contribution to the forthcoming {\it Cosmicflows} update.  The methodology for obtaining galaxy distances involves the correlation between galaxy luminosities and rotation rates, known as the Tully-Fisher Relation \citep[TFR; ][]{1977A&A....54..661T}.  Thanks to the completion of the Arecibo Legacy Fast ALFA Survey (ALFALFA; \citealt{2011AJ....142..170H, 2018ApJ...861...49H}) the sky has now been covered in the declination range $0<\delta<+38$ with sufficient sensitivity to have detected many thousands of galaxies extending to $\sim 15,000$~\kms.  Concurrently, photometry over most of this same sky has been made available by the Sloan Digital Sky Survey \citep[SDSS; ][]{2015ApJS..219...12A} providing imaging in the five optical bands $u$,{\it g},{\it r},{\it i} and {\it z}.  Complementary infrared photometry is available from the all-sky observations of the Wide-field Infrared Satellite Explorer \citep[WISE; ][]{2010AJ....140.1868W}, imaging in the {\it W1} and {\it W2} (3.4 and 4.6 $\mu$m) bands.

The availability of seven-band photometry spanning a decade in wavelengths ($0.4-4\mu$m) permits a considerably refined calibration of the TFR.  In \citet{2019ApJ...884...82K} (hereafter K19) there was exploration of two important ingredients: the definition of inclinations and the properties of internal reddening as a function of inclination. Then, in \citet{2020ApJ...896....3K} (hereafter K20) the TFR was calibrated in the seven photometric bands with slopes determined from $\sim 600$ galaxies in 20 clusters and the zero-point set by 64 galaxies with Cepheid and/or tip of the red giant branch distances.
In the studies of both these papers, the calibrations benefited from information provided by such distance-independent parameters as colors, surface brightness, and relative \hi to optical-infrared fluxes.

In this paper, we use the luminosity$-$line-width correlations to calculate the distances of almost 10,000 spirals. This effort involves a mix of SDSS optical and WISE infrared photometry. We investigate discrepancies and color-dependent systematics and uncertainties that are inherent in the utilization of TFRs across multiple bands. 

\section{Data}\label{sec:data}

There are roughly power-law relations between the absolute luminosities of spiral galaxies at optical and near infrared wavelengths and their 
rotation rates probed through the width of the 21cm emission line of their neutral hydrogen (\hi) content. 
Therefore, the existence of high signal-to-noise \hi data is one of the essentials in the compilation of our catalog. 
Next, we need to know the inclinations of target spirals. Systems with spatial orientations approaching face-on are not useful, because of the ambiguity in deprojecting to full rotation rates. Finally, we need high-quality imaging data to measure apparent magnitudes and other photometric metrics.

The following conditions are adopted to initially select a set of 19,905 potential candidates, all of which have radial velocities within 15,000~\kms: (1) morphological types {\it Sa} or later; (2) inclinations estimated to be greater than $45^{\circ}$ from face-on, based on axial ratios cataloged in HyperLEDA\footnote{\url{http://leda.univ-lyon1.fr/}} \citep{2003A&A...412...45P}; (3) high-quality \hi measurements as explained in \S\ref{sec:HI}; (4) no suggestion of tidal distortion, \hi confusion, or gross anomaly.
These limitations, plus an assessment of the quality condition of the optical/infrared photometry, reduces the number of candidates to 13,434. Further pruning based on the results of our more accurate inclination measurements, described in \S\ref{sec:inclination}, leaves us with 10,737 galaxies.

\subsection{\texorpdfstring{\hi}\ Data} \label{sec:HI}

We accept the \hi line widths and fluxes from four resources: (1) Our primary source (78\% of cases) is the All Digital \hi catalog (ADHI), which has been collected over the course of the $Cosmicflows$ program \citep[A. Dupuy et al., in preparation]{2009AJ....138.1938C, 2011MNRAS.414.2005C} and is accessible online at the Extragalactic Distance Database (EDD) website\footnote{\url{http://edd.ifa.hawaii.edu}; catalog ``All Digital \hi''.}
(2) Most of the remainder (19\% of cases) are given by the Arecibo Legacy Fast ALFA Survey (ALFALFA; \citealt{2011AJ....142..170H, 2018ApJ...861...49H}) with coverage over the declination range $0<\delta<+38 $ degrees. Neutral hydrogen spectral information provided by the ALFALFA 40\% early data release is already included in ADHI but information only made available with the 100\% ALFALFA data release has not yet been ingested into ADHI. (3) In a small number of cases (50 galaxies, i.e. 0.5\% of the total cases), line widths are uniquely provided by the {\it Springob/Cornell} \hi catalog \citep{2005ApJS..160..149S}. (4) For 3\%, the source is the {\it Pre Digital \hi} catalog that is available in EDD, which provides information from early analog \hi line profiles, either from single beam or interferometric observations \citep{1981ApJS...47..139F, 1989gcho.book.....H}. Generally, this old material must be used in the cases of nearby galaxies that are much larger than the beam sizes of currently operational radio telescopes.

ADHI provides a measure of the \hi line widths, $W_{mx}$, that robustly encodes the rotation rates of spirals along the line of sight. The value of $W_{mx}$ is derived from the observable quantity $W_{m50}$, the width of the \hi line profile at 50\% of the average \hi flux within the range that covers 90\% the total \hi flux, and is adjusted for spectral resolution and redshift. 
Contributions to ADHI derive from observations with diverse facilities (those at Green Bank Observatory, Arecibo, Parkes, Nancay, and Effelsberg) but no matter which sources they are from, they are carried through our pipeline that takes account of differences in spectral resolution and smooths consistently \citep{2009AJ....138.1938C, 2011MNRAS.414.2005C}. As mentioned, ADHI includes material from the ALFALFA 40\% release, analyzed by us in our standard way.  The consequence is a large overlap in measurements of ALFALFA profiles between ADHI and those of the ALFALFA team that permits a reliable transformation of the ALFALFA 100\% line widths into the ADHI system. 

To be compatible with the ADHI $W_{mx}$ values, we transform ALFALFA line widths, $W_{alf}$, using $W_{mx}=W_{alf}-6$ \kms, which is derived for galaxies covered by both catalogs.
{\it Springob/Cornell} provides values for $W_{M50}$, that are adjusted using $W_{m50} - W_{M50} = 1.015W_{m50}-11$ \kms\ and are then converted into $W_{mx}$ values based on the ADHI standard procedure explained by \citet{2009AJ....138.1938C}. 
The {\it Pre Digital \hi} catalog provides $W_{20}$, the width at 20\% of the \hi profile maximum. We translate $W_{20}$ values to $W_{mx}$ based on the relation described by \citet{2009AJ....138.1938C}. 

In the regime of dwarf galaxies, the TFR scatter increases substantially. Such faint galaxies are only accessible nearby and are considered less useful for our purposes.  Hence, we impose a low-luminosity cutoff at $M_i=-17$.  Accordingly, we can safely reject galaxies with $W_{mx}$ less than $64$ \kms\ because such cases will inevitably lie faintward of the luminosity cut.

\hi detections of poor quality are rejected. ADHI $W_{mx}$ values are considered if their associated uncertainties are less than or equal to $20$ \kms.
We discard candidates with anomalous, confused, or low signal-to-noise line profiles.
In the cases of ALFALFA line widths, we set the threshold of $S/N>10$, which is reasonably compatible with our condition for accepting ADHI data.

To extract \hi information, we assign the highest priority to ADHI and ALFALFA catalogs and the lowest priority to {\it Pre Digital \hi} catalog. For galaxies that are listed in both ADHI and ALFALFA catalogs, we take the average of \hi flux and line width values. Out of 10,737 candidates that meet all of our requirements, 8333 spirals have \hi data in the ADHI catalog, 5120 galaxies are introduced by ALFALFA, and 3255 galaxies have \hi measurements in both catalogs. The \hi data for 236 galaxies is provided by {\it Springob/Cornell} catalog, and for 302 galaxies we use information from the {\it Pre Digital \hi} catalog.
We convert the \hi flux within the 21~cm line profiles, $F_{HI}$, given in the units Jy$\cdot$\kms, to an \hi magnitude, $m_{21}$, using $m_{21} = -2.5 {\rm log} F_{HI} + 17.40$.

\subsection{Imaging Data} \label{sec:imaging}

For the optical photometry of our galaxies, we use the {\it SDSS DR12} data release \citep{2000AJ....120.1579Y}.
For each galaxy with available SDSS data, we download all the single exposure cutouts at {\it u,g,r,i} and {\it z} bands\footnote{Our data acquisition pipeline is accessible online at \url{https://github.com/ekourkchi/SDSS\_get}}, which are drizzled and combined using {\tt MONTAGE}, an astronomical application to assemble images \citep{2010ascl.soft10036J}. Our pipeline provides galaxy cutouts at all {\it ugriz} passbands with a spatial resolution of 0.4'' pixel$^{-1}$.
For the infrared part, we obtain the {\it W1} (3.4$\mu m$) and {\it W2} (4.6$\mu m$) images of the WISE survey \citep{2010AJ....140.1868W}, from the NASA/IPAC infrared science archive (IRSA). We generate the cutouts of galaxy images by drizzling single exposure frames using version 3.8.4 of the Image Co-addition with Optional Resolution Enhancement ({\tt ICORE}) software \citep{2009ASPC..411...67M, 2013ascl.soft02010M}. Our final co-added infrared images have a spatial scale of 1'' pixel$^{-1}$. All of our optical and infrared images are calibrated to produce magnitudes in the AB system. For more detailed information on how we conduct our image preprocessing, please refer to \S 2.2 and 2.3 of K19.

\begin{figure*}
\centering
\includegraphics[width=0.52\linewidth]{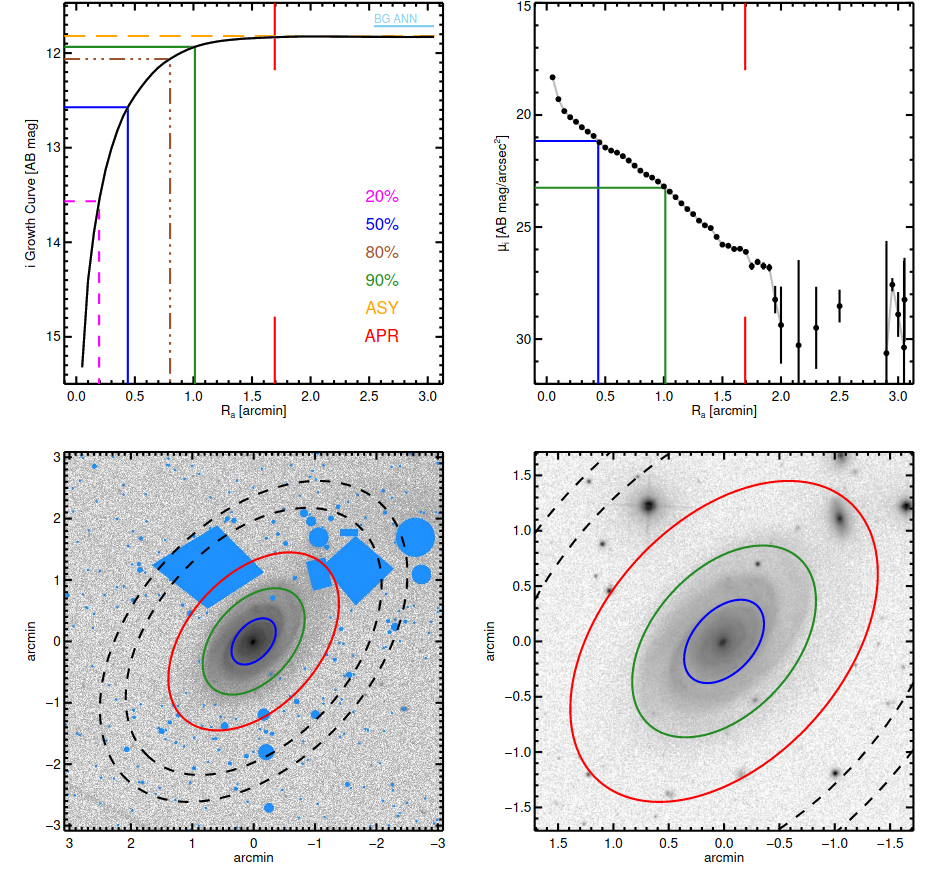}
\hspace{5mm}
\includegraphics[width=0.42\linewidth]{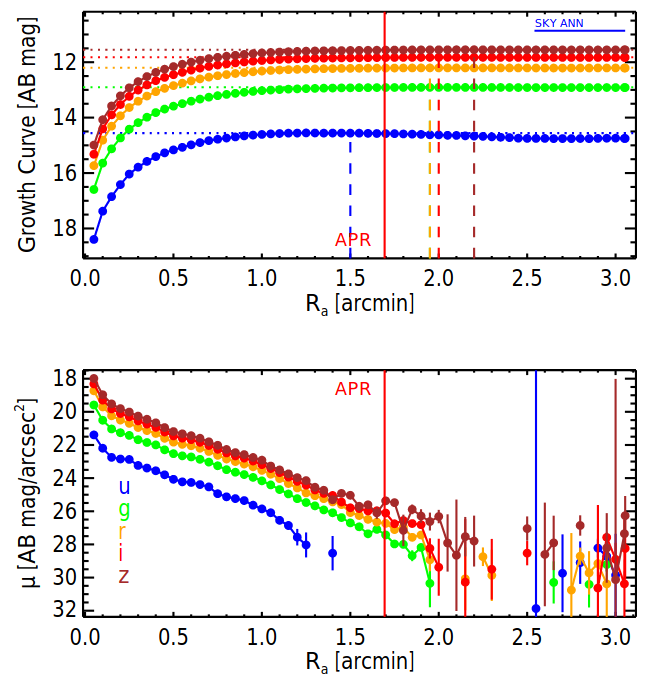}
\caption{Example photometry results for the galaxy NGC 881 (PGC 8822) at optical SDSS bands. 
Four panels on the left show the photometry results at {\it i} band. Top left panel plots the the growth curve of {\it i}-band luminosity, calculated within concentric elliptical apertures with semimajor axes of $R_a$. Top middle panel displays the evaluated average surface brightness as a function of $R_a$. 
Bottom left/middle panels show the galaxy cutout image, with red ellipse being the chosen photometry aperture and black dashed annulus being the region of sky background determination. Blue and green colors represent apertures enclosing 50\% and 90\% of the total galaxy light, respectively. The blue patches on the bottom left panel image show the masked regions.
Top panel on the right illustrates the luminosity growth curves at all SDSS {\it ugriz} bands, color-coded differently, and on the bottom right we have the corresponding surface brightness profiles. Vertical solid red line marks the photometry aperture, $R_a$, and dashed vertical lines are drawn where growth curves become asymptotically flattened.
\label{fig:NGC0881:OP}}
\end{figure*}

\subsubsection{Photometry} \label{sec:photometry}

For the surface photometry of our galaxies, we use the photometry pipeline that was originally developed to assemble the WISE Nearby Galaxy Atlas (WNGA; M. Seibert et al. 2020, in preparation).
We added flexibility to the WNGA pipeline and improved the efficiency of its user interface, providing easily accessible tools that facilitate the manual procedures required in our photometry program.
In the photometry process, the galaxy light profile is derived within concentric elliptical apertures, with geometrical information such as the center, size, and axial ratio initially taken from the HyperLEDA\footnote{\url{http://leda.univ-lyon1.fr/}} \citep{2003A&A...412...45P} catalog.
The aperture is later repeatedly adjusted either by visual inspections, or with the aid of {\tt SExtractor} \citep{1996A&AS..117..393B} and/or the galaxy isophots visualized by {\tt DS9} \citep{2003ASPC..295..489J}.

The sky background is evaluated within a large annulus far from the photometry aperture. All foreground stars are initially masked automatically---however, further manual masking is required for companion extended objects, point sources that are not automatically recognized, and other features such as diffraction spikes. On the other hand, the software tends to mask the blue star forming clumps in spiral arms, which needed to be unmasked. The resulting light curve is calculated radially in increments of 3'' for {\it u,g,r,i,z} bands and 6'' for {\it W1} and {\it W2} bands. 

The quality of the generated light profiles and growth curves are visually inspected. If necessary, further masking/unmasking and adjustments of the aperture and the background estimation annulus are applied iteratively until the growth curve converges.
At the end of each iteration, the resulting luminosity growth curve and surface brightness profile is evaluated for abnormal changes in luminosity due to unmasked objects or poor subtraction of the sky level. 
 
\begin{figure*}
\begin{minipage}{1.\textwidth}
\centering
\includegraphics[width=0.52\linewidth]{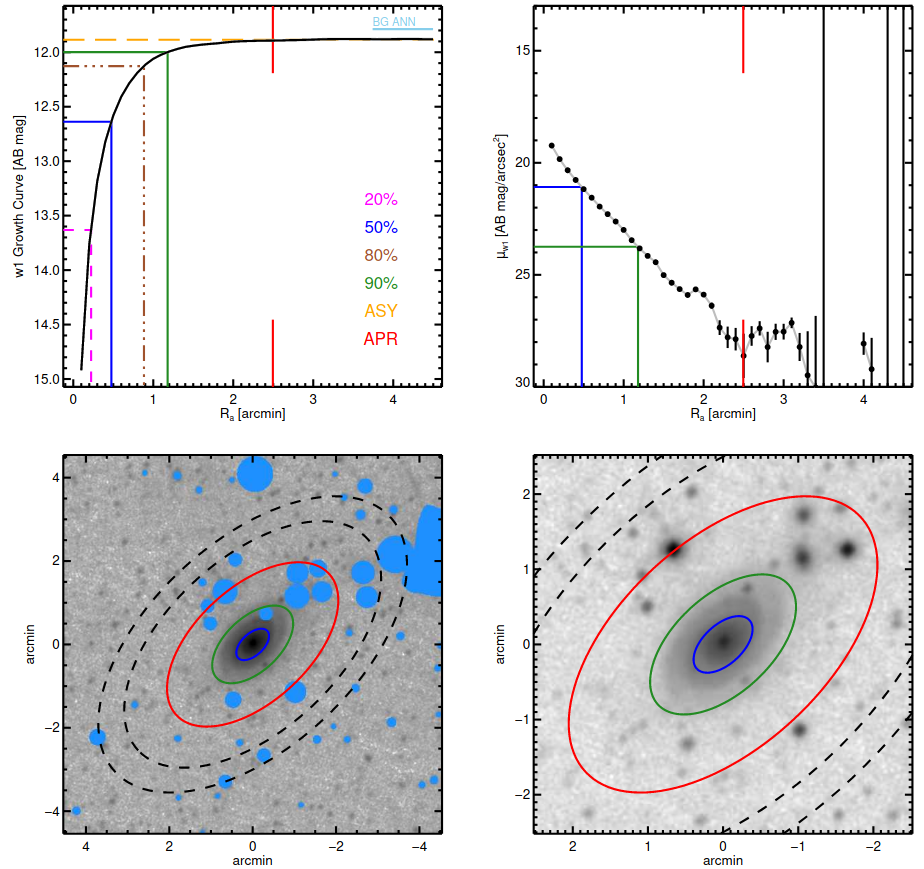}
\hspace{5mm}
\includegraphics[width=0.42\linewidth]{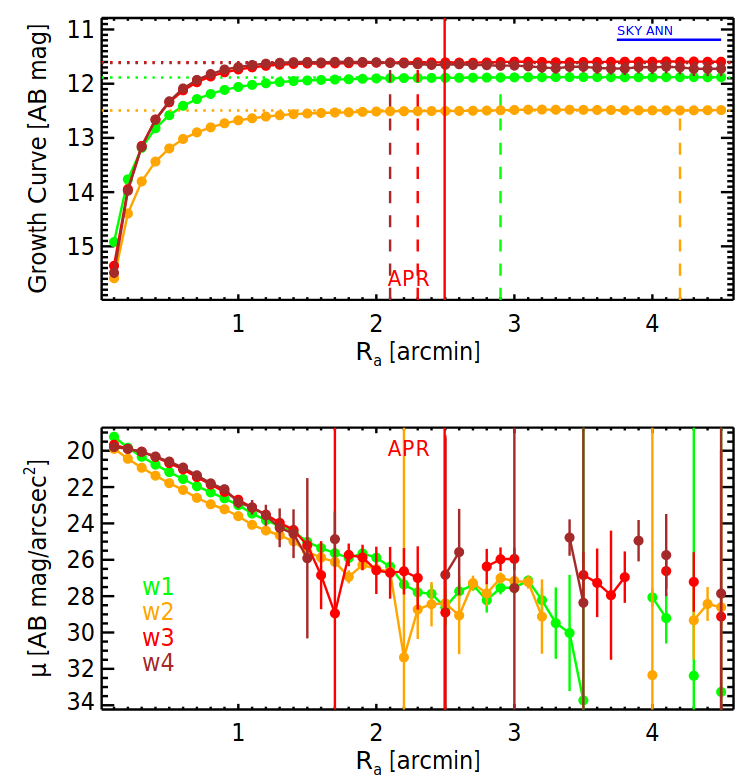}
\caption{Photometry results for NGC 881 at infrared {\it W1} and {\it W2} bands. Details are similar to Fig. \ref{fig:NGC0881:OP}.
\label{fig:NGC0881:IR}}
\end{minipage}
\end{figure*}

In some cases, the galaxy image is superimposed with many other sources in crowded regions, such as near the Galactic plane, which makes masking and sky-subtracting very challenging. 
Furthermore, pollution by unresolved background objects may affect our evaluated sky level. 
We attempt to address these complexities by means of visual inspections and playing with the masking thresholds. 
Tackling this issue is simpler for our SDSS cutouts, owing to their high spatial resolution (0.4'' pixel$^{-1}$) and the minimal sky level that remains from the initial SDSS sky subtraction.
The large-resolution elements of the WISE images ($\sim$6") make it more difficult to accurately evaluate the sky background. 
At infrared bands, our standard routine produces reasonable results. However, for galaxies with a nonconvergent curve of growth, we need to interactively alter the background level by a few percent of the initial estimations in order to eliminate surface brightness anomalies and force the curve of growth to converge. 

As examples, Figures \ref{fig:NGC0881:OP} and \ref{fig:NGC0881:IR} display photometry apertures, masks, and light curves for NGC 881 (PGC 8822). Our photometry results, including light curves and all measured quantities, are publicly available online through the {\it EDD}\footnote{
To query the results of our SDSS photometry, go to \url{http://edd.ifa.hawaii.edu/cf4_photometry/get_sdss_cf4.php}, and enter the galaxy name or its PGC number.
In a similar fashion, you can access to our WISE photometry results at \url{http://edd.ifa.hawaii.edu/cf4_photometry/get_wise_cf4.php}.
}.

Our photometry pipeline calculates two versions of ``total magnitude''. 
(1) The asymptotic magnitude is derived within the aperture radius, beyond which the curve of growth is flat (the horizontal dotted lines in the top right panel of Figures \ref{fig:NGC0881:OP} and \ref{fig:NGC0881:IR}).
Asymptotic radii are robust parameters that are independent of the user's choice of aperture. 
(2) An isophotal magnitude calculated within 25.5 mag arcsec$^{-2}$ with augmentation calculated from extrapolating the extension of an exponential fit of the galaxy disk to infinity \citep{1996AJ....112.2471T, 2014ApJ...792..129N}. We found that the average discrepancy between these two types of magnitude is no worse than $0.02$ mag in all bands for the brightest objects, and it always remains below $0.05$ mag for fainter galaxies at all passbands---except for the {\it u} band, which has lower quality ($\sigma \sim 0.1$ mag). 
Our final results are insensitive to the magnitude choice, and we choose to use asymptotic magnitudes that are derived with no assumptions about galaxy type. 

\begin{figure*}[t]
\centering
\includegraphics[width=\linewidth]{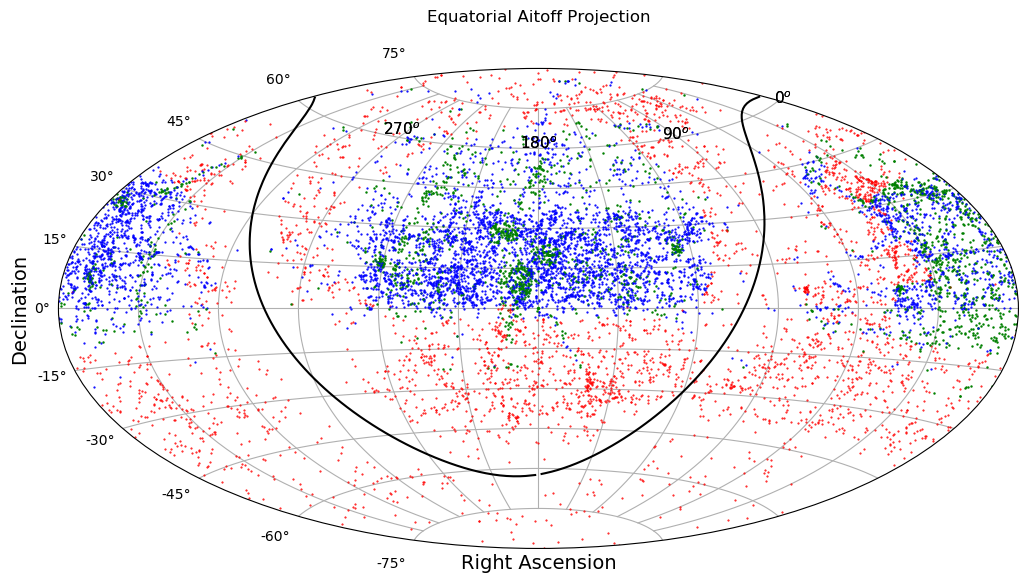}
\caption{
Aitoff equatorial projection of spiral galaxies in this study. 
Red points are spirals with only WISE photometry (3234 galaxies). Blue points represent spirals that only have photometry information from the SDSS imaging (5258 galaxies), and green points have photometry coverage from both SDSS and WISE (2244 galaxies). Black solid curve is the projection of the Milky Way plane.}
\label{fig:aitoff_equatorial}
\end{figure*} 
\subsubsection{Adjusting apparent magnitudes} \label{sec:mag:adjustment}

At any passband $\lambda$, the raw total apparent magnitude of each galaxy, $m^{total}_{\lambda}$, is corrected to account for dust obscuration in the Milky Way, $A^{\lambda}_b$, and dust attenuation in the target galaxy along the line of sight, $A^{\lambda}_i$, and the effect of spectral redshift on the galaxy luminosity at each band ($k$-correction), $A^{\lambda}_k$. The adjusted magnitude is given as
\begin{equation}
\label{Eq:bkai}
m^*_\lambda = m^{(total)}_{\lambda} - A^{\lambda}_b - A^{\lambda}_k  - A^{\lambda}_a - A^{\lambda}_i.
\end{equation}
Here, $A^{\lambda}_a$ is an adjustment that accounts for the diffuse scattered light from extended objects (galaxies) lost from the fixed-size apertures used in establishing the photometric calibration with point sources (stars).
There is a detailed discussion of the derivation of $A^{\lambda}_b$, $A^{\lambda}_k$, and $A^{\lambda}_a$ in \S 2.5 of K19.

In K19, it is shown that the amplitude of dust obscuration in the host galaxy, $A^{\lambda}_i$, can be modeled as a function of (1) the galaxy physical properties that are probed by the galaxy observables and (2) the galaxy spatial inclination. The description of $A^{\lambda}_i$ is given as
\begin{equation}
\label{eq:dust}
    A^{\lambda}_i=\gamma_{\lambda} \mathcal{F}_{\lambda}(\dot{\imath})~,
\end{equation}
where $\mathcal{F}_{\lambda}(i)$ is a monotonically increasing function of the inclination angle from face-on, {\it i}\footnote{Symbol {\it i} should not to be confused with the photometric passband.} (see \S \ref{sec:inclination} for the full discussion on the measurement of inclinations), described by 
\begin{equation}
\label{Eq:Fli}
\mathcal{F}_\lambda(i)={\rm log}\Big[\cos ^2(i)+q_\lambda^2\sin^2(i)\Big]^{-1/2} ~,
\end{equation}
where $q_\lambda$ is a wavelength-dependent parameter of the obscuration model.
The dust attenuation factor, $\gamma_{\lambda}$ is calculated using a third-degree polynomial function of $P_{1,W2}$, the main principal component of the galaxy observables, constructed based on the \hi line width adjusted for inclination using $W^i_{mx}=W_{mx}/{\rm sin}(i)$, a pseudo-color calculated based on the \hi 21 cm and {\it W2} magnitudes, $C_{21W2}=m_{21}-\overline{W}2$, and the average surface brightness of galaxy at {\it W2} band within the effective radius that is corrected for the geometric effect of inclination, $\langle \mu_2 \rangle^{(i)}_e$.
The relation to calculate $P_{1,W2}$ is expressed as 
\begin{equation}
\label{Eq:P1}
\begin{split}
    P_{1,W2} = 0.524({\rm log}W^i_{mx}-2.47)/0.18 \\
    + 0.601(C_{21W2}-1.63)/1.15 \\
    - 0.603(\langle \mu_2 \rangle^{(i)}_e-23.35)/1.38 ~ ,
    \end{split}
\end{equation}
and $\langle \mu_2 \rangle^{(i)}_e$ is given by 
\begin{equation}
\label{Eq:Mu50_i}
\langle \mu_2 \rangle^{(i)}_e=\langle \mu_2 \rangle_e+0.5 \log_{10} (a/b)~, 
\end{equation} 
where $a$ and $b$ are the semimajor and semiminor axes of the photometry aperture. It should be kept in mind that the galaxy effective surface brightness is derived from its total magnitude and the effective radius, $R_{\lambda e}$, following
\begin{equation}
\label{Eq:Mu50_e}
\langle \mu_\lambda \rangle_e = m^{(total)}_{\lambda} +2.5 {\rm log_{10}}(2\pi R^2_{\lambda e}) ~.
\end{equation} 
There is a full discussion in K19 regarding the calculation of $A^{\lambda}_i$.

A problem arises if there is missing information.  SDSS photometry is only available across part of the sky. WISE photometry is available, in principal, across the entire sky, but the time-consuming effort to acquire WISE photometry for all potential targets has not been completed. In the Appendix \ref{dust:prediction} there is discussion of a predictive algorithm that provides acceptable substitutions for calculations of attenuation.

Figure \ref{fig:aitoff_equatorial} illustrates the distribution of our galaxies color coded based on their photometry coverage. The ALFALFA survey running $0<\delta<+38$ excluding the zone of avoidance, strongly overlaps with the SDSS footprint. 

\begin{figure*}
\centering
\includegraphics[width=0.80\linewidth]{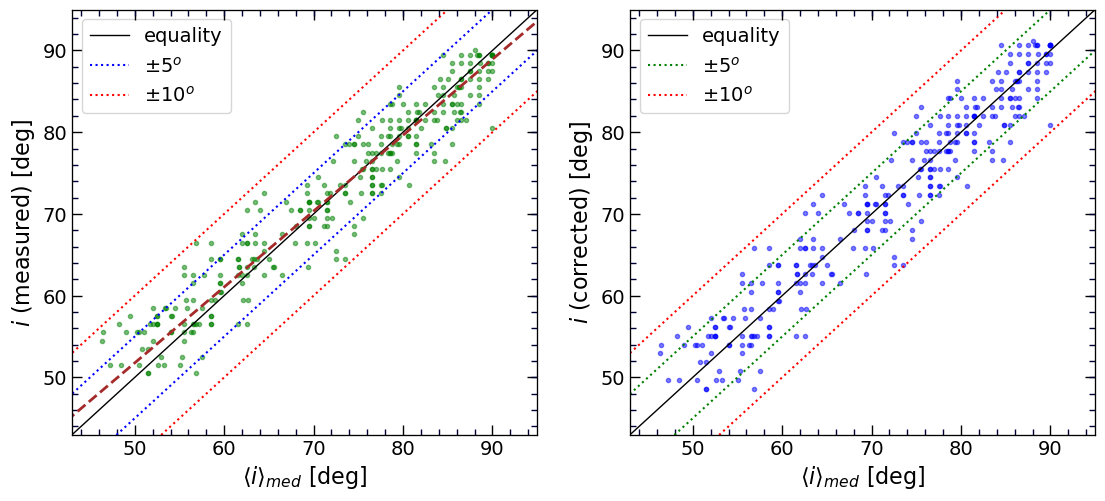}
\caption{
Evaluated inclinations by an user versus the median of all measurements given by all users, $\langle i \rangle_{med}$. Each point represents a galaxy. 
Black line shows the locus of equality of the values. Blue and red dotted lines illustrate deviations from equality by $\pm5$\dg and $\pm10$\dg, respectively. {\bf Left:} green points are the unadjusted inclinations reported by the user.
Maroon dashed line fits the green points found by minimizing the mean squared of the residuals along the vertical axis.
{\bf Right} corrected inclinations using the linear fit presented in the left panel.
}
\label{fig:inc_corrections}
\end{figure*} 

\subsection{Inclination} \label{sec:inclination}

Inclinations are a significant source of error in the TFR.  Inclinations enter into the determinations of both magnitudes and line widths. The lesser problem is with magnitudes; the issues in that regard are discussed in \S \ref{sec:mag:adjustment} and Appendix \ref{dust:prediction}.  Corrections to line widths introduce greater uncertainties and potential systematics.

If the image of a spiral galaxy is the projection of a perfect disk with an oblate spheroidal shape, then the inclination, {\it i}, can be derived from the axial ratio, $b/a$, following the formulation ${\rm cos}^2~i=[(b/a)^2-q_o^2]/(1-q_o^2)$, 
where $q_o$ is the thickness of the spiral disk as observed edge-on. 
The choice of $q_o=0.2$ has been used in previous studies \citep{2013AJ....146...86T}.
We adopted the analogous relation in \S 4 of K19 to formulate the internal dust attenuation of spiral galaxies, where we let $q_o$ be a free parameter that depends on the passband. In that formulation, $q_o$ is not constant, but instead is treated just as a hyperparameter without direct geometrical implications.

Ellipticity-derived inclinations can be misleading for various reasons.
The disks of spirals might be warped, or axially asymmetric due to tidal interactions. 
The bulges of large spirals dominate their disks, inflating the observed ratio $b/a$, especially at redward passbands.
Extreme contrast in the surface brightness of disk components, such as bars, arms, and irregularities, can alter the shape of the photometry aperture and lead to large errors.  Otherwise, confusion can simply originate from the nontrivial orientation of strong spiral components relative to the major axis of projection on the sky (spiral arms opening onto the minor vs the major axis). Previously, the statistical determinations of the inclinations of large samples of spirals have been unsatisfactory.  

In this study, we pursue a different approach that relies on the judgment of the human eye to evaluate galaxy inclinations. 
Fortunately, a large fraction of spirals have well-defined inclinations defined by their axial ratios.
These good cases provide a baseline with sufficient coverage of morphological types and inclinations over the range of $45^{\circ}-90$\dg that concerns us.

The details of our procedures are described in \S 2.5 of K19, where there is a discussion of the online graphical tool, {\it Galaxy Inclination Zoo (GIZ)}\footnote{\url{http://edd.ifa.hawaii.edu/inclination/index.php}}.  Users of the interface are asked to situate a target galaxy within a lattice of galaxies with established inclinations.  The interface was initially validated by two of us (E.K. gave attention to the entire sample and R.B.T. looked at about half).  The site was then opened up to colleagues and students, with initially intensive and progressively more relaxed training.

\begin{figure*}
\centering
\includegraphics[width=0.45\linewidth]{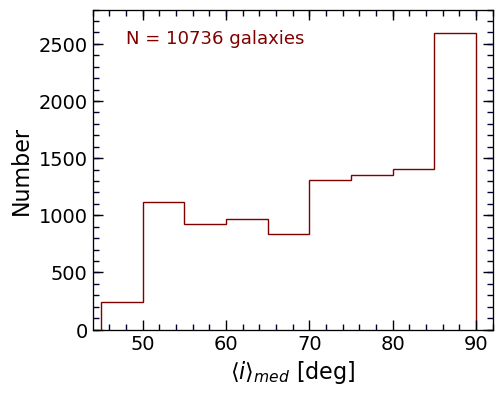}
\includegraphics[width=0.45\linewidth]{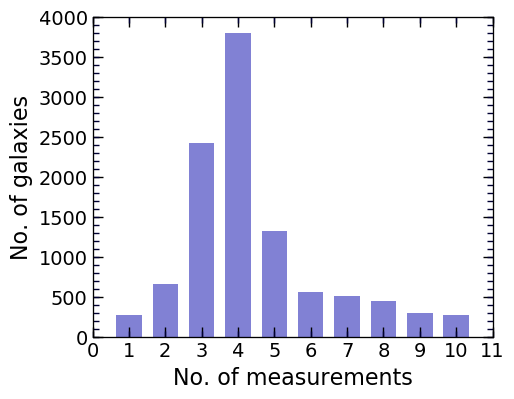}
\caption{
{\bf Left:} histogram of the evaluated inclinations for 10,737 galaxies that turned out to be more edge-on than $45$\dg.
{\bf Right:} histogram of the number of measured inclinations by different users.
}
\label{fig:inc_hist}
\end{figure*} 

\subsubsection{Evaluating User's Measurements} \label{sec:evaluation}

Initially, the performance of the tool was evaluated, for a significant fraction of galaxies, by two of the authors.
First, E.K. and R.B.T. gave careful attention to more than $\sim$2000 spirals involved in the sample used for the study of the dust attenuation of K19 and the TF calibrators study of K20. The results are in agreement, with root-mean-squared (rms) differences between E.K. and R.B.T. of $3.3$\dg with no meaningful systematics. 
Subsequently, E.K. used the tool to measure the inclinations of our $\sim$20,000 tentative candidates. 

In order to improve the accuracy of our final estimations, we opened the tool to 10 different users, including graduate and undergraduate students. We provided them with multiple sets of galaxies with sufficient overlaps that allowed us to assess the consistency of the measurements and correct for user-dependent systematics. Later, we opened the online tool to a larger number of citizen scientist and amateur astronomers across the world, whose participation helped us to improve our number statistics and efficiency.

The downside of adding the multiple measurements of different users is the possibility of introducing biases due to human mistakes. 
To remove the user-driven systematics, we build up our compendium of good measurements by adding the measured inclinations of each user individually.
In the beginning, we adopt the measurements of E.K. and R.B.T. as acceptable inclinations, and for each galaxy, we use the median of the measured values as a benchmark to assess new measurements.
For the results of a user to be accepted, we first compare with the median of the previous good measurements in our ensemble for galaxies in common. In a case of large discrepancies, we reject all the user measurements. Small discrepancies are modeled by fitting linear relations, allowing us to correct the new measurements prior to integrating them into our ensemble. As an example, Fig.~\ref{fig:inc_corrections} plots the results of a user for 310 galaxies versus the median value of all inclinations measured by different users except for the chosen user. In the left panel of this figure, it is seen that the raw measurements are clearly overestimated at the face-on end.
The dashed line shows our linear fit to this inclination-dependent bias, used to produce the corrected inclinations shown in the right panel. While users continue their work, the number of cross-evaluations increases, and so the number of reliable measurements does as well. We iteratively modify the linear-correcting relations, until reaching convergence. We also acknowledge the differences in the users' performances by considering weighted numbers toward the calculation of the median inclinations. We use integer weight numbers that are no greater than 4. We derive weight number in an iterative process based on user experience, the level of needed corrections (according to the slope of the fitted line), and/or the rms scatter of differences between raw measurements and the median values. 

The left panel of Fig. \ref{fig:inc_hist} plots the distribution of the median evaluated inclinations for all the 10,737 accepted spirals in our program. There is an excess in the most edge-on bin because, in the portion of the \hi sample derived from individual pointings (the non-ALFALFA portion), edge-on systems have been favored.
The distribution of the number of measured inclinations for these spirals is illustrated in the right panel of Fig. \ref{fig:inc_hist}.
We ended up with less than three measurements in 925 cases, as users (other than E.K. and R.B.T.) rejected galaxies for invalid reasons.

\begin{figure}[t]
\centering
\includegraphics[width=0.950\linewidth]{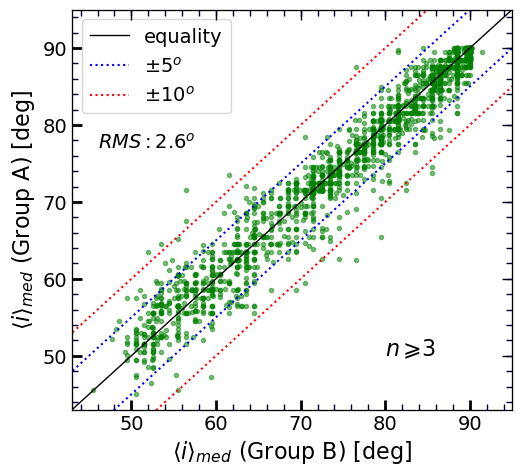}
\caption{
Median of the evaluated inclinations by two different groups of users for $\sim$1800 galaxies. Each point represents a galaxy with at least three independent measurements by each group. The rms of deviations from equality is $2.6$\dg. Other details are similar to those of Figure \ref{fig:inc_corrections}. Each group consists of 15 users, and $n$ is the number of measurements.
}
\label{fig:inc_groupAB}
\end{figure}

Increasing the number of individual measurements for each galaxy improves the results of the evaluated inclinations. Nevertheless, the manual evaluation is a tedious task and requires long hours of visual inspections by multiple users. 
In our program, we required each galaxy to be evaluated by three different users.  
To quantify the consistency of the final results, we randomly divide all participants in two groups, A and B, in such a way that the total number of measurements done by each group is almost the same. 
Fig. \ref{fig:inc_groupAB} compares the median evaluated inclinations of the galaxies inspected by both groups.
Each galaxy has at least three separate measurements by the users of each group.
The $RMS$ scatter of discrepancies across the entire range of inclinations is better than $3$\dg and no inclination dependent systematic is apparent. 

\begin{figure}[t]
\centering
\includegraphics[width=1.0\linewidth]{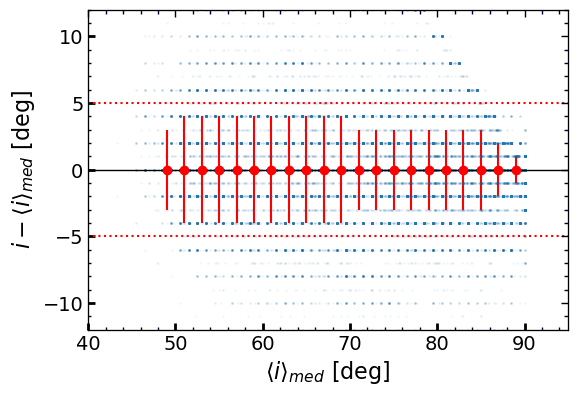}
\caption{
Differences between individual evaluated inclinations, {\it i}, and the median of all measurements, $\langle i \rangle_{med}$. Each blue point represents a single estimated inclination for a galaxy. Red dotted horizontal lines are drawn at the level of $\pm5$\dg deviations. Red filled circles and their error bars illustrate the median and $1\sigma$ standard deviations of the blue points within the $1$\dg
bins. }
\label{fig:inc_scatter}
\end{figure}

\begin{figure*}
\centering
\includegraphics[width=0.85\linewidth]{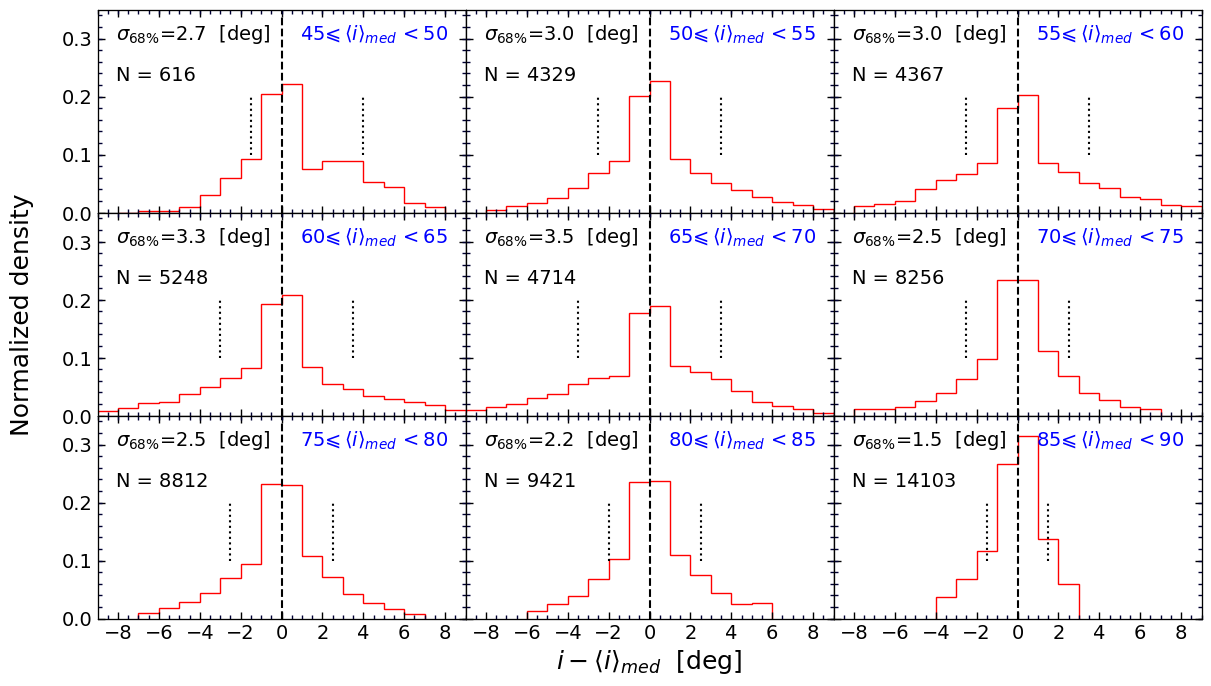}
\caption{
Distribution of the deviations of individual evaluations of inclinations, {\it i}, from the median value of all measurements, $\langle i \rangle_{med}$. Each panel covers a $5$\dg interval of inclination, and $N$ represents the number of individual measurements used in each panel. Vertical dotted lines identify the exclusion of $34$\% of data points on opposite sides of the histograms.  
}
\label{fig:inc_scatter_hist}
\end{figure*} 

\subsubsection{Uncertainties} \label{sec:uncertainties}

Fig. \ref{fig:inc_groupAB} reveals that the scatter of deviations from the equality line is smaller for the edge-on galaxies and it increases toward the face-on spirals. To quantify the average uncertainty of the users measurements as a function of inclination, Fig. \ref{fig:inc_scatter} plots the deviation of all user-evaluated inclinations for all individual galaxies from the median estimated inclinations. In this figure, each blue point represents one galaxy whose inclination $i$ is evaluated by one user, with $\langle i \rangle_{med}$ being the median of all measurements for that galaxy. 
In this analysis, we assume that the $\langle i \rangle_{med}$ values are the correct estimations of inclination, and hence we use them as a reference to evaluate the level of uncertainty introduced by users. 
In Fig. \ref{fig:inc_scatter}, red filled points display the median and 1$\sigma$ standard deviation of blue points within $1$\dg bins of $\langle i \rangle_{med}$. In agreement with our observation in Fig. \ref{fig:inc_groupAB}, the scatter of the users' measurements is small at large inclinations and gradually increases as the inclination angle decreases. For spirals more edge-on than 88\dg, the user-evaluated inclinations are consistent within 1\dg error bars, whereas for galaxies with inclinations more face-on than 50\dg, the scatter of the measured inclinations is $\sim 4$\dg. 

Providing a more detailed examinations, Fig.~\ref{fig:inc_scatter_hist} plots the distribution of measurement discrepancies, $i - \langle i \rangle_{med}$, within inclination intervals of 5\dg.  In each panel of Fig. \ref{fig:inc_scatter_hist}, the black vertical dotted lines give the boundaries that exclude 34\% of the measurements at each extremity from the median, which is zero by definition. 

Except for the top left panel of this figure, $45^{\circ} \leq \langle i \rangle_{med} < 50^{\circ}$, the other distributions look almost symmetrical. Therefore, we calculate $\sigma_{68\%}$ by averaging the 1$\sigma$ right and left wings of histograms. 

The inferred inclination-dependence of user measurements from Figures \ref{fig:inc_scatter} and \ref{fig:inc_scatter_hist} are consistent. 
For spirals that are more edge-on than 88\dg, the error is about 1\dg. It is no worse than $\sim$2\dg for inclinations larger than 85\dg. The ambiguity is about 3\dg for measurements in the range 70\dg$-$85\dg, and it is no worse than 4\dg for spirals that are more edge-on than 50\dg. Our error evaluation for inclinations between 45\dg and 50\dg is not robust, owing to the small number of galaxies in our sample and the consequently small number of measurements in this range. We adopt a conservative uncertainty of 5\dg for spirals more face-on than 50\dg. We adopt these uncertainty values as a {\it minimum floor} for errors of reported inclinations. For an individual galaxy, the quoted error is the larger of these minimum values and the standard deviation of all averaged measured inclinations for that galaxy.

\begin{figure}
\centering
\includegraphics[width=1.0\linewidth]{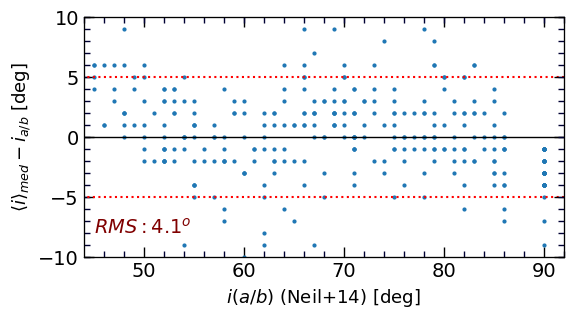}
\caption{
Difference of the median measured inclinations in this study, $\langle i \rangle_{med}$, and those derived based on axial rations, $i_{a/b}$. Red dotted lines identify the $\pm5$\dg region. The rms of deviations equals $4.1$\dg.
\label{fig:inc_literature}}
\end{figure}

In Fig. \ref{fig:inc_literature}, using a subset of common galaxies, we compare our derived inclinations with those tabulated in \citet{ 2014ApJ...792..129N} that are derived based on axial ratios. Both measurements seems to be in good agreement on average, within an rms uncertainty of about $4$\dg. However, as discussed earlier, axial ratios are not good inclination estimators, as they might be subject to morphological peculiarities. Thus, there is no expectation that the two different methods yield exactly the same results.

Our collection of carefully measured inclinations provides a rich data set for training a machine-learning algorithm, such as the Convolutional Neural Network (CNN) \citep{cccfa4f7238441b0a9021bb9f917e8ed, NIPS2012_4824}, to replace the human eye in the future projects. To successfully instruct such a network to produce satisfactory results, a training set of order $10^4$ representative galaxies is required. Our entire sample is of such a size, and hence suitable for exploring machine-learning capabilities. Moreover, $n$-body cosmological simulations such as {\it Illustris} \citep{2015A&C....13...12N} provide exquisite images of projected spiral galaxies with known 3D orientations that could be of potential interest as training sets for inclination studies.

\subsection{Data Catalog}

Table \ref{tab_data} gathers together directly observed and adjusted data and inferred parameters for 10,737 spiral galaxies used in this study. 
Columns provide the following information. 

(1) The ID number of the galaxy in the Principal Galaxy Catalog (PGC; \url{http://leda.univ-lyon1.fr/}). 

(2) Common name. 

(3) The measured inclination of the galaxy in degrees with error (see \S \ref{sec:inclination}). 

(4) Heliocentric velocity from \hi observations.

(5) The \hi line width adjusted to approximate twice the maximum rotation velocity, $W_{mx}$, with error, in \kms. 

(6) The logarithm of the inclination-corrected \hi line width, calculated from $W^i_{mx}=W_{mx}/\sin(i)$, where {\it i} is the inclination angle presented in column~3, with error. 

(7) The \hi 21 cm magnitude calculated from the \hi flux, $F_{HI}$, using $m_{21}=-2.5 \log F_{HI}+17.4$, with error. 

(8-12) The {\it SDSS u,g,r,i} and {\it z} total raw magnitudes in the AB system. 

(13-14) The WISE {\it W}1 and {\it W}2 total raw magnitudes in the AB system. 

(15) The $b/a$ axial ratios of the elliptical photometry apertures used for the photometry of SDSS images, where $a$ and $b$ are the semi-major and semi-minor axes of the elliptical aperture, respectively. 

(16) Analogous to 15 but for the photometry apertures of the WISE images. 

(17-23) The semi-major axes of apertures that enclose half the total light of galaxies at optical/infrared bands in arcmin.

(24-30) Optical/infrared magnitudes corrected for Milky Way obscuration, redshift $k-$correction, and aperture effects, based on the corresponding raw magnitudes listed in columns (8-14). 

(31-37) The dust attenuation corrections at optical/infrared bands, calculated from Equation\ref{eq:dust}, with errors. 

(38-44) The magnitudes after correcting for the effect of global dust obscuration, 
$m^*_\lambda = m^{(total)}_{\lambda} - A^{\lambda}_b - A^{\lambda}_k  - A^{\lambda}_a - A^{\lambda}_i$.

(45) The predicted value for the {\it W2}-band magnitude based on the optical photometry and other observable features of the galaxy (discussed in the Appendix \ref{dust:prediction}). 

(46-52) Host dust attenuation factors as defined in Equation~\ref{eq:dust}, with blank entries for galaxies with missing infrared photometry. 

(53-59) The predicted dust extinction factors calculated using $W_{2p}^*$, the predicted {\it W}-band magnitude defined in Appendix \ref{sec:optimization}.

(60-66) The predicted dust extinction of hosts corresponding to columns (52-58).

(67) The main principal component based on the {\it W}2-band photometry calculated as given by Equation~\ref{Eq:P1}, with error. 

(68) The main principal component derived based on the predicted value for {\it W}2-magnitude. 

(69) The main principal component derived for the {\it W}1 band. This parameter is useful for galaxies with poor {\it W}2-band photometry. Here, $P_{1,W2}$ is approximated by $1.021 P_{1,W1}-0.094$ (K19).

(70-71) The manually assigned quality for the photometry of SDSS and WISE images. The quality grade ranges from $0$ for the poorest quality (or missing data) to $5$ for the best quality. 

(72) An integer number between 0 and 2 that specifies what parameters are used for calculation of the dust obscuration in the host galaxies ($0$ if the predicted $W_{2p}^*$ is used, $1$ if the observed {\it W}1-band magnitude is used and $2$ if the $W$2-band magnitude is used through the standard attenuation formulation as discussed in \S \ref{sec:mag:adjustment}).


\begin{turnpage}
\setlength{\tabcolsep}{0.1cm}
\begin{deluxetable*}{rrcccccccccccccccccccccc}
\tablecaption{Data Catalog\tablenotemark{*}
\label{tab_data}}
\tabletypesize{\scriptsize}
\tablehead{ \\
PGC & 
Name & 
$Inc.$ &
$V_{h}$ & 
$W_{mx}$ & $\log (W^i_{mx})$ & $m_{21}$ & 
$u$ & {\it g} & {\it r} & 
{\it i} & {\it z} & 
{\it W1} & {\it W2} & 
$(b/a)_S$ & $(b/a)_W$ &
$R_{eu}$ &
$R_{eg}$ &
$R_{er}$ &
$R_{ei}$ &
$R_{ez}$ &
$R_{eW1}$ &
$R_{eW2}$
\\
 &  & 
(deg) & (\kms) &
(\kms) &  &
(mag) & (mag) & (mag) & 
(mag) & (mag) & (mag) & 
(mag) & (mag) &
 &  & 
($'$) & ($'$) & ($'$) & 
($'$) & ($'$) & ($'$) & 
($'$) 
\\
(1) & (2) & (3) & 
(4) & (5) & (6) & 
(7) & (8) & (9) & 
(10) & (11) & (12) & 
(13) & (14) &
(15) & (16) &
(17) & (18) & (19) &
(20) & (21) &
(22) & (23)
}
\startdata
2 & UGC12889 & 52$\pm$4 & 5004 & 437$\pm$16 & 2.744$\pm$0.029 & 15.713$\pm$0.185 &   &   &   &   &   & 11.86 & 12.49 &   & 0.66 &   &   &   &   &   & 0.51 & 0.51 \\ 
4 & PGC000004 & 85$\pm$2 & 4458 & 154$\pm$5 & 2.189$\pm$0.014 & 16.688$\pm$0.076 & 17.57 & 16.43 & 15.91 & 15.57 & 15.42 & 16.06 & 16.42 & 0.24 & 0.33 & 0.25 & 0.23 & 0.21 & 0.23 & 0.27 & 0.27 & 0.33 \\ 
12 & PGC000012 & 82$\pm$3 & 6548 & 400$\pm$19 & 2.606$\pm$0.021 & 16.071$\pm$0.185 &   &   &   &   &   & 13.60 & 14.20 &   & 0.36 &   &   &   &   &   & 0.25 & 0.25 \\ 
16 & PGC000016 & 65$\pm$4 & 5667 & 297$\pm$14 & 2.515$\pm$0.025 & 17.207$\pm$0.185 & 16.08 & 14.83 & 14.21 & 13.86 & 13.61 & 13.98 & 14.67 & 0.52 & 0.64 & 0.22 & 0.21 & 0.19 & 0.19 & 0.19 & 0.21 & 0.20 \\ 
55 & UGC12898 & 80$\pm$3 & 4779 & 179$\pm$10 & 2.260$\pm$0.025 & 15.757$\pm$0.076 & 16.82 & 16.08 & 15.71 & 15.50 & 15.40 & 16.18 & 16.60 & 0.32 & 0.46 & 0.28 & 0.23 & 0.23 & 0.22 & 0.22 & 0.22 & 0.26 \\ 
68 & ESO538-017 & 57$\pm$4 & 7664 & 206$\pm$18 & 2.390$\pm$0.043 & 16.883$\pm$0.185 & 16.31 & 15.24 & 14.86 & 14.69 & 14.49 & 14.99 & 15.49 & 0.66 & 0.66 & 0.16 & 0.17 & 0.16 & 0.15 & 0.15 & 0.19 & 0.19 \\ 
70 & UGC12900 & 90$\pm$1 & 6800 & 433$\pm$5 & 2.636$\pm$0.005 & 15.052$\pm$0.076 & 16.70 & 15.41 & 14.66 & 14.25 & 13.87 & 13.55 & 14.08 & 0.08 & 0.14 & 0.61 & 0.55 & 0.51 & 0.48 & 0.46 & 0.43 & 0.43 \\ 
76 & UGC12901 & 68$\pm$4 & 6920 & 390$\pm$5 & 2.624$\pm$0.013 & 15.778$\pm$0.076 & 15.83 & 14.39 & 13.65 & 13.26 & 13.00 & 13.21 & 13.77 & 0.45 & 0.47 & 0.34 & 0.29 & 0.27 & 0.26 & 0.25 & 0.30 & 0.32 \\ 
92 & PGC000092 & 80$\pm$3 & 5376 & 144$\pm$5 & 2.165$\pm$0.016 & 15.956$\pm$0.023 & 17.48 & 16.39 & 15.84 & 15.51 & 15.28 &   &   & 0.34 &   & 0.15 & 0.16 & 0.16 & 0.17 & 0.18 &   &   \\ 
94 & UGC12905 & 90$\pm$1 & 4098 & 188$\pm$14 & 2.274$\pm$0.032 & 16.006$\pm$0.185 &   &   &   &   &   & 15.82 & 16.49 &   & 0.29 &   &   &   &   &   & 0.32 & 0.30 \\ 
96 & UGC12903 & 81$\pm$3 & 14743 & 559$\pm$9 & 2.753$\pm$0.008 & 16.075$\pm$0.076 & 16.68 & 15.18 & 14.45 & 13.99 & 13.73 &   &   & 0.26 &   & 0.32 & 0.30 & 0.26 & 0.25 & 0.23 &   &   \\ 
102 & IC5376 & 82$\pm$3 & 5048 & 427$\pm$7 & 2.635$\pm$0.007 & 15.411$\pm$0.076 & 15.81 & 14.28 & 13.45 & 13.02 & 12.73 & 12.85 & 13.45 & 0.25 & 0.27 & 0.43 & 0.38 & 0.33 & 0.30 & 0.27 & 0.33 & 0.34 \\ 
124 & UGC12913 & 86$\pm$2 & 6350 & 252$\pm$6 & 2.402$\pm$0.010 & 15.782$\pm$0.076 & 17.23 & 15.85 & 15.26 & 14.96 & 14.76 & 15.18 & 15.75 & 0.23 & 0.29 & 0.29 & 0.32 & 0.31 & 0.30 & 0.30 & 0.32 & 0.34 \\ 
128 & PGC000128 & 59$\pm$4 & 12559 & 218$\pm$8 & 2.406$\pm$0.024 & 16.850$\pm$0.076 & 17.11 & 16.19 & 15.96 & 15.70 & 15.47 &   &   & 0.69 &   & 0.14 & 0.14 & 0.13 & 0.12 & 0.13 &   &   \\ 
146 & UGC12916 & 50$\pm$4 & 6360 & 162$\pm$5 & 2.325$\pm$0.029 & 16.716$\pm$0.076 & 16.80 & 15.72 & 15.22 & 14.92 & 14.82 & 15.43 & 16.02 & 0.86 & 0.74 & 0.24 & 0.22 & 0.21 & 0.21 & 0.20 & 0.24 & 0.27 \\ 
\nodata \\
\enddata
\tablenotetext{*}{The complete version of this table is available online.}
\end{deluxetable*}

\addtocounter{table}{-1}
\begin{deluxetable*}{c||cccccccccccccccccccc}
\tablecaption{Data Catalog (continued)\tablenotemark{*}
}
\tabletypesize{\scriptsize}
\tablehead{ \\
PGC & 
\colhead{$\overline{u}$} & \colhead{$\overline{g}$} & \colhead{$\overline{r}$} & 
\colhead{$\overline{i}$} & \colhead{$\overline{z}$} & 
\colhead{$\overline{W1}$} & \colhead{$\overline{W2}$} &
\colhead{$A^{(i)}_u$} & \colhead{$A^{(i)}_g$} & \colhead{$A^{(i)}_r$} & 
\colhead{$A^{(i)}_i$} &  \colhead{$A^{(i)}_z$} &  
\colhead{$A^{(i)}_{W1}$} &  \colhead{$A^{(i)}_{W2}$} & 
\colhead{$u^*$} & \colhead{$g^*$} & \colhead{$r^*$} &  
\colhead{$i^*$} &  \colhead{$z^*$}  \\
 ~ &  
(mag) & (mag) & (mag) & 
(mag) & (mag) & 
(mag) & (mag) &
(mag) & (mag) & (mag) & 
(mag) & (mag) & 
(mag) & (mag) & 
(mag) & (mag) & (mag) & 
(mag) & (mag) & 
\\
(1) &  (24) & 
(25) & (26) & (27) & 
(28) & (29) & (30) & 
(31) & (32) & (33) & 
(34) & (35) &
(36) & (37) &
(38) & (39) & (40) &
(41) & (42) 
}
\startdata
2 &   &   &   &   &   & 11.90 & 12.53 &   &   &   &   &   & 0.01$\pm$0.00 & 0.01$\pm$0.00 &   &   &   &   &   \\ 
4 & 17.11 & 16.11 & 15.69 & 15.41 & 15.30 & 16.08 & 16.44 & 0.81$\pm$0.12 & 0.52$\pm$0.08 & 0.36$\pm$0.06 & 0.29$\pm$0.05 & 0.23$\pm$0.04 & 0.01$\pm$0.00 & 0.00$\pm$0.00 & 16.30 & 15.59 & 15.33 & 15.12 & 15.07 \\ 
12 &   &   &   &   &   & 13.64 & 14.24 &   &   &   &   &   & 0.05$\pm$0.01 & 0.01$\pm$0.01 &   &   &   &   &   \\ 
16 & 15.86 & 14.69 & 14.12 & 13.79 & 13.56 & 14.02 & 14.71 & 0.46$\pm$0.08 & 0.33$\pm$0.06 & 0.26$\pm$0.05 & 0.21$\pm$0.04 & 0.16$\pm$0.03 & 0.03$\pm$0.00 & 0.01$\pm$0.00 & 15.40 & 14.36 & 13.86 & 13.58 & 13.40 \\ 
55 & 16.53 & 15.90 & 15.58 & 15.42 & 15.33 & 16.21 & 16.63 & 0.59$\pm$0.10 & 0.37$\pm$0.06 & 0.26$\pm$0.04 & 0.20$\pm$0.04 & 0.16$\pm$0.03 & 0.01$\pm$0.00 & 0.00$\pm$0.00 & 15.94 & 15.53 & 15.32 & 15.22 & 15.17 \\ 
68 & 16.12 & 15.15 & 14.79 & 14.67 & 14.46 & 15.04 & 15.54 & 0.31$\pm$0.05 & 0.22$\pm$0.04 & 0.18$\pm$0.03 & 0.15$\pm$0.03 & 0.12$\pm$0.02 & 0.01$\pm$0.00 & 0.00$\pm$0.00 & 15.81 & 14.93 & 14.61 & 14.52 & 14.34 \\ 
70 & 16.27 & 15.10 & 14.46 & 14.10 & 13.76 & 13.59 & 14.12 & 1.62$\pm$0.03 & 1.23$\pm$0.03 & 1.00$\pm$0.03 & 0.86$\pm$0.03 & 0.74$\pm$0.02 & 0.05$\pm$0.00 & 0.01$\pm$0.00 & 14.65 & 13.87 & 13.46 & 13.24 & 13.02 \\ 
76 & 15.52 & 14.18 & 13.51 & 13.16 & 12.93 & 13.25 & 13.82 & 0.53$\pm$0.09 & 0.38$\pm$0.07 & 0.30$\pm$0.05 & 0.25$\pm$0.04 & 0.20$\pm$0.03 & 0.02$\pm$0.00 & 0.01$\pm$0.00 & 14.99 & 13.80 & 13.21 & 12.91 & 12.73 \\ 
92 & 17.06 & 16.09 & 15.64 & 15.37 & 15.17 &   &   & 0.69$\pm$0.11 & 0.45$\pm$0.08 & 0.34$\pm$0.06 & 0.28$\pm$0.05 & 0.23$\pm$0.04 &   &   & 16.37 & 15.64 & 15.30 & 15.09 & 14.94 \\ 
94 &   &   &   &   &   & 15.85 & 16.52 &   &   &   &   &   & 0.01$\pm$0.00 & 0.00$\pm$0.00 &   &   &   &   &   \\ 
96 & 16.33 & 14.93 & 14.30 & 13.88 & 13.66 &   &   & 0.98$\pm$0.17 & 0.71$\pm$0.12 & 0.55$\pm$0.10 & 0.45$\pm$0.08 & 0.34$\pm$0.06 &   &   & 15.35 & 14.22 & 13.75 & 13.43 & 13.32 \\ 
102 & 15.42 & 14.00 & 13.27 & 12.88 & 12.63 & 12.87 & 13.48 & 1.04$\pm$0.19 & 0.76$\pm$0.14 & 0.61$\pm$0.11 & 0.51$\pm$0.09 & 0.40$\pm$0.07 & 0.04$\pm$0.01 & 0.02$\pm$0.01 & 14.38 & 13.24 & 12.66 & 12.37 & 12.23 \\ 
124 & 17.04 & 15.74 & 15.18 & 14.91 & 14.72 & 15.22 & 15.79 & 0.98$\pm$0.16 & 0.66$\pm$0.11 & 0.48$\pm$0.08 & 0.40$\pm$0.07 & 0.32$\pm$0.06 & 0.01$\pm$0.00 & 0.00$\pm$0.00 & 16.06 & 15.08 & 14.70 & 14.51 & 14.40 \\ 
128 & 16.76 & 16.00 & 15.81 & 15.67 & 15.40 &   &   & 0.31$\pm$0.06 & 0.22$\pm$0.04 & 0.18$\pm$0.03 & 0.15$\pm$0.03 & 0.12$\pm$0.02 &   &   & 16.45 & 15.78 & 15.63 & 15.52 & 15.28 \\ 
146 & 16.59 & 15.60 & 15.13 & 14.87 & 14.78 & 15.47 & 16.06 & 0.19$\pm$0.04 & 0.14$\pm$0.02 & 0.10$\pm$0.02 & 0.09$\pm$0.02 & 0.07$\pm$0.01 & 0.00$\pm$0.00 & 0.00$\pm$0.00 & 16.40 & 15.46 & 15.03 & 14.78 & 14.71 \\ 
\nodata \\
\enddata
\tablenotetext{*}{The complete version of this table is available online.}
\end{deluxetable*}

\addtocounter{table}{-1}
\begin{deluxetable*}{c||ccccccccccccccccc}
\tablecaption{Data Catalog (continued)\tablenotemark{*}
}
\tabletypesize{\scriptsize}
\tablehead{ \\
PGC & 
\colhead{$W1^*$} &  \colhead{$W2^*$}  &
\colhead{$W2p^*$} &
\colhead{$\gamma_u$} &  \colhead{$\gamma_g$}  &
\colhead{$\gamma_r$} &  \colhead{$\gamma_i$}  &
\colhead{$\gamma_z$} &  
\colhead{$\gamma_{w1}$} &  \colhead{$\gamma_{w2}$}  &
\colhead{$\gamma_u^{(p)}$} &  \colhead{$\gamma_g^{(p)}$}  &
\colhead{$\gamma_r^{(p)}$} &  \colhead{$\gamma_i^{(p)}$}  &
\colhead{$\gamma_z^{(p)}$} &  
\colhead{$\gamma_{w1}^{(p)}$} &  \colhead{$\gamma_{w2}^{(p)}$}  \\
 ~ &  
(mag) & (mag) & (mag) & 
(mag) & (mag) & 
(mag) & (mag) &
(mag) & (mag) & (mag) & 
(mag) & (mag) & 
(mag) & (mag) & 
(mag) & (mag) & (mag) 
\\
(1) &  (43) & 
(44) & (45) & (46) & 
(47) & (48) & (49) & 
(50) & (51) & (52) & 
(53) & (54) &
(55) & (56) &
(57) & (58) & (59) 
}
\startdata
2 & 11.89 & 12.52 & 12.53 & 1.06 & 0.70 & 0.51 & 0.37 & 0.20 & 0.05 & 0.02 & 1.06 & 0.70 & 0.51 & 0.37 & 0.20 & 0.05 & 0.02 \\ 
4 & 16.07 & 16.44 & 16.55 & 0.79 & 0.50 & 0.34 & 0.28 & 0.22 & 0.01 & 0.00 & 0.79 & 0.50 & 0.34 & 0.28 & 0.22 & 0.01 & 0.00 \\ 
12 & 13.59 & 14.23 & 14.24 & 1.24 & 0.90 & 0.71 & 0.60 & 0.47 & 0.07 & 0.02 & 1.24 & 0.90 & 0.71 & 0.60 & 0.47 & 0.07 & 0.02 \\ 
16 & 13.99 & 14.70 & 14.43 & 1.23 & 0.89 & 0.70 & 0.58 & 0.44 & 0.07 & 0.02 & 1.23 & 0.88 & 0.69 & 0.57 & 0.43 & 0.07 & 0.02 \\ 
55 & 16.20 & 16.63 & 16.81 & 0.78 & 0.49 & 0.34 & 0.27 & 0.22 & 0.01 & 0.00 & 0.77 & 0.47 & 0.32 & 0.26 & 0.21 & 0.01 & 0.00 \\ 
68 & 15.03 & 15.54 & 15.5 & 1.16 & 0.84 & 0.66 & 0.56 & 0.46 & 0.06 & 0.02 & 1.17 & 0.84 & 0.67 & 0.57 & 0.46 & 0.06 & 0.02 \\ 
70 & 13.54 & 14.11 & 14.42 & 1.11 & 0.79 & 0.62 & 0.52 & 0.43 & 0.05 & 0.02 & 1.08 & 0.76 & 0.59 & 0.50 & 0.42 & 0.04 & 0.01 \\ 
76 & 13.23 & 13.81 & 13.86 & 1.24 & 0.90 & 0.71 & 0.59 & 0.46 & 0.07 & 0.02 & 1.23 & 0.89 & 0.70 & 0.58 & 0.44 & 0.07 & 0.02 \\ 
92 &   &   & 16.54 &   &   &   &   &   &   &   & 0.91 & 0.61 & 0.45 & 0.37 & 0.31 & 0.02 & 0.01 \\ 
94 & 15.84 & 16.52 & 16.52 & 0.78 & 0.48 & 0.33 & 0.27 & 0.21 & 0.01 & 0.00 & 0.78 & 0.48 & 0.33 & 0.27 & 0.21 & 0.01 & 0.00 \\ 
96 &   &   & 14.34 &   &   &   &   &   &   &   & 1.23 & 0.88 & 0.69 & 0.56 & 0.42 & 0.07 & 0.02 \\ 
102 & 12.83 & 13.46 & 13.4 & 1.24 & 0.90 & 0.71 & 0.59 & 0.47 & 0.07 & 0.02 & 1.23 & 0.89 & 0.70 & 0.58 & 0.44 & 0.07 & 0.02 \\ 
124 & 15.21 & 15.79 & 15.84 & 0.88 & 0.58 & 0.42 & 0.35 & 0.29 & 0.02 & 0.01 & 0.89 & 0.59 & 0.43 & 0.36 & 0.30 & 0.02 & 0.01 \\ 
128 &   &   & 16.84 &   &   &   &   &   &   &   & 1.09 & 0.77 & 0.60 & 0.51 & 0.42 & 0.05 & 0.01 \\ 
146 & 15.47 & 16.06 & 16.13 & 0.99 & 0.69 & 0.52 & 0.44 & 0.36 & 0.03 & 0.01 & 1.02 & 0.71 & 0.54 & 0.46 & 0.38 & 0.04 & 0.01 \\  
\nodata \\
\enddata
\tablenotetext{*}{The complete version of this table is available online.}
\end{deluxetable*}

\addtocounter{table}{-1}
\begin{deluxetable*}{c||cccccccccccccc}
\tablecaption{Data Catalog (continued)\tablenotemark{*}
}
\tabletypesize{\scriptsize}
\tablehead{ \\
PGC & 
\colhead{$A^{(p)}_u$} & \colhead{$A^{(p)}_g$} & \colhead{$A^{(p)}_r$} & 
\colhead{$A^{(p)}_i$} &  \colhead{$A^{(p)}_z$} &  
\colhead{$A^{(p)}_{W1}$} &  \colhead{$A^{(p)}_{W2}$} & 
\colhead{$P_{1,W2}$} & \colhead{$P_{1,W2p}$} &
\colhead{$P_{1,W1}$} &
\colhead{$Q_{S}$} &
\colhead{$Q_{W}$} &
\colhead{$R_{src}$} \\
 ~ &  
(mag) & (mag) & (mag) & 
(mag) & (mag) & 
(mag) & (mag) &
~ & ~ & 
~ &
~ & 
~ & 
~ & 
\\
(1) &  (60) & 
(61) & (62) & (63) & 
(64) & (65) & 
(66) & (67) &  
(68) & (69) &
(70) & (71) &
(72) 
}
\startdata
2 & 0.22 & 0.15 & 0.11 & 0.08 & 0.04 & 0.01 & 0.00 & 2.02$\pm$0.14 & 2.02 & 2.14 & 0 & 5 & 2 \\ 
4 & 0.82 & 0.52 & 0.36 & 0.29 & 0.24 & 0.01 & 0.00 & -2.75$\pm$0.10 & -2.73 & -2.69 & 5 & 4 & 2 \\ 
12 & 1.04 & 0.76 & 0.60 & 0.51 & 0.40 & 0.05 & 0.02 & 0.56$\pm$0.15 & 0.56 & 0.66 & 0 & 5 & 2 \\ 
16 & 0.46 & 0.33 & 0.26 & 0.21 & 0.16 & 0.02 & 0.01 & 0.92$\pm$0.15 & 1.02 & 1.04 & 5 & 5 & 2 \\ 
55 & 0.58 & 0.36 & 0.24 & 0.20 & 0.16 & 0.01 & 0.00 & -2.83$\pm$0.12 & -3.03 & -2.73 & 5 & 4 & 2 \\ 
68 & 0.31 & 0.22 & 0.18 & 0.15 & 0.12 & 0.01 & 0.00 & -0.35$\pm$0.18 & -0.30 & -0.35 & 5 & 5 & 2 \\ 
70 & 1.58 & 1.19 & 0.96 & 0.83 & 0.71 & 0.04 & 0.01 & -0.74$\pm$0.14 & -0.91 & -0.69 & 5 & 5 & 2 \\ 
76 & 0.52 & 0.38 & 0.30 & 0.25 & 0.19 & 0.03 & 0.01 & 0.75$\pm$0.09 & 0.89 & 0.88 & 5 & 5 & 2 \\ 
92 & 0.69 & 0.46 & 0.34 & 0.28 & 0.23 & 0.01 & 0.00 & -1.90$\pm$0.11 & -1.90 &   & 5 & 0 & 0 \\ 
94 & 1.14 & 0.76 & 0.53 & 0.44 & 0.36 & 0.01 & 0.00 & -2.91$\pm$0.17 & -2.91 & -2.79 & 0 & 4 & 2 \\ 
96 & 0.98 & 0.70 & 0.55 & 0.45 & 0.34 & 0.05 & 0.02 & 1.06$\pm$0.11 & 1.06 &   & 5 & 0 & 0 \\ 
102 & 1.04 & 0.75 & 0.59 & 0.49 & 0.37 & 0.05 & 0.02 & 0.60$\pm$0.11 & 0.93 & 0.73 & 5 & 5 & 2 \\ 
124 & 0.99 & 0.67 & 0.49 & 0.41 & 0.34 & 0.02 & 0.01 & -2.07$\pm$0.11 & -2.00 & -1.93 & 5 & 5 & 2 \\ 
128 & 0.31 & 0.22 & 0.17 & 0.15 & 0.12 & 0.01 & 0.00 & -0.84$\pm$0.13 & -0.84 &   & 5 & 0 & 0 \\ 
146 & 0.20 & 0.14 & 0.10 & 0.09 & 0.07 & 0.01 & 0.00 & -1.40$\pm$0.11 & -1.26 & -1.20 & 4 & 4 & 2  \\
\nodata \\
\enddata
\tablenotetext{*}{The complete version of this table is available online.}
\end{deluxetable*}

\end{turnpage}

\clearpage


\section{Distance Measurements}

In this section, we summarize our process for measuring the distances of our sample galaxies by deriving their absolute luminosities, $M$, from the luminosity$-$line width relations (TFR) that were calibrated in K20. The distance modulus of each galaxy is then given by $DM_{\lambda}=m_{\lambda}-M_{\lambda}$, where $m_{\lambda}$ and $M_{\lambda}$ are the apparent and absolute magnitudes at wave band $\lambda$, respectively. In \S \ref{sec:TFRs}, we explain how we calculate the absolute luminosity of a spiral galaxy given its \hi line width.

\begin{table*}[t]
\setlength{\tabcolsep}{0.2cm}
\centering
\caption{TFR Parameters Before and After Corrections} 
\label{tab:revised_ITFR}
\begin{tabular}{cl | cccc | c c}
\tablewidth{0pt}
\hline \hline
 \multirow{2}{*}{Band} &  TFR &  \multicolumn{4}{|c}{TFR Parameters} & \multicolumn{2}{|c}{Curvature} \\
 \cline{3-6}  \cline{7-8}
    & Code &  Slope &  ZP &  $C_{zp}$ & rms  & Break Point & $A_2$ \\
\hline 
\decimals
{\it r}  & TF$_r$    & -7.96$\pm$0.13 & -20.57$\pm$0.10 & -0.08 & 0.49 & log$(W^i_{mx})$=2.5 & 4.56$\pm$0.89 \\
{\it i}  & TF$_{i}$  & -8.32$\pm$0.13  & -20.80$\pm$0.10 & -0.04 & 0.49 & 2.5 & 5.34$\pm$0.91 \\
{\it z}  & TF$_{z}$  & -8.46$\pm$0.13  & -20.89$\pm$0.10 & -0.08 & 0.50 & 2.5 & 5.81$\pm$0.91 \\
{\it W1} & TF$_{W1}$ & -9.47$\pm$0.14  & -20.36$\pm$0.07 &       & 0.58 & 2.4 & 3.81$\pm$0.42 \\
\hline \hline 
\end{tabular}
\end{table*} 

\subsection{Luminosity$-$line width Correlations} \label{sec:TFRs}

TFRs are power-law relations between the absolute luminosity and \hi line width of spiral galaxies. 
In the process of calibrating the TFRs, the residuals from the fitted relation are minimized along the direction of line width to approximately nullify the Malmquist bias that is the consequence of the asymmetrical scatter of galaxies along the luminosity axis. 
This procedure is called ``inverse TFR" (ITFR) and was employed in K20 to calibrate the luminosity$-$line width relations at multiple bands.

Following the formalism explained in K20, the absolute luminosity at the wave band $\lambda$ is given as
\begin{equation}
\label{Eq:adjustedTFR}
M_{\lambda}=Slope \big({\rm log}W^i_{mx}-2.5 \big)+\overline{ZP} ~. 
\end{equation}
where $\overline{ZP} = ZP + C_{zp}$ is the zero point, with a potential offset, $C_{zp}$, that can be applied to ensure that the measured distance moduli at different passbands are statistically consistent (see \S 3.4 of K20 for further discussion).

In K20, the curvature of the ITFR at the high-luminosity end is modeled using a deviation from the linear ITFR that is given as
\begin{equation}
\label{Eq:TFRcurve}
    \Delta M_{\lambda} = A_2 X^2+A_1 X+A_0~,
\end{equation}
where $X=\rm{log}(W^i_{mx})-2.5$. If the line width is smaller than the break point, then the absolute luminously is obtained using the linear part of the ITFR (Equation \ref{Eq:adjustedTFR}). For line widths that are larger than the break point, the absolute luminosity equals $M_{\lambda}+ \Delta M_{\lambda}$. We require that the curved and linear parts of the ITFR share the same slope at the break point, $\Delta M_{\lambda}=0$, which leaves $A_2$ as the free parameter that is determined in the fitting process. Table \ref{tab:revised_ITFR} lists parameters of the calibrated ITFRs that are used in this study.

The slopes of the ITFRs are steeper at longer wave bands, since there is a trend toward redder colors in massive galaxies with faster rotation.  Since red and blue galaxies at a given observed linewidth are displacing with respect to each other in different passbands, this implies the possibility of adopting additional parameters such as color or surface brightness to improve the consistency of the TFRs and/or to decrease their scatter.
If optical photometry is available, we preferably adopt observations in the {\it r}, {\it i} and {\it z} bands, where the imaging quality is best and the slopes and zero points of the ITFRs are relatively in the same regime. For comparison purposes, we use the average of the {\it r-, i-} and {\it z-}band material as a reference.

For those cases that do not have optical coverage, the use of infrared ITFRs is required, although the quality of the data is inferior and the ITFR scatter is greater. We prefer the {\it W}1 band ITFR for distance measurements due to the better imaging quality and smaller average RMS scatter at this band compared to the {\it W}2-band. In K20, we demonstrated that {\it W}1-band ITFR scatter is improved with adjustments based on optical-infrared colors. However, lack of SDSS coverage precludes color adjustments. In any event, in the presence of optical photoemtry, it is hard to justify the use of infrared ITFRs to measure distances even after optical-infrared color adjustments. If only infrared photometry is available, we look to reduce scatter from information such as surface brightness.

There are redundancies between results at different passbands, but the diversity of our photometric and kinematic information with statistically significant samples allows us to uncover secondary parameter influences.  Our investigations of relevance will be discussed following the important diversion of the next section.

\subsection{Residual Malmquist bias} \label{sec:residual_bias}

\begin{figure*}[ht]
\centering
\includegraphics[width=0.45\linewidth]{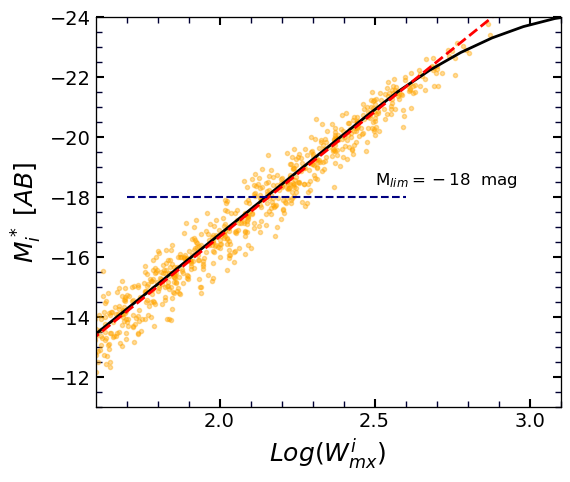}
\includegraphics[width=0.45\linewidth]{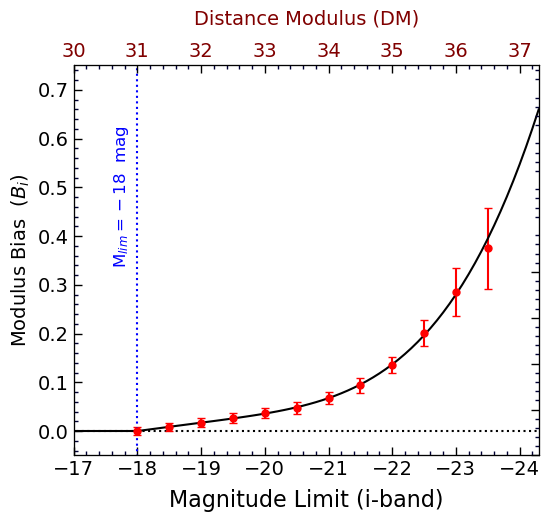}
\includegraphics[width=0.45\linewidth]{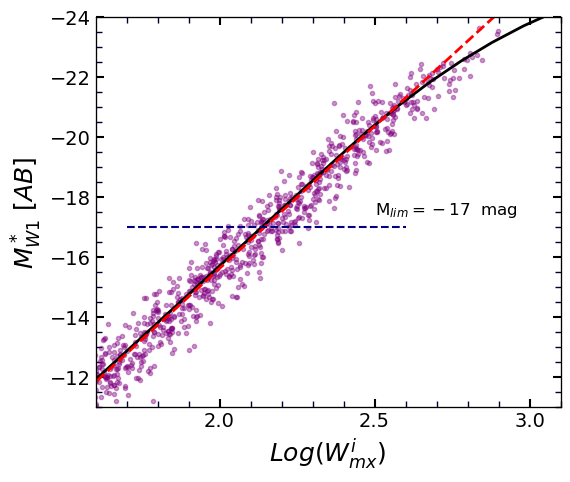}
\includegraphics[width=0.45\linewidth]{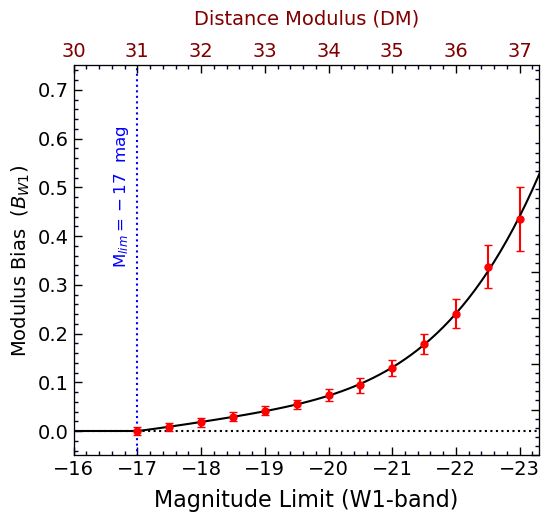}
\caption{{\bf Left:} 1000 simulated galaxies along the curved TFR drawing from the appropriate Schechter luminosity function and scattered along the magnitude axis based on the rms scatter model presented in Figure 9 of K20. Horizontal dashed lines represent the statistical limiting magnitude of our sample at the {\it i} and {\it W}1 bands at the distance of the Virgo cluster ($DM=31$) mag. {\bf Right:} distance bias versus the limiting magnitude limit of the simulated sample. Top axes of the right panels are the distance modulus at which our sample is complete up to the limiting absolute magnitude given on the bottom axes.}
\label{fig:Slpoe_calib}
\end{figure*}

\begin{table*}[t]
\setlength{\tabcolsep}{0.15cm}
\centering
\caption{Parameters of the third-degree polynomial function, $\mathcal{B}_{\lambda}(DM_{\lambda}) = 0.01\times \sum_{n=0}^{3} B^{(\lambda)}_n (DM_{\lambda}-31)^n$, that fits the distance modulus bias at different wave bands. The last two columns are the fitted parameters of luminosity function as described by the Schechter function. } 
\label{tab:bias_params}
\begin{tabular}{cl|cccc|cc}
\tablewidth{0pt}
\hline \hline
 Band & TFR & \multicolumn{4}{c|}{Modulus Bias Function} &  \multicolumn{2}{c}{LF Parameters} \\
  & Code & $B_3$ & $B_2$ & $B_1$ & $B_0$ & $M^{\star}$  & $\alpha$  \\
 \hline 
\decimals
{\it r}	 & TF$_{r}$	 & 0.07$\pm$0.00 & -0.58$\pm$0.05 & 1.48$\pm$0.16 & 0.44$\pm$0.16 & -21.8$\pm$0.2 & -1.0$\pm$0.1	 \\
{\it i}	 & TF$_{i}$	 & 0.05$\pm$0.01 & -0.12$\pm$0.13 & 0.06$\pm$0.36 & 1.72$\pm$0.32 & -22.0$\pm$0.1 & -1.0$\pm$0.1	 \\
{\it z}	 & TF$_{z}$	 & 0.08$\pm$0.01 & -0.28$\pm$0.10 & 0.31$\pm$0.28 & 1.58$\pm$0.23 & -22.1$\pm$0.2 & -1.0$\pm$0.1	 \\
{\it W1}	 & TF$_{W1}$ & 0.04$\pm$0.01 & -0.15$\pm$0.07 & 0.31$\pm$0.22 & 1.65$\pm$0.23 &  -21.9$\pm$0.1	 & -1.0$\pm$0.1	 \\
\hline
\end{tabular}
\end{table*}

Although the inverse fitting process to calibrate the luminosity$-$line width correlations minimizes the amplitude of the Malmquist bias, it does not completely remove it. 
Our sample does not cover the entire absolute luminosity range at all distances. At greater distances, fewer small galaxies are picked. 
The residual bias that we investigate here arises from features in the luminosity function of \hi-rich galaxies described by the Schechter formalism \citep{1976ApJ...203..297S} and the sampling of that luminosity function with the change of limiting absolute magnitude as a function of distance.

The residual bias is studied following a simulation procedure similar to that described in \S 3.8 of K20. A synthetic sample is generated with a random draw of 20,000 absolute magnitudes from the Schechter luminosity function of each band. We denote the simulated absolute magnitudes as $M_{fid}$ (for fiducial).
A line width is assigned to each synthetic luminosity based on the curved ITFR. We require that the ${\rm log}W^i_{mx}$ value of all synthetic galaxies range between $1.5$ and $3$, the domain that is covered by our sample (see Fig. \ref{fig:P1_gamma_logW}). The simulated absolute magnitude of each synthetic galaxy is then statistically dispersed following a normal distribution that is centered on the original synthetic magnitude, with the standard deviation taken from the scatter models presented in Figure 9 of K20. 
The left panels of Figure \ref{fig:Slpoe_calib} illustrate subsets of 1000 simulated galaxies chosen randomly from the whole ensemble.
The dispersed absolute magnitudes $M_{obs}$ are related to the measured values $m_{obs}$ registered by an observer. 

Each simulated galaxy deviates from the fiducial relation by $offset = M_{fid} - M_{obs}$.  This offset can be positive or negative.  However, there will tend to be more positive offsets, for two reasons. First, the dominant effect is caused by the exponential cutoff at the bright end of the luminosity function.  Fewer galaxies reach a given $M_{obs}$ by faintward scatter from the fiducial relation than from upward scatter.  Upward-scattered cases land to the left of the TFR, i.e., they have positive offsets.  The second, less important effect results from the increased scatter toward fainter magnitudes.  The upward-scattered cases, landing preferentially to the left of the TFR, are drawn from a more dispersed population.

The tendency for there to be more positive than negative offsets is a bias. This bias depends on distance. Nearby, where galaxies over the full practical range of the TFR are being sampled, the bias is small, but it becomes more acute at distances where the effective limiting magnitude for inclusion, $M_{lim}$, is comparable to the break in the luminosity function characterized by the Schechter parameter $M^{\star}$ (where $M_i^{\star}=-22$).\footnote{See K20 Table 3 for Schechter $M^{\star}$ and $\alpha$ parameters at different passbands.}

We are interested in the statistical expectation value of the bias. We define the residual bias as the ensemble average of the offset values of all synthetic galaxies that are brighter than the distance-dependant limiting magnitude:
\begin{equation}
\label{Eq:bias}
\begin{split}
   \mathcal{B}(DM)=\langle M_{fid} &- M_{obs} \rangle~,  \\
      & M_{obs}>M_{lim}.
\end{split}
\end{equation}

There is a clear distinction between the terms ``limiting magnitude" used here and ``cutoff magnitude" introduced in K20.  The ITFR are calibrated to a faint limit cutoff of $M_i = -17.0$ or $M_{W1} = -16.1$.  These cutoffs were imposed because dispersion in the ITFR becomes increasingly large in the dwarf regime (in the absence of forming the baryonic TF variant \citep{2000ApJ...533L..99M}) and because our particular interest is in deriving distances beyond the reach of dwarf galaxies.

Here in our field sample, the \hi flux limitation of the dominant ALFALFA contributions give rise to corresponding limiting optical-infrared magnitudes.  Through empirical experimentation, we find the optimal correspondences illustrated in the right panels of Figure~\ref{fig:Slpoe_calib} scaled to the case of the Virgo cluster at $DM=31$.
The residual bias is a function of the limiting absolute magnitude, $M^{(i)}_{lim}=-18-(DM-31)$ and $M^{(W1)}_{lim}=-17-(DM-31)$. 
At $DM<31$ mag, we assume that our sample is complete and the bias is consequently negligible. This condition is satisfied by normalizing the residual bias to zero at $DM=31$ mag. 

We have calculated the bias for a set of discrete absolute magnitude limits using 50 different random ensembles. Red points and their error bars in the right panels of Figure \ref{fig:Slpoe_calib} display the average and 1$\sigma$ scatter of the results. We model the bias using a third-degree polynomial fitted to the discrete points of the form, $\mathcal{B}_{\lambda}(DM_{\lambda}) = 0.01\times \sum_{n=0}^{3} B^{(\lambda)}_n (DM_{\lambda}-31)^n$. Table~\ref{tab:bias_params} lists the coefficients of the polynomial function, $B^{(\lambda)}_n$ for all passbands of interest ($\lambda=r, i, z$ and {\it W1}).

In summary, the adjusted distance modulus for the residual Malmquist bias has the form $DM_{\lambda}+\mathcal{B_\lambda}(DM_{\lambda})$, where $DM_{\lambda}$ is the raw distance modulus that is computed from the apparent magnitude (Equation~\ref{Eq:bkai}) and the curved luminosity$-$line width correlation (Equation~\ref{Eq:adjustedTFR} and \ref{Eq:TFRcurve}) following $DM_{\lambda}=m_{\lambda}^*-(M_{\lambda}+\Delta M_{\lambda})$.

\subsection{Color-dependent Systematics} \label{sec:col_dep_sys}

In \S 4 of K20, it was shown that the slope of the TFR is wavelength$-$dependent, implicitly the effect of third parameters, and therefore we attempted to incorporate various color terms and other structural parameters, such as the average surface brightness, into our formalism. We showed that the scatter about the adjusted relations is significantly reduced in the extreme cases of blueward $u$ and $g$ bands and redward {\it W}1 and {\it W}2 bands.  This issue is further explored in this section.

Using the information provided in \S \ref{sec:TFRs} and \ref{sec:residual_bias}, it is possible to determine the bias-adjusted distance modulus of an individual galaxy, preferentially in the SDSS $r$, $i$, and $z$ bands.  The values derived in the three bands differ through the photometry measurements but are highly correlated through the common line width and inclination parameters.  Mean distances vary at the level of $\pm1.2\%$, providing a minimal estimate of systematic uncertainties.  

\begin{figure*}
\centering
\includegraphics[width=\linewidth]{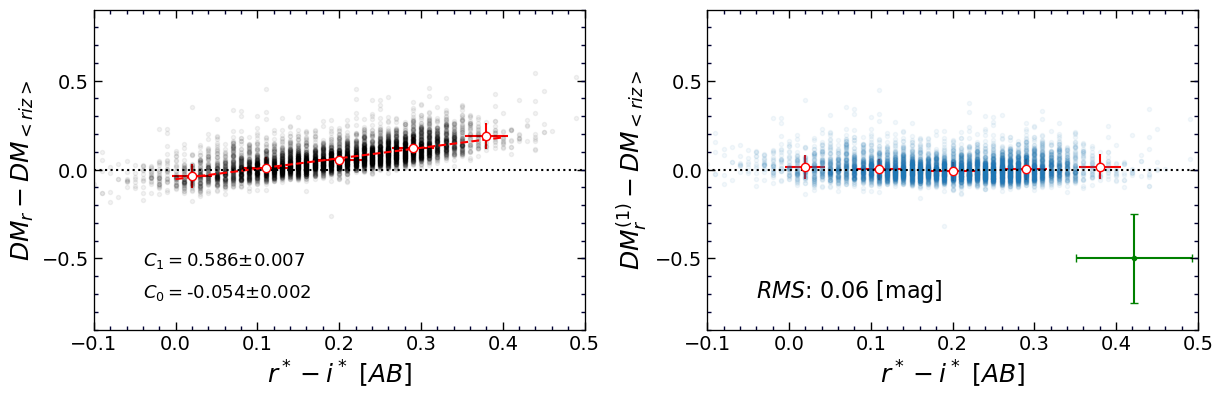} \\
\includegraphics[width=\linewidth]{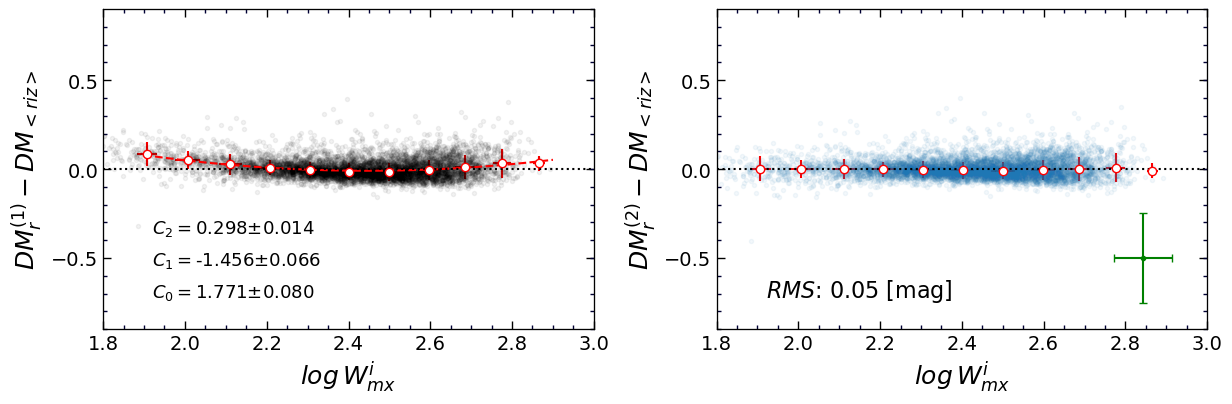}
\caption{
{\bf Left:} differences between distance moduli measured using TF$_r$ and the average moduli, $DM_{\langle riz \rangle}$, versus the $r^*-i^*$ color (top) and line width (bottom). Each black point represents a galaxy. Correlations are modeled using least square linear fits of the form $DM_r-DM_{\langle riz \rangle}=C_1 \Theta + C_0$, where $\Theta$ is either $r^*-i^*$ or ${\rm log} W^i_{mx}$.
{\bf Right:} similar to the left panels but for the adjusted distance moduli, $DM^{1}_r$, using the illustrated correlations in the corresponding left panels. Green error bar in bottom left corner exhibits the typical uncertainty of an individual distance modulus measurement. Open red points show the average of black/blue data points within the horizontal bins of the same size. Here, rms is the root mean square scatter of the moduli differences in the right panels.
\label{fig:DMr12}}
\end{figure*} 

\begin{figure*}
\centering
\includegraphics[width=\linewidth]{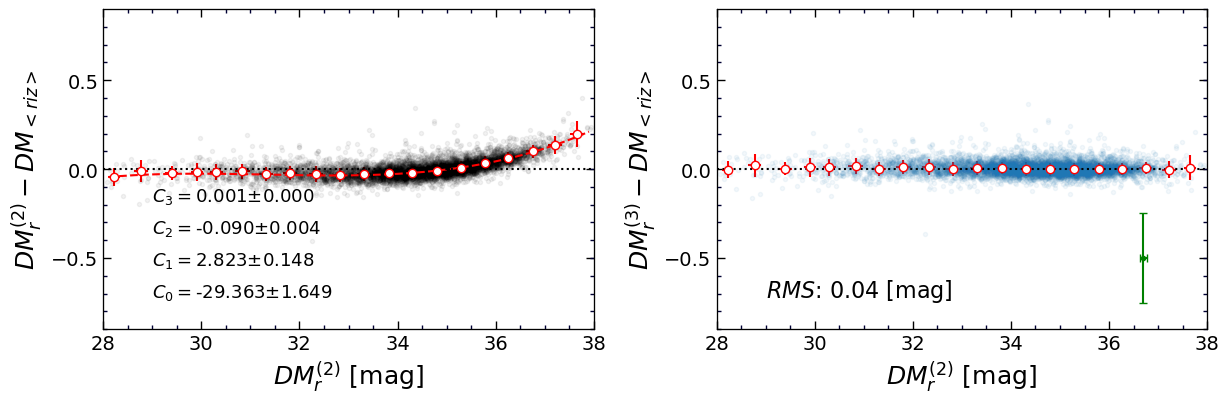} \\
\includegraphics[width=\linewidth]{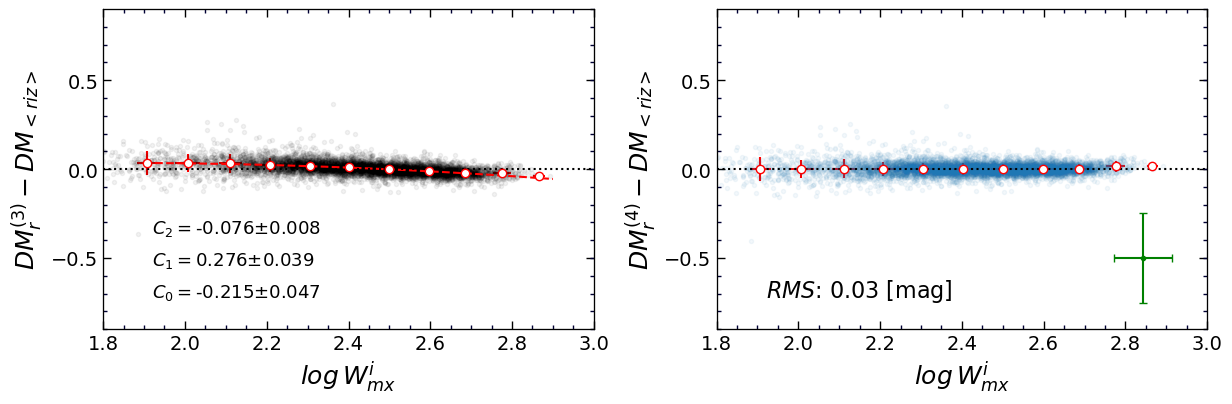} \\
\includegraphics[width=\linewidth]{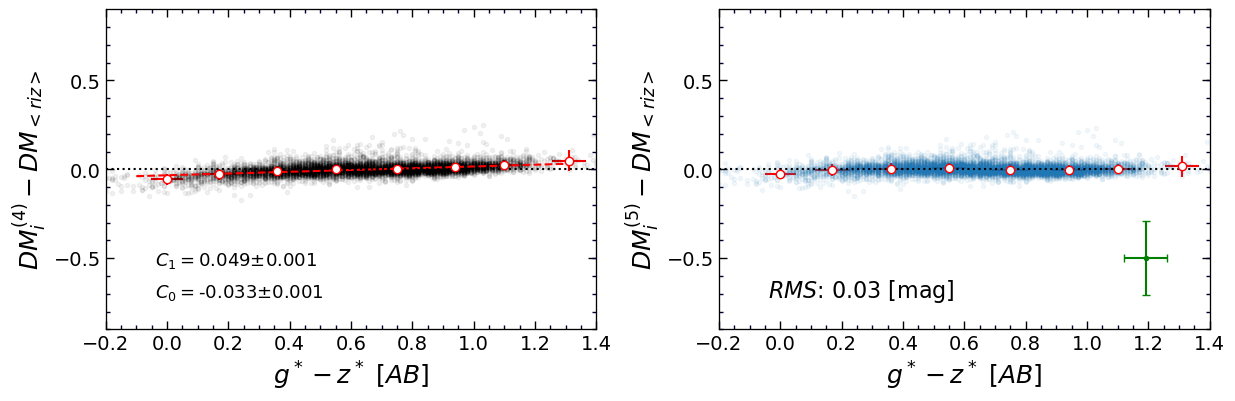}
\caption{
Similar to Figure \ref{fig:DMr12}. Removing systematics by applying $DM_i^{(2)}$,  ${\rm log} W^i_{mx}$, and $g^*-z^*$, respectively.
\label{fig:DMr345}}
\end{figure*} 

To better understand the source of this category of systematics, assume that there are two calibrated ITFRs at two wavelengths, $\lambda1$ and $\lambda2$
\begin{subequations}
\label{eq:TFR2}
	\begin{align}
	\begin{split}
		M^*_{\lambda1} = & S_1({\rm log}W^i_{mx} - 2.5) + Z_1 ~, 
	\end{split} \\
	\begin{split}
		M^*_{\lambda2} = & S_2({\rm log}W^i_{mx} - 2.5) + Z_2, 
	\end{split}
	\end{align}
\end{subequations}
where $S$ and $Z$ are the slopes and zero points of the ITFRs. The galaxy apparent magnitudes at these two wave bands are $m_{\lambda1}$ and $m_{\lambda2}$, whence the measured distance modulus at each band is given by
\begin{subequations}
\label{eq:DM2}
	\begin{align}
	\begin{split}
		DM_{\lambda1} = & m_{\lambda1} - M^*_{\lambda1} ~, 
	\end{split} \\
	\begin{split}
		DM_{\lambda2} = & m_{\lambda2} - M^*_{\lambda2} . 
	\end{split}
	\end{align}
\end{subequations}

Rearranging equations \ref{eq:TFR2} and \ref{eq:DM2}, the modulus difference between bands is
\begin{equation}
\label{eq:DM1-DM2}
\begin{split}
DM_{\lambda1}&-DM_{\lambda2} = (m_{\lambda1}-m_{\lambda2}) \\ 
  & + (S_2-S1)({\rm log}W^i_{mx} - 2.5) + (Z_2-Z_1)~.
\end{split}
\end{equation}
The distance modulus is wavelength-independent, which requires $DM_{\lambda1}-DM_{\lambda2}=0$. Setting the left side of Equation \ref{eq:DM1-DM2} to zero, we find the color index of the galaxy
\begin{equation}
\label{eq:m12}
m_{\lambda2}-m_{\lambda1} = (S_2-S_1)({\rm log}W^i_{mx} - 2.5) + (Z_2-Z_1)~.
\end{equation}
Obviously, the colors of the spirals with a given line width do not all rigorously follow this linear correlation. There are deviations from this linear relation that imply $DM_{\lambda1} \neq DM_{\lambda2}$. Equation \ref{eq:DM1-DM2} implies that, at a constant line width and in the absence of any other adjustments, redder galaxies will have smaller measured distances relative to blue galaxies at longer passbands. 

\begin{figure}
\centering
\includegraphics[width=\linewidth]{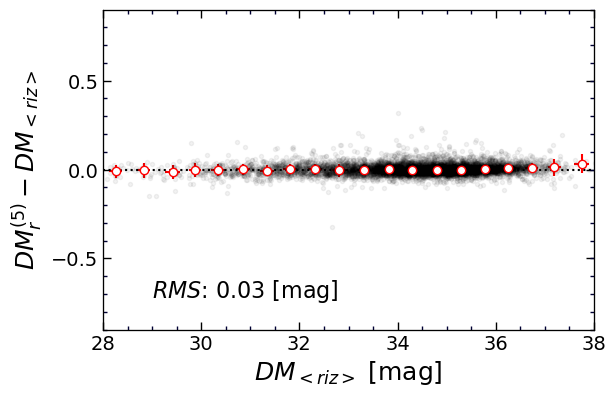} \\
\includegraphics[width=\linewidth]{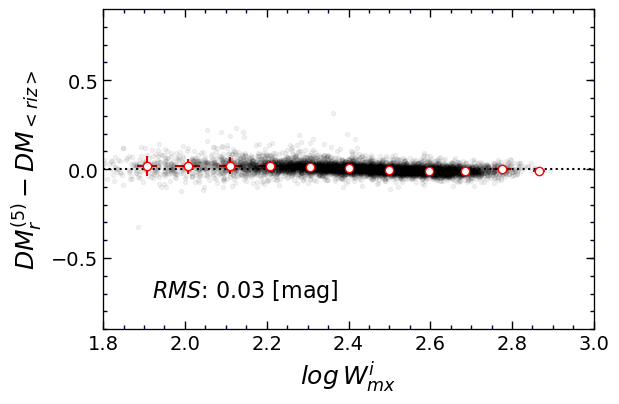} \\
\includegraphics[width=\linewidth]{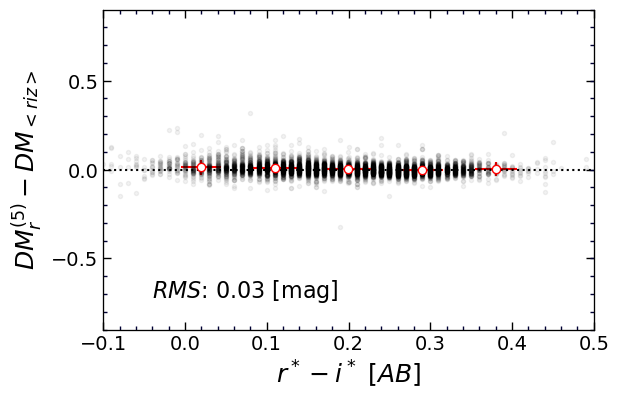}
\caption{
Similar to the right panels of Figure \ref{fig:DMr12}.
Deviations of the quintuply adjusted {\it r}-band moduli, $DM^{(5)}_r$, from the $<riz>$ moduli, versus various parameters.  
\label{fig:DMr5_test}}
\end{figure} 

There can be reasonable compensation for color effects if the analysis is constrained to the optical $r$, $i$, $z$ bands as will be discussed in the following subsections.  There is greater concern regarding the integration of the infrared analysis.   
Galaxies with optical and/or infrared photometry coverage are unevenly distributed across the sky (see Figure \ref{fig:aitoff_equatorial}). Color-dependant systematics might have serious implications for an analysis of the peculiar velocities of a sample of galaxies that combines distances that are measured at multiple passbands. Our statistical analysis in appendix \ref{sec:OPIR-OP-IR} reveals that the subsample of spirals with both SDSS and WISE photometry (OP+IR) on average consists of redder galaxies than the subsample of galaxies with only SDSS photometry (OP$-$IR). In Figure \ref{fig:aitoff_equatorial} blue and green points display the spatial distributions of these two subsamples, and red points locate cases with only WISE photometry, showing that the three components of our study are distinctively patterned. 
Color-dependant systematics that vary in geometrical distribution could  introduce false galaxy-flow patterns.

\begin{figure*}
\centering
\includegraphics[width=\linewidth]{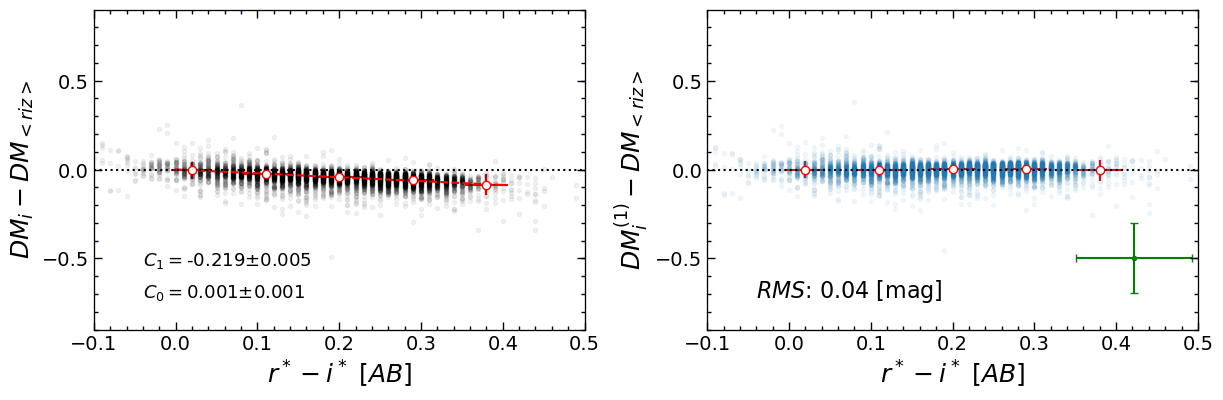} \\
\includegraphics[width=\linewidth]{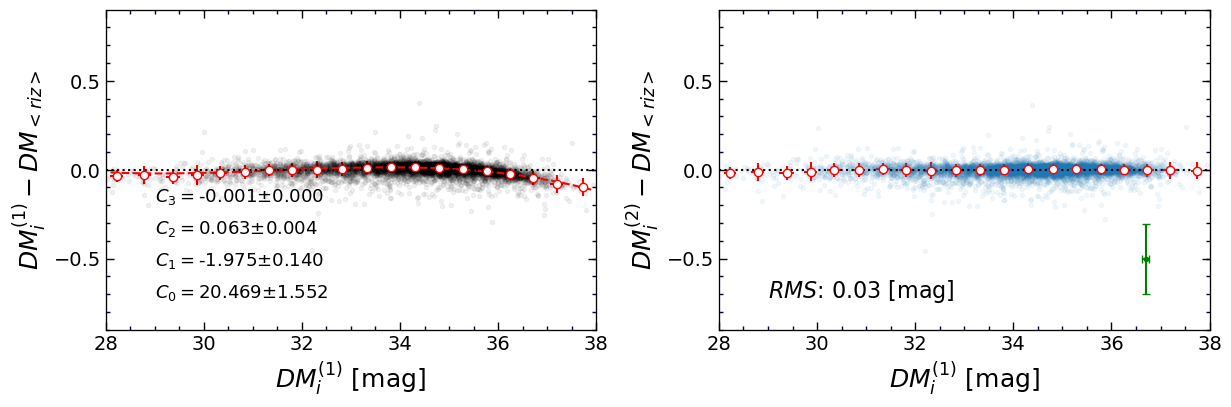} \\
\includegraphics[width=\linewidth]{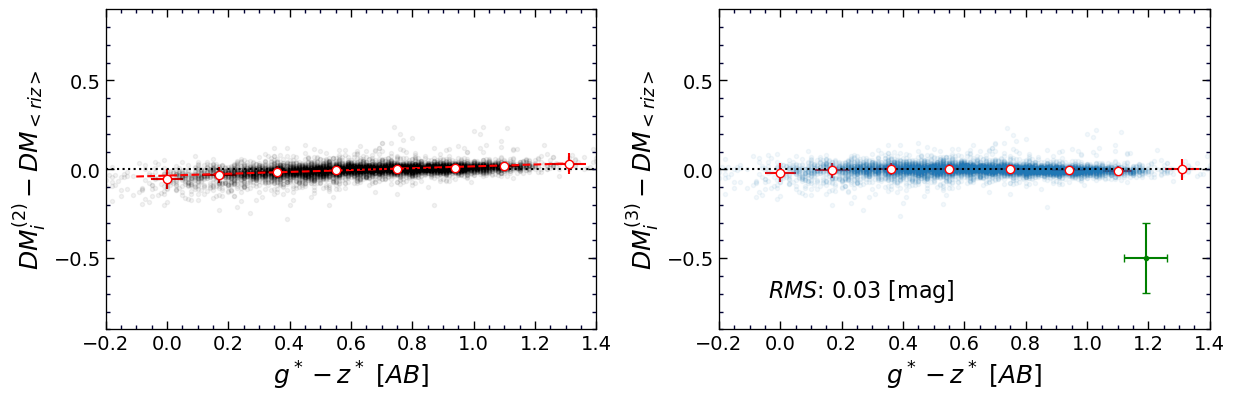}
\caption{
Similar to Figure \ref{fig:DMr12}. Each step illustrates one step of the adjustment process of the {\it i}-band distance moduli.
\label{fig:DMi123}}
\end{figure*} 

Our method to address the remaining band-to-band systematics involves creating a composite of $r$, $i$, and $z$ moduli as the reference.
For each galaxy with optical photometry, we define the average of the distance moduli at these three bands as

\begin{equation}
DM_{\langle riz \rangle} = (DM_r+DM_i+DM_z)/3 ~.
\label{eq:DMriz}
\end{equation}
The averaged modulus is less affected by color-dependent systematics, and therefore the discussion of systematics that follows is founded on the combined $\langle riz \rangle$ distance moduli as the reference of comparison. Only galaxies with photometry quality grades, $Q_s$ and/or $Q_w$ (columns 70 and 71 of Table~\ref{tab_data}) better than 3 are considered. 
Moduli are adjusted for the residual Malmquist bias.
In the following subsections, we investigate systematics by comparing moduli $DM_\lambda$, $\lambda=r,i,z,W$1, against the composite moduli $DM_{<riz>}$ and provide alleviating formalisms. 

We plot deviations of $DM_{\lambda}$ from $DM_{\langle riz \rangle}$ versus various distance-independent observables, $\Theta$, with 
 $\mathcal{F}^{(t)}_{\lambda}(\Theta)=DM^{(t)}_{\lambda}-DM_{\langle riz \rangle}$, where $\mathcal{F}^{(t)}$ is modeled by a polynomial function of the form $\sum_{n=0}^{N} C^{(t)}_n \Theta^n$, and $\Theta$ is a distance independent galaxy observable such as color, \hi line width, surface brightness, etc. The degree of the polynomial function, $N$, is chosen based on the shape of the correlation between the moduli differences and the selected parameter for the adjustment. We perform the adjustments in a series of steps, where at each step, $t$, the output of the previous step, $DM_{\lambda}^{(t-1)}$, is adjusted following $DM_{\lambda}^{(t)} = DM_{\lambda}^{(t-1)} - \mathcal{F}^{(t)}(\Theta)$, where $DM_{\lambda}^{(0)}$ is set to $DM_{\lambda}$. After performing the adjustments at each step, we evaluate the rms of the moduli discrepancies to measure the efficiency of the adjustments. We continue the adjustment process until no further improvements can be made.

\subsubsection{Adjusting the {\it r}-band distances} \label{sec:adjsut_rband}

\begin{figure}
\centering
\includegraphics[width=\linewidth]{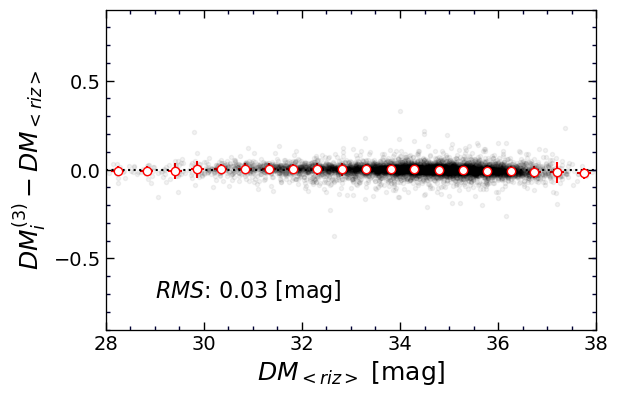}
\includegraphics[width=\linewidth]{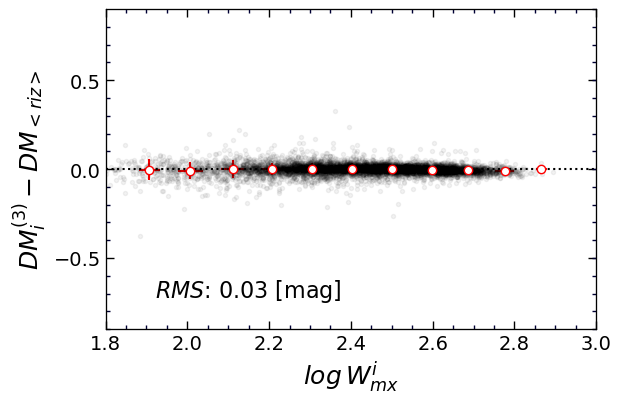}
\caption{
Analogous to Figure \ref{fig:DMr12}. Assessments of the adjusted {\it i}-band moduli.
\label{fig:DMi2_test}}
\end{figure} 

We correct the systematics of the {\it r} band moduli in five steps: (1) The top left panel of Figure \ref{fig:DMr12} displays the deviations of $DM_r$ from $DM_{\langle riz \rangle}$ as a function of $r^*-i^*$ color. The color dependency of the deviations can be described by a linear function, $\mathcal{F}^{(1)}$. 
The adjusted moduli are then derived from $DM_r^{(1)} = DM_r - \mathcal{F}^{(1)}(r^*-i^*)$.
The top right panel of Figure~\ref{fig:DMr12} plots the resultant $DM_r^{(1)}-DM_{\langle riz \rangle}$, leaving negligible residual correlation with $r^*-i^*$. The rms of discrepancies is $0.06$ mag at the end of step 1. (2) We continue our process by adopting line widths for corrections. Line width and color are not completely correlated, so this second round of adjustments can reduce discrepancies by offering extra information. The bottom left panel of Figure \ref{fig:DMr12} displays the correlation of $DM_r^{(1)}-DM_{<riz>}$ with ${\rm log}W^i_{mx}$, which is modeled by a quadratic function. Subsequent to the adjustments, the rms of $DM_r^{(2)}-DM_{\langle riz \rangle}$ is reduced to $0.05$ mag. (3) In the third step, the adjusted moduli of the previous step, $DM_r^{(2)}$, is adopted to advance our operation to improve the remaining systematics. The top panels of Figure \ref{fig:DMr345} illustrate how the moduli deviations, before and after adjustments, are correlated to $DM_r^{(2)}$.  The formulation $\mathcal{F}^{(2)}_r(DM_r^{(2)})$ uses a third-degree polynomial function. The third level of adjustments lowers the rms of moduli deviations to $0.04$ mag, revealing the effectiveness of the distance parameter in resolving systematics. (4) The adjustments in the previous step introduces an extra systematic that is correlated to line width. Hence, we consider using line width, again despite the fact it was used before in step 2. After the adjustments, the rms of the deviations is down to $0.03$ mag. (5) In the last step, we base an adjustment on $g^*-z^*$ (bottom panels of Figure \ref{fig:DMr345}). After applying the adjustments, the rms deviations are not significantly improved, and therefore we stop the adjustments chain. Ultimately, we test $DM^{(5)}_r$ against $DM_{\langle riz \rangle}$ to look for potential remaining systematics. Figure \ref{fig:DMr5_test} plots $DM^{(5)}_r-DM_{\langle riz \rangle}$ versus $DM_{\langle riz \rangle}$, line width and color. In all cases, there is no indication of a significant remaining systematic that can be further improved through more steps of adjustments. 

In practice, we have explored various adjusting various parameters and different permutations. The order of adjustments and parameters is chosen to achieve the smallest moduli deviations possible.
Choosing a different set of parameters and shuffling the order of adjustments may require more steps to get the same results. At some point, increasing the number of steps does not particularly improve the rms of deviations once it reaches approximately the level of our photometry accuracy of $\sim 0.05$ mag. 

As an alternative approach, one can carry out a principal component analysis to generate a set of fully independent parameters to adjust distances. However, the superiority of our method is in its flexibility and iterative nature, which allows us to examine as many parameters as are relevant and to address nonlinear correlations whenever needed.
 
\begin{figure*}
\centering
\includegraphics[width=\linewidth]{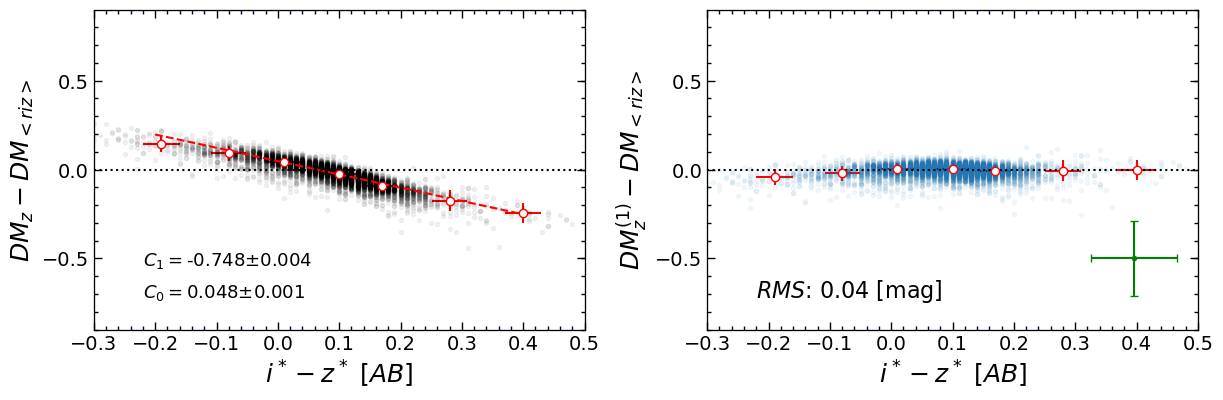} \\
\includegraphics[width=\linewidth]{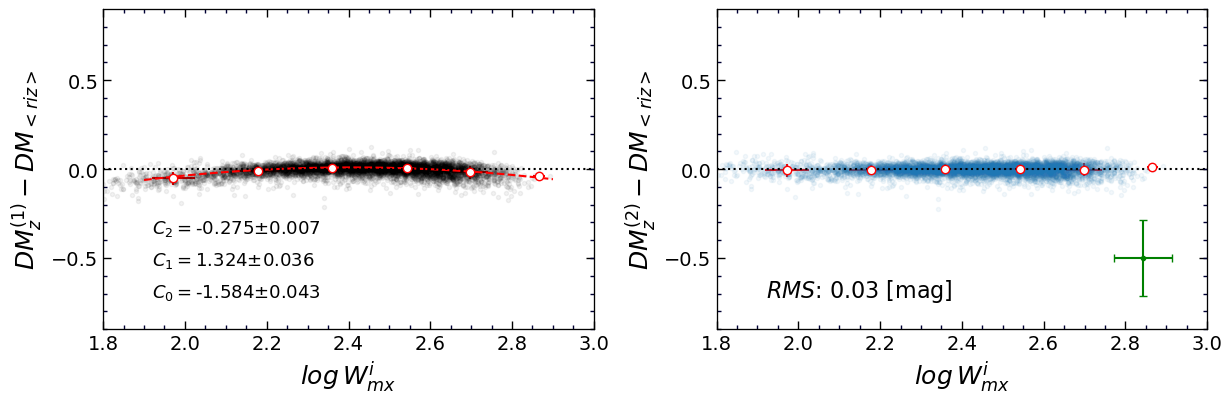} \\
\includegraphics[width=\linewidth]{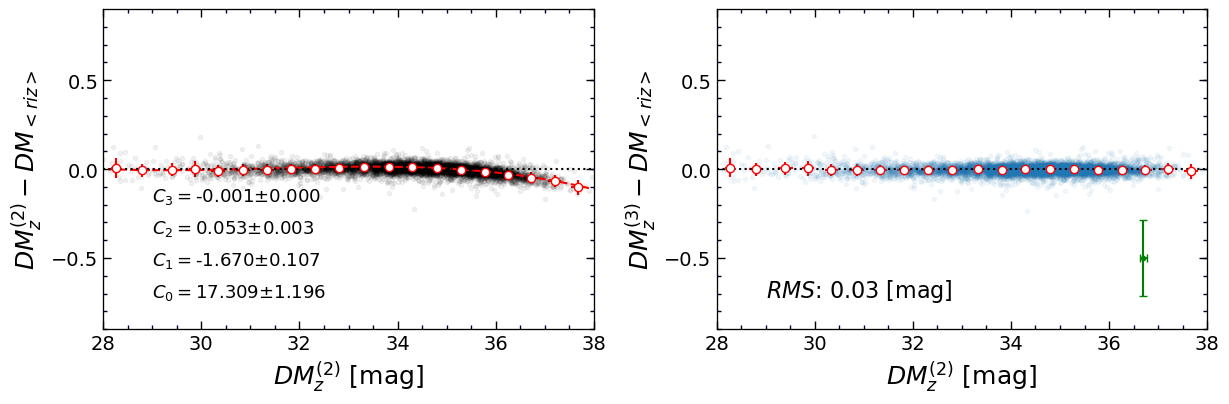}
\caption{
Similar to Figure \ref{fig:DMr12}. Each step illustrates one step of the adjustment process of the {\it z}-band distance moduli.
\label{fig:DMz123}}
\end{figure*} 

\begin{figure}
\centering
\includegraphics[width=\linewidth]{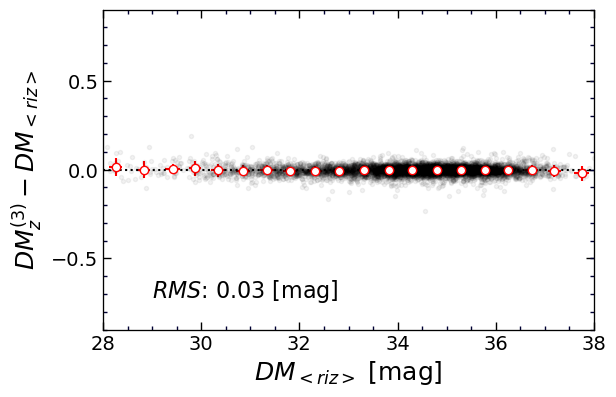}
\includegraphics[width=\linewidth]{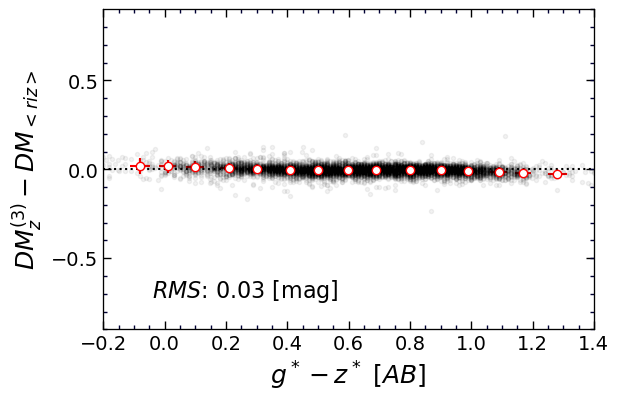}
\caption{
Analogous to Figure \ref{fig:DMz123}. Assessments of the adjusted {\it z}-band moduli.
\label{fig:DMz2_test}}
\end{figure} 

\begin{figure*}
\centering
\includegraphics[width=\linewidth]{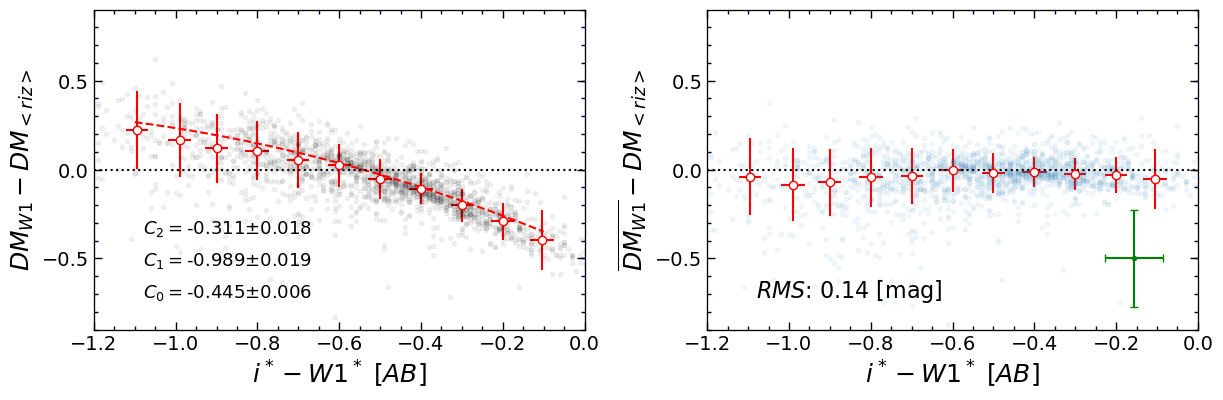} \\
\caption{
Similar to Figure \ref{fig:DMr12} for spirals in our sample with both optical and infrared photometry coverage, (OP+IR subsample). Here, $\overline{DM_{W1}}$ is the adjusted modulus that equals $DM_{W1}-\overline{\mathcal{F}}(i^*-W1^*)$, where $\overline{\mathcal{F}}=\sum_{n=0}^{2} C_n (i^*-W1^*)^n$.
\label{fig:DMw1_bar}}
\end{figure*}

\subsubsection{Adjusting {\it i}-band distances} \label{sec:adjsut_iband}

We follow the same methodology as described in \S \ref{sec:adjsut_rband} to correct the color systematics of the {\it i}-band distances. As illustrated in Figure \ref{fig:DMi123}, we first start the adjustment process with the $r^*-i^*$ color. After the corresponding adjustment, the rms  of $DM_r-DM_{\langle riz \rangle}$ is $0.04$ mag. Two further sequential adjustments using $DM_i^{(1)}$ and $g^*-z^*$ reduce the rms to $0.03$ mag, a value similar to that achieved after the entire adjustment process of the {\it r}-band distances. Here, after the first step, we observe no correlation between moduli discrepancies and the \hi line width. Therefore, the adjustment process requires fewer steps to produce acceptable results. It is shown in Figure \ref{fig:DMi2_test} that the differences of $DM_i^{(3)}$ from the average {\it r,i,z} moduli are not noticeably correlated to $DM_{\langle riz \rangle}$ and line width.

\subsubsection{Adjusting {\it z}-band distances} \label{sec:adjsut_zband}

In a similar fashion, we first adopt the $i^*-z^*$ color for adjustments (see Figure \ref{fig:DMz123}). We observe an inverse correlation between the moduli differences and color that is in agreement with our conclusion earlier in this section that the unadjusted moduli of redder spirals are smaller at longer passbands. In the second step, we use line width for corrections, with the correlation that is described by a quadratic relation. In the last step, we find a slight correlation of discrepancies with the doubly adjusted moduli, $DM^{(2)}_z$.  
In a search for remaining hidden systematics, Figure \ref{fig:DMz2_test} plots the differences between the triply adjusted moduli and $DM_{\langle riz \rangle}$ versus the average {\it r,i,z} distances and the $g^*-z^*$ colors. This figure, together with examinations with other parameters, fails to reveal further systematics that can be significantly improved by introducing more adjustment steps. The rms scatter of moduli differences is $0.03$ mag, which is similar to that obtained in \S \ref{sec:adjsut_rband} and \ref{sec:adjsut_iband} for the {\it r} and {\it i} bands. We conclude that we are reaching the statistical limit of adjustment capabilities.

\begin{figure*}
\centering
\includegraphics[width=\linewidth]{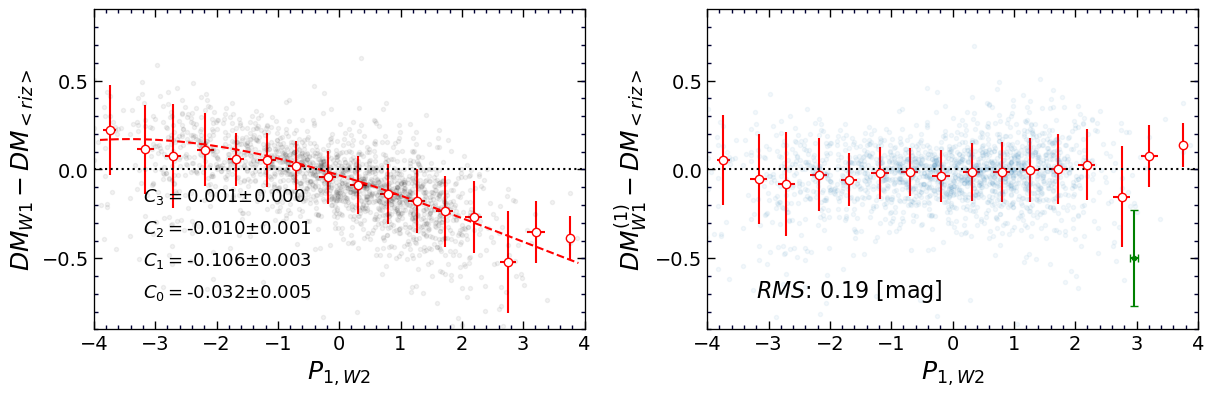} \\
\includegraphics[width=\linewidth]{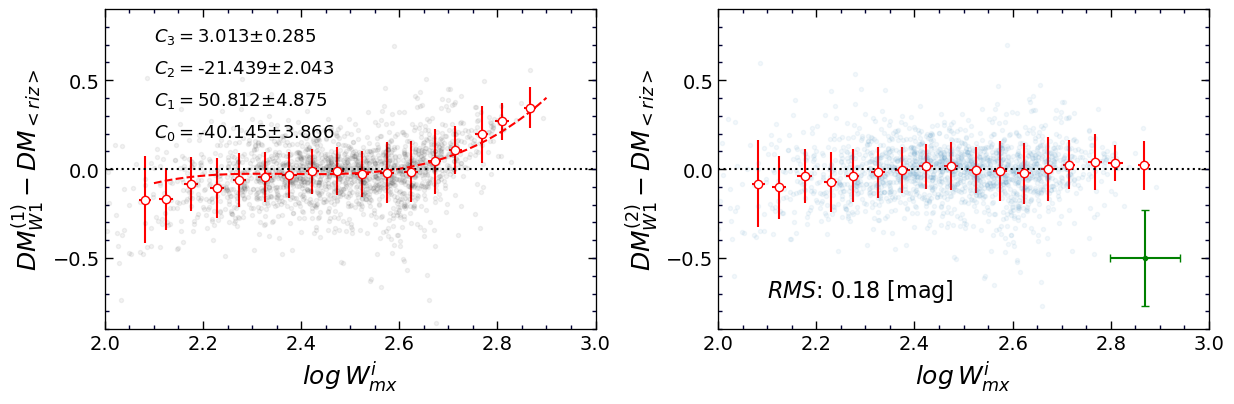} \\
\caption{
Analogous to Figure \ref{fig:DMr12}, adjustments of distance moduli at {\it W1} band for spirals in our sample with infrared WISE photometry data. Adjustments are carried out in two steps using the main principal component at {\it W2} band, $P_{1,W2}$, and \hi line width, ${\rm log} W^i_{mx}$. 
\label{fig:DMw1_12}}
\end{figure*}

\begin{figure}
\centering
\includegraphics[width=\linewidth]{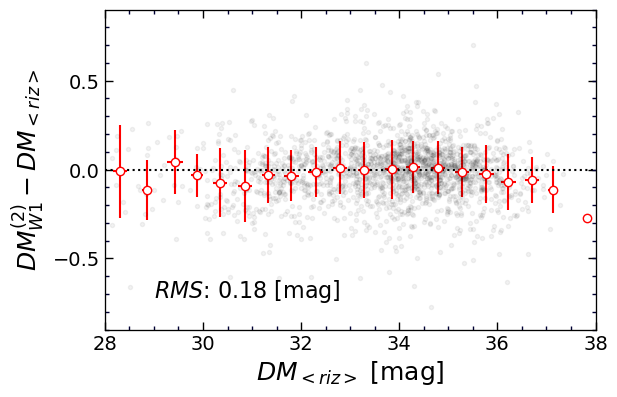} \\
\includegraphics[width=\linewidth]{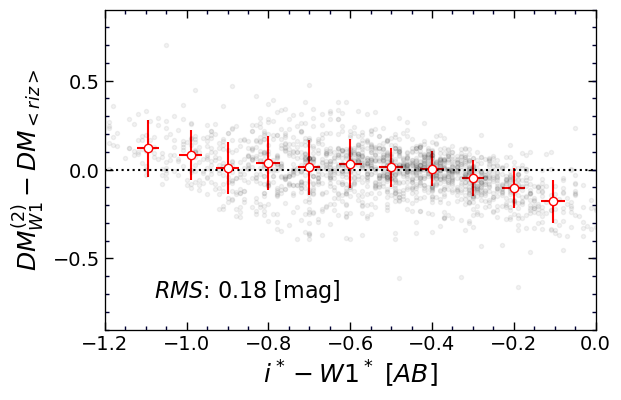}
\caption{
Similar to Figure \ref{fig:DMz123} for the deviations of the adjusted {\it W1}-band moduli from the $<riz>$ moduli.
\label{fig:DMw1_test}}
\end{figure} 

\subsubsection{Adjusting {\it W1}-band distances} \label{secadjsut_w1band}

If both optical and infrared information are available, an adjustment sequence such as that begun in Figure~\ref{fig:DMw1_bar} can be initiated.  Adjustment for the pronounced $i^*-W1^*$ color dependency reduces scatter to $0.14$ mag.
In practice, we avoid using {\it W}1 distances when we have optical photometry, given the larger scatter of the $TF_{W1}$ relation.  A more serious issue arises because a $i^*-W1^*$ adjustment is impossible if optical photometry is not available, as is the case for $\sim 2/3$ of our infrared sample. The $W1^*-W2^*$ color is not a useful parameter to incorporate into our analysis, because it is very weakly correlated to the other features of spiral galaxies, such as size, line width, surface brightness, absolute luminosity, and so on. Therefore, we need to start our adjustment process with some other distance-independent parameter that is applicable to the entire infrared sample and entirely relies on the infrared information.

A most promising parameter is the main principal component that was originally introduced for the calculation of dust attenuation in spirals, $P_{1,W2}$ (see Equation~\ref{Eq:P1}), which carries crucial information about galaxy characteristics through packaging some of their most important distance-independent features. The high correlation of $P_{1,W2}$ with other spiral features allows this parameter to be utilized in our analysis as an effective substitute for optical$-$infrared colors. Accordingly, as seen in Figure~\ref{fig:DMw1_12}, in the first step we adopt $P_{1,W2}$ for the adjustment of {\it W}1 moduli. After employing the first round of adjustments, the rms of discrepancies is $0.19$ mag. The scatter is slightly reduced in the second step, which uses line width as the adjusting parameter. The small effect of the second round of corrections is not surprising, given that $P_{1,W2}$ contains ${\rm log} W^i_{mx}$ as a component, ergo line width alone does not introduce much more knowledge to the process. 

Unfortunately, we are unable to advance our adjustment procedure any further, due to the lack of any additional distance-independent parameter. Inspired by the illustrated correlation in the top left panel of Figure~\ref{fig:DMr345}, we explore the capability of distance moduli to further reduce discrepancies. However, it is seen in the top panel of Figure~\ref{fig:DMw1_test} that there is no useful correlation between $DM_{W1}^{(2)}-DM_{<riz>}$ and $DM_{<riz>}^{(2)}$ for the initiation of another round of adjustments. As illustrated in the bottom panel of Figure \ref{fig:DMw1_test}, in those cases where optical information is available, there is still a residual color-dependent systematic in $DM_{W1}^{(2)}$ that cannot be efficiently removed with infrared information alone.

\begin{figure}
\centering
\includegraphics[width=\linewidth]{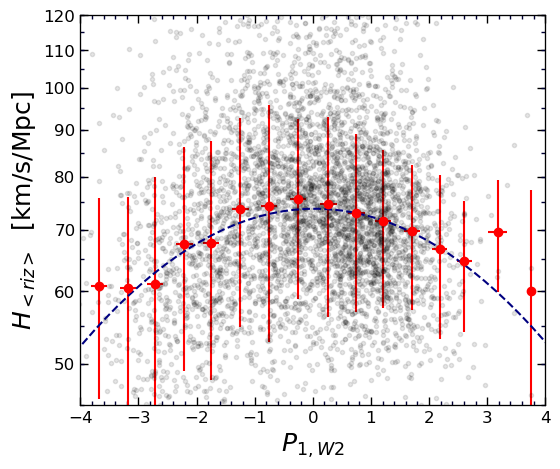}
\includegraphics[width=\linewidth]{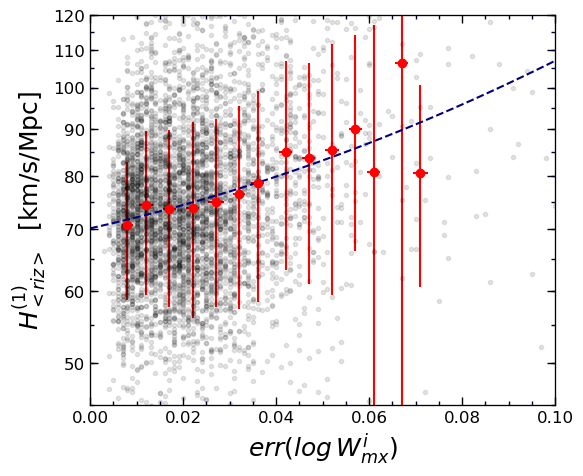}
\caption{
{\bf Top:} Hubble parameter based on merged $\langle riz \rangle$ photometry parameter vs. principal component, $P_{1,W2}$. 
{\bf Bottom:} Hubble parameter, after distance regularization described by the dashed curve in top panel, vs. error on line width. 
Each black point represents a galaxy. Red points exhibit the average of the data points within bins of constant size, with error bars showing the 1$\sigma$ scatter in data.
In both panels, navy dashed curves have quadratic forms.
Only galaxies with Local Sheet referenced velocities larger than $4000$ \kms\ are plotted.
\label{fig:Hi_P1logW}}
\end{figure}

\begin{figure}
\centering
\includegraphics[width=\linewidth]{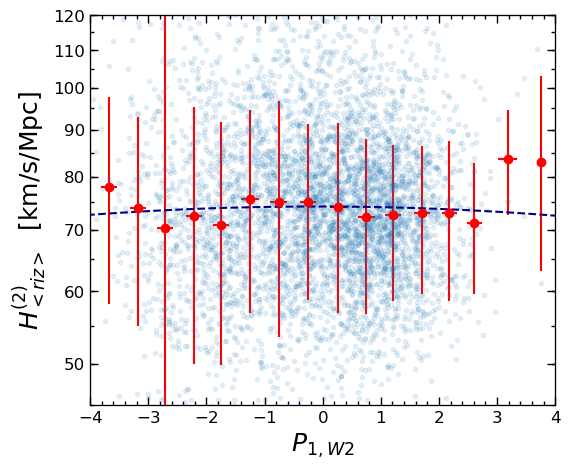}
\includegraphics[width=\linewidth]{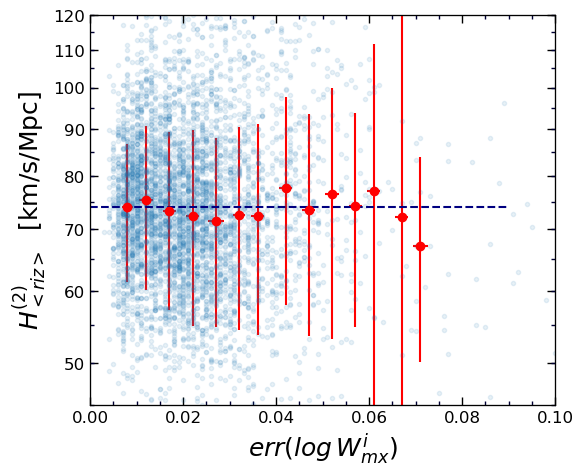}
\caption{
The $<riz>$ Hubble parameter after applying distance regularizations that eliminate the illustrated systematics in 
Figure \ref{fig:Hi_P1logW}. Other details are analogous to those in Figure \ref{fig:Hi_P1logW}.
\label{fig:Hi_test}}
\end{figure} 

\begin{table*}[t]
\begin{center}
\scriptsize
\caption{Distance Catalog}
\begin{tabular}{rcccccccccccc} 
\hline
PGC & $V_{LS}$ & $V_{CMB}$ & $f$ & $DM_{best}$ & $DM_r$ & rms$_r$ & $DM_i$ & rms$_i$ & $DM_z$ & rms$_z$ & $DM_{W1}$ & rms$_{W1}$ \\
    &  \kms     &   \kms    &  & mag  &  mag   & mag  &  mag   & mag  &  mag   & mag  &  mag   & mag  \\
 (1) & (2) & (3) & (4) & (5) & (6) & (7) & (8) & (9) & (10) & (11)  & (12) & (13)   \\ 
\hline \hline
2 & 5296 & 4726 & 1.013 & 34.24$\pm$0.29 &  &  &  &  &  &  & 34.24$\pm$0.29 & 0.48  \\
4 & 4706 & 4109 & 1.011 & 33.29$\pm$0.17 & 33.31$\pm$0.16 & 0.56 & 33.29$\pm$0.17 & 0.60 & 33.28$\pm$0.17 & 0.59 & 33.25$\pm$0.17 & 0.66  \\
12 & 6685 & 6195 & 1.016 & 35.03$\pm$0.22 &  &  &  &  &  &  & 35.03$\pm$0.22 & 0.48  \\
16 & 5809 & 5312 & 1.014 & 34.70$\pm$0.24 & 34.70$\pm$0.23 & 0.39 & 34.70$\pm$0.24 & 0.40 & 34.69$\pm$0.24 & 0.40 & 34.71$\pm$0.25 & 0.53  \\
55 & 5052 & 4454 & 1.012 & 34.00$\pm$0.24 & 34.00$\pm$0.23 & 0.56 & 34.00$\pm$0.24 & 0.59 & 34.01$\pm$0.24 & 0.59 & 34.09$\pm$0.25 & 0.65  \\
68 & 7740 & 7338 & 1.019 & 34.81$\pm$0.37 & 34.80$\pm$0.36 & 0.52 & 34.82$\pm$0.37 & 0.52 & 34.81$\pm$0.38 & 0.52 & 34.72$\pm$0.41 & 0.61  \\
70 & 7040 & 6447 & 1.017 & 35.12$\pm$0.12 & 35.10$\pm$0.12 & 0.35 & 35.12$\pm$0.12 & 0.39 & 35.12$\pm$0.12 & 0.40 & 35.08$\pm$0.10 & 0.48  \\
76 & 7183 & 6583 & 1.017 & 34.73$\pm$0.16 & 34.73$\pm$0.16 & 0.35 & 34.73$\pm$0.16 & 0.39 & 34.74$\pm$0.16 & 0.40 & 34.77$\pm$0.15 & 0.48  \\
92 & 5592 & 5015 & 1.013 & 33.25$\pm$0.18 & 33.27$\pm$0.17 & 0.56 & 33.24$\pm$0.18 & 0.60 & 33.23$\pm$0.18 & 0.59 &  &   \\
94 & 4367 & 3995 & 1.011 & 33.89$\pm$0.32 &  &  &  &  &  &  & 33.89$\pm$0.32 & 0.65  \\
96 & 14934 & 14380 & 1.038 & 36.17$\pm$0.13 & 36.19$\pm$0.13 & 0.35 & 36.16$\pm$0.13 & 0.39 & 36.16$\pm$0.13 & 0.40 &  &   \\
102 & 5323 & 4726 & 1.013 & 34.20$\pm$0.13 & 34.19$\pm$0.13 & 0.35 & 34.20$\pm$0.13 & 0.39 & 34.20$\pm$0.13 & 0.40 & 34.41$\pm$0.11 & 0.48  \\
124 & 6529 & 5988 & 1.016 & 34.49$\pm$0.14 & 34.49$\pm$0.14 & 0.51 & 34.49$\pm$0.14 & 0.51 & 34.49$\pm$0.14 & 0.51 & 34.54$\pm$0.13 & 0.61  \\
128 & 12829 & 12230 & 1.032 & 35.90$\pm$0.23 & 35.90$\pm$0.22 & 0.51 & 35.92$\pm$0.23 & 0.51 & 35.89$\pm$0.23 & 0.51 &  &   \\
146 & 6590 & 6004 & 1.016 & 34.39$\pm$0.27 & 34.40$\pm$0.26 & 0.55 & 34.38$\pm$0.27 & 0.57 & 34.38$\pm$0.27 & 0.56 & 34.24$\pm$0.29 & 0.64  \\
155 & 8156 & 7582 & 1.020 & 34.51$\pm$0.37 & 34.52$\pm$0.35 & 0.54 & 34.52$\pm$0.37 & 0.55 & 34.50$\pm$0.37 & 0.55 & 34.33$\pm$0.40 & 0.63  \\
165 & 7878 & 7280 & 1.019 & 34.69$\pm$0.13 & 34.69$\pm$0.13 & 0.48 & 34.69$\pm$0.13 & 0.48 & 34.69$\pm$0.13 & 0.47 & 34.72$\pm$0.12 & 0.58  \\
171 & 2725 & 2326 & 1.006 & 33.29$\pm$0.29 &  &  &  &  &  &  & 33.29$\pm$0.29 & 0.57  \\
176 & 6612 & 6109 & 1.016 & 34.97$\pm$0.26 & 34.96$\pm$0.25 & 0.35 & 34.98$\pm$0.26 & 0.39 & 34.98$\pm$0.26 & 0.40 & 34.94$\pm$0.28 & 0.48  \\
179 & 5736 & 5175 & 1.014 & 34.04$\pm$0.27 & 34.04$\pm$0.26 & 0.46 & 34.05$\pm$0.27 & 0.46 & 34.04$\pm$0.28 & 0.45 & 34.07$\pm$0.30 & 0.57  \\
\nodata \\
\hline
\end{tabular}
\label{tab:distance_catal}
\end{center}
\begin{quote}
The complete version of this table is available online.
\end{quote}
\end{table*}

\subsection{Distance Regularization with \hnut} \label{sec:H0_regularization}

We have been using the combination of optical bands $\langle riz \rangle$ as a reference for adjustments to assure consistency across all the bands useful for determining distances (Equation \ref{eq:DMriz}). However, there is no guarantee that $DM_{\langle riz \rangle}$ is immune from systematics. Here, we make checks based on external information.

At substantial redshifts, peculiar velocities are small compared with Hubble expansion velocities.  As peculiar velocities become only a minor cause of scatter, within the framework of $\Lambda$CDM cosmology, the average Hubble parameter should be roughly constant. 
For each galaxy, we can construct its Hubble parameter, $H=fV_{cmb}/d_{\langle riz \rangle}$, where $d_{\langle riz \rangle}$ is the composite $\langle riz \rangle$ distance to a galaxy with velocity in the CMB frame, $V_{cmb}$, modified by a small cosmological correction.\footnote{The cosmological correction term is
$$
f = 1 + \frac{1}{2} [1 - q_0]z - \frac{1}{6} [1 - q_0 -3 q_0^2 + j_0]z^2~,
$$
where $j_0 = 1$, and $q_0 = {\frac{1}{2}} ( \Omega_m -2 \Omega_{\Lambda} ) = -0.595$ assuming $\Omega_m=0.27$, $\Omega_m+\Omega_{\Lambda}=1$, and $z=V_{LS}/c$ \citep{2006PASP..118.1711W}.
}

Figure \ref{fig:Hi_P1logW} plots the composite $\langle riz \rangle$ Hubble parameter versus the main principal component, $P_{1,W2}$, and the \hi line width error.  The Hubble parameter is plotted in the logarithm because the dominant errors are Gaussian-distributed in distance modulus (observed velocity errors are minor).  At small distances, peculiar velocities can be a significant fraction of the Hubble expansion rate, so we exclude galaxies with radial velocities in the frame of the Local Sheet, $V_{LS}$, smaller than 4000 \kms.
Red points illustrate averages of the logarithm of the Hubble parameter in equally spaced horizontal bins. Looking at the top panel of Figure \ref{fig:Hi_P1logW}, there is a manifestation of a remaining systematic that is modeled by the dashed navy curve that has a quadratic form, $F^{(1)}_i = \sum_{n=0}^{2} C^{(i)}_n P^n_{1,W2}$.

This systematic can be corrected via a process similar to the one we explained in \S \ref{sec:adjsut_rband}. Accordingly, the adjusted Hubble parameter is derived using ${\rm log} H^{(1)}_{riz}= {\rm log} H_{riz} - F^{(1)}_{riz}(P_{1,W2}) + \langle {\rm log} H_{riz} \rangle$, where
$\langle {\rm log} H_{riz} \rangle$ is the median of the error-weighted Hubble parameter in log space, which preserves the average value of the Hubble parameter.
The regularized distance modulus is $DM^{(1)}_{\langle riz \rangle} = DM_{\langle riz \rangle} + 5F^{(1)}_{riz}(P_{1,W2}) - 5 \langle {\rm log} H_{riz} \rangle$.

It can be seen in the bottom panel of Figure \ref{fig:Hi_P1logW} that, after the first round of regularization, there is still a remaining systematic that is correlated to the line width error. 
The reasons for this residual systematic are complex. The principal component parameter $P_{1,W2}$ used in step one is a composite of line width, ratio of \hi to infrared fluxes, and infrared surface brightness (K19, K20).  Massive, high surface brightness, relatively gas deficient galaxies have relatively positive $P_{1,W2}$ values, while small gas rich galaxies have relatively negative values.  Step one drives each extreme in $P_{1,W2}$ toward lower $H_0$. The correlation in step two partially redresses this effect.  At the high-mass end, our flux-limited \hi observations favors inclusion of smaller over larger line widths (higher $S/N$ due to compressed flux), i.e. those tending to the left of the TFR but with substantial errors because they are near the $S/N$ limit. At the small galaxy end, fractional errors in line width are largest for the smallest line widths, i.e, again those tending to the left of the TFR.  
This hidden systematic is ameliorated using the line width error, $err({\rm log ~ W^i_{mx}})$, to achieve the distance regularization that results in $H^{(2)}_{\langle riz \rangle}$ (see Figure \ref{fig:Hi_test}).

Now that the composite $\langle riz \rangle$ distances are regularized, we apply the same sequence of regularizations on the adjusted distances at other wave bands; $DM^{(5)}_r$, $DM_i^{(3)}$, $DM^{(3)}_{z}$ and $DM^{(2)}_{W1}$. Ultimately, to ensure that the measured distances in all passbands are in agreement with each other, we repeat the adjustments as described in sections \ref{sec:adjsut_rband}, \ref{sec:adjsut_zband} and \ref{secadjsut_w1band}, followed by the regularizations we explained in this section.

\begin{figure}[t]
\centering
\includegraphics[width=1.0\linewidth]{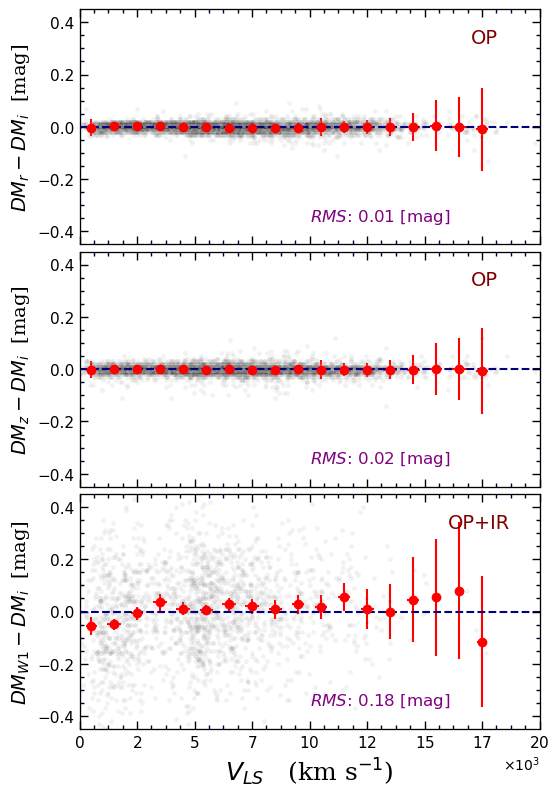}
\caption{ Offset of distance moduli measured at the {\it r}, {\it z} and {\it W1} bands from those measured at the {\it i} band, as a function of radial velocity. Each gray point represents a galaxy. Red points show average of the data points within the bins of equal size, with their error bars representing the 1$\sigma$ scatter of data. In each panel, RMS is the root mean square of the moduli offset. Maroon labels of the top right corners denote the considered galaxy subsample, with ``OP" and ``IR" standing for optical and infrared photometry coverage, respectively.
}
\label{fig:DM_compare}
\end{figure} 
\subsection{Final comparisons} \label{sec:dist_evals}

We check that there are no remaining systematics between bands correlated with the independent variable redshift. 
The final values of the measured distance moduli at {\it i} band are taken as a reference. Figure \ref{fig:DM_compare} illustrates the $DM_\lambda-DM_i$ differentials as a function of their radial velocities in the frame of Local Sheet, with distance moduli taken from Table \ref{tab:distance_catal}\footnote{ 
The {\it ``CF4 TF-distances"} table on the the Extragalactic Distance Database (\url{http://edd.ifa.hawaii.edi}) includes all the information of Table \ref{tab:distance_catal} and also provides more information on each individual galaxy. While the machine-readable version of Table \ref{tab:distance_catal} is fixed at publication, the table at the Extragalactic Distance Database is expected to receive updates.}. 
The adjustment process successfully enforces the means of the moduli differences to be zero. As expected, the optical moduli are in better agreement with each other because the availability of color terms has provided enough information to remove the color-dependent systematics. The rms of the optical moduli offset is no worse than $0.03$ mag, as also manifested in Figures \ref{fig:DMr5_test}, \ref{fig:DMi2_test} and \ref{fig:DMz2_test}. 

It can be seen in the bottom panel of Figure~\ref{fig:DM_compare} that, while there is no strong correlation with redshift, the $DM_{W1}-DM_i$ offsets exhibit larger scatter. This increase is due to the lack of optical photometry in the construction of the adjusted infrared moduli. 

A minor contributor to scatter arises because dust attenuation in spirals is negligible at WISE bands compared to that at optical passbands. Any uncertainty in the evaluation of dust attenuation influences the measured distance moduli at all optical passbands in the same direction, with a minimal impact upon discrepancies between optical distances. At the {\it W}1 band, dust attenuation has a minor effect on the measured distance moduli.  In comparisons with optical moduli, inaccuracies in measurements of dust obscurations at optical bands translate to statistical scatter in infrared$-$optical moduli offsets.

\section{Distance Catalog} \label{sec:distance_catal}

The finalized distance moduli of 9792 spiral galaxies are given in Table \ref{tab:distance_catal}.  Potential targets have been excluded if their inferred absolute magnitudes are fainter than $M_i = -17$ or $M_{W1} = -16.1$ or they are extreme outliers as evaluated from inferred Hubble parameter values\footnote{Cases with deviant Hubble parameter values greater than $3.5\sigma$ were rejected if anomalous in any way. A small number of such deviant cases have been retained because no basis was discerned for exclusion.}, including cases arising in connection with the comparisons discussed in \S~\ref{sec:literature_comp}. 
All distances are corrected for the band-to-band color-dependant systematics discussed in \S \ref{sec:col_dep_sys} and regularized following the process we explained in \S \ref{sec:H0_regularization}. 

Descriptions of the columns are as follows:
Column (1) gives the PGC ID of the galaxy. Columns (2) and (3) list radial velocities in the Local Sheet and CMB rest frames, respectively. Column (4) gives the cosmological correction factors defined in \S \ref{sec:H0_regularization}. Column (5) provides our best distance modulus; by preference, the average of {\it r, i, z} moduli or, if missing SDSS photometry, the {\it W}1-band modulus. 
For a given galaxy, the measured distances at different passbands are not completely independent of each other.  Analyses at all bands are based on the same \hi linewidth measurement. Moreover, the adjustment procedures combine the photometry information of different passbands. We adopt the median of the {\it r, i, z} moduli uncertainties as the uncertainty of the best distance modulus for cases where the average of {\it r, i, z} moduli is adopted.
Column (6) tabulates the measured distance moduli at {\it r} band after applying all the corrections.
The moduli uncertainties are calculated by the Gaussian propagation of the uncertainties in the associated measured quantities (\hi line-widths, color indices, $P_{1,W2}$) and in the optimized parameters of the adopted TFR relations given in Table \ref{tab:revised_ITFR} and the adjusting/regularizing relations (\S \ref{sec:col_dep_sys} and \S \ref{sec:H0_regularization}). See Appendix \ref{sec:uncertainties} for further discussions.
Column (7) lists the rms scatter about the $TF_r$ relation, rms$_r$, which is a function of the {\it r}-band absolute magnitude (see Figure 9 of K20). In a similar fashion, columns (8), (10), and (12) list fully modified distance moduli at the {\it i}, {\it z} and {\it W}1 bands, with the rms scatters of the corresponding TFRs listed in columns (9), (11), and (13).
The entries associated with missing optical or infrared photometry data are left blank.

\section{Comparisons with Alternate Sources of TFR Distances} \label{sec:literature_comp}

\begin{figure}[t]
\centering
\includegraphics[width=1.0\linewidth]{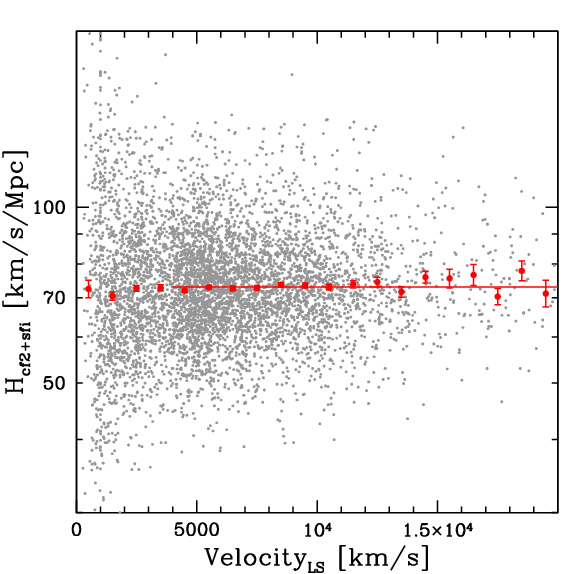}
\caption{Individual Hubble parameter values for galaxies in the combined ``cf2" and ``sfi" samples are plotted against their velocities in the Local Sheet frame as gray points.  Values averaged in 1000~\kms\ intervals are plotted in red with standard deviation error bars. Red line is at the average value over the range $4000-20,000$~\kms\ of $H_0=73$~\kmsMpc. 
}
\label{fig:cf2}
\end{figure} 

\begin{figure}[t]
\centering
\includegraphics[width=1.0\linewidth]{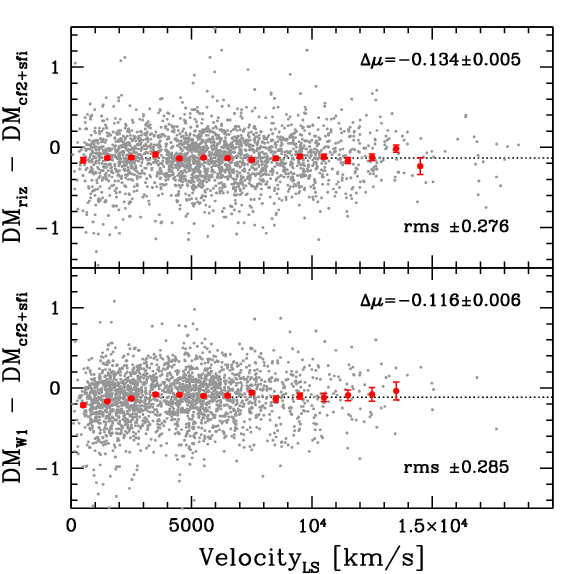}
\caption{Differences in distance moduli between values from the current study (combined {\it riz} and {\it W1} bands in top and bottom panels, respectively) and combined values from the ``cf2" and ``sfi" samples as a function of velocity.  Averaged values in velocity bins are plotted in red.  Labels record zero-point offsets and dispersions.
}
\label{fig:delmodcf2}
\end{figure} 

\begin{figure}[htp]
\centering
\includegraphics[width=1.0\linewidth]{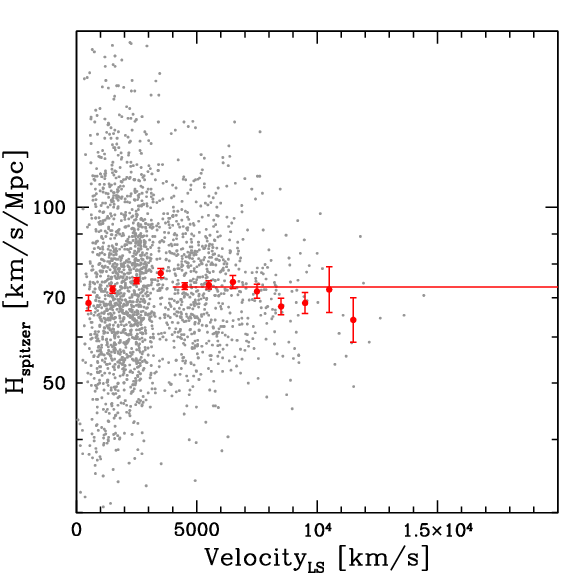}
\caption{Individual Hubble parameter values for galaxies in the ``spitzer" sample are plotted against their velocities in the Local Sheet frame as gray points.  Values averaged in 1000~\kms\ intervals are plotted in red with standard deviation error bars.  Red line is at the average value over the range $4000-20,000$~\kms\ of $H_0=73$~\kmsMpc. 
}
\label{fig:spit}
\end{figure} 

\begin{figure}[t]
\centering
\includegraphics[width=1.0\linewidth]{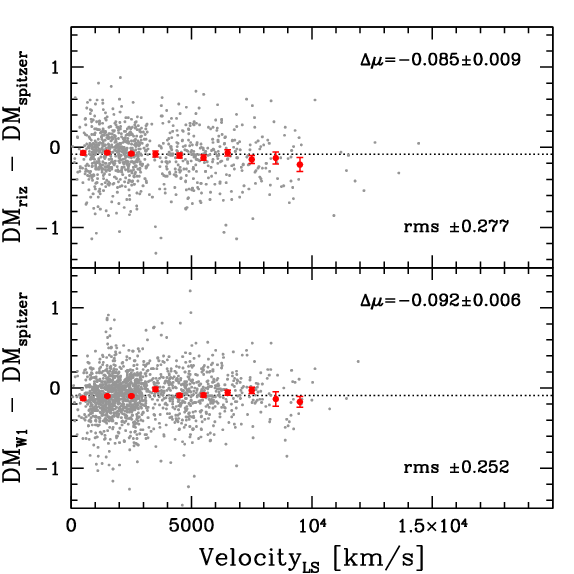}
\caption{Differences in distance moduli between values from the current study (combined {\it riz} and {\it W}1 bands are shown in top and bottom panels, respectively) and values from the ``spitzer" sample as a function of velocity.  Averaged values in velocity bins are plotted in red.  Labels record zero-point offsets and dispersions.
}
\label{fig:delmodspt}
\end{figure} 

\begin{figure}[t]
\centering
\includegraphics[width=1.0\linewidth]{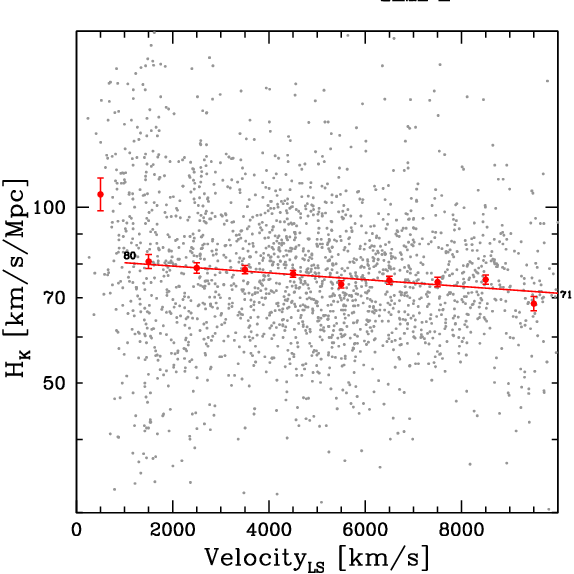}
\caption{Individual Hubble parameter values for galaxies in the ``2mtf" sample are plotted against their velocities in the Local Sheet frame as gray points.  Values averaged in 1000~\kms\ intervals are plotted in red with standard deviation error bars. Sloping red line is a fit to the binned values over the range $1000-10,000$~\kms. 
}
\label{fig:2mtf}
\end{figure} 

\begin{figure}[t]
\centering
\includegraphics[width=1.0\linewidth]{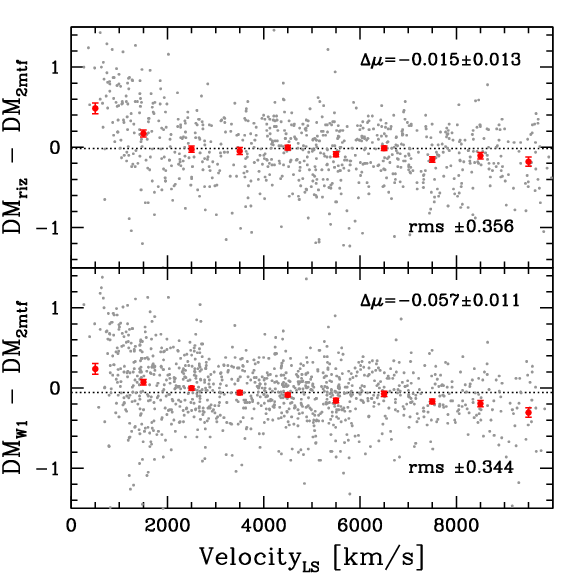}
\caption{Differences in distance moduli between values from the current study (combined {\it riz} and {\it W}1 bands in top and bottom panels, respectively) and values from the ``2mtf" sample at $K$ band as a function of velocity.  Averaged values in velocity bins are plotted in red. Dotted lines are at the overall mean difference. Labels record zero-point offsets and dispersions.
}
\label{fig:delmod2mtf}
\end{figure} 

\begin{figure}[t]
\centering
\includegraphics[width=1.0\linewidth]{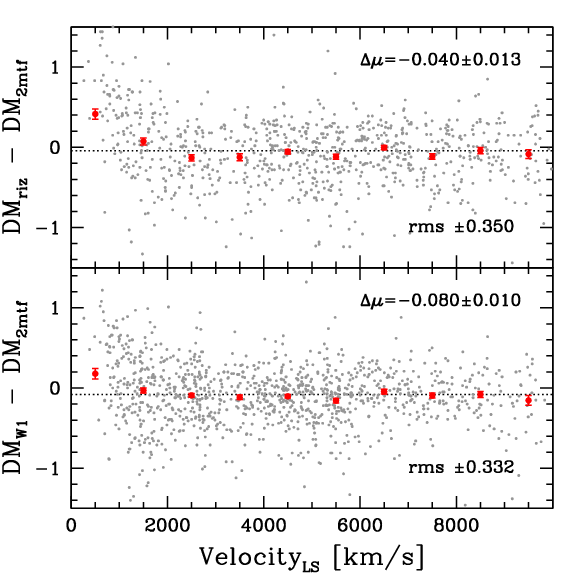}
\caption{Same as in Fig.~\ref{fig:delmod2mtf} after adjustments to 2MTF moduli.
}
\label{fig:delmod2mtfadj}
\end{figure} 

Aside from zero-point scaling, alternative TFR distance measurements of a specific target should agree.  Separate programs observe the same intrinsic kinematics and inclination by slightly different means.  Luminosities might be measured in different passbands, but magnitudes at a given line width and inclination are highly correlated on the TFR between bands.  Intercomparisons, then, provide a mechanism to match alternative sources to a common system, to monitor potential biases, and to filter out egregiously bad data.

We identify four collections that are sufficiently extensive and well-characterized for purposes of comparison.  Two of these were already included in {\it Cosmicflows-2}.  The one we call ``cf2" is our own compilation based on pointed observations for photometry at Cousins {\it I} band \citep{2011MNRAS.415.1935C}\footnote{The ``cf2" sample-includes a zero point calibration from Spitzer $3.6\mu$m photometry.} and for line widths from \hi profiles \citep{2009AJ....138.1938C, 2011MNRAS.414.2005C}. The other, which we call ``sfi" was assembled by the Cornell group \citep{2007ApJS..172..599S}, also employing {\it I} band photometry and \hi global profiles from pointed observations.  There is a substantial overlap in the raw observational materials used by these two sources, but the paths to generate distances are distinct.  A third study, which we call ``spitzer", again undertaken by our collaboration \citep{2013ApJ...765...94S, 2014MNRAS.444..527S, 2014ApJ...792..129N}, was introduced in {\it Cosmicflows-3}.  The major innovation was the use of photometry at $3.6\mu$m from pointed observations with the Spitzer Space Telescope.  The fourth contribution, which we call ``2mtf" draws on $J$, $H$, $K$ photometry from the Two Micron All Sky Survey (2MASS) and pointed \hi profile information analyzed following the procedures introduced by the Cornell group \citep{2019MNRAS.487.2061H}.  We give consideration to each of these four sources in turn.

With each sample, we apply two tests.  The first test stands alone with the source.  At large redshifts, peculiar velocities become extremely subdominant to Hubble expansion. Hence, averaged values of the Hubble parameter, $H=fV_{cmb}/d$, should be roughly constant with redshift.  

The second test involves a comparison of distance moduli between the literature sample and those of the current study.  Again, we look for trends with redshift.  Moreover, we give attention to strongly deviant cases.  Data that deviate by $3\sigma$ in {\it both} the Hubble parameter plot and the plot of distance modulus differences are rejected. 

Detailed inspection confirms that the ``cf2" and ``sfi" samples are coherently matched, which was the intent when they were merged in the construction of the {\it Cosmicflows-2} catalog.  The two samples are combined in the following discussion.  It is shown in Figure~\ref{fig:cf2} that the combination of the full ``cf2" and ``sfi" samples passes the first test.  Data scatter around a constant mean value of the Hubble parameter as a function of systemic velocity.  These samples equally pass the second test, as seen in Figure~\ref{fig:delmodcf2}.  Here, distance modulus comparisons are made alternately with the combined {\it riz} and {\it W}1 band values of this paper, with the combined ``cf2" and ``sfi" samples.  Giving attention to the double $3\sigma$ rejection test, comparisons between CF4 {\it riz} and ``cf2"+``sfi" result in six and ten rejections, respectively, of 2517 matches.  In comparisons with CF4 {\it W}1 band, with ``cf2"+``sfi" there are 11 and 12 rejections, respectively, of 2404 matches.  Far less than 1\% are rejected in all cases.  We conclude that the ``cf2" and ``sfi" samples can be successfully merged with the present sample after suitably accounting for zero-point offsets.

Our tests show the ``spitzer" sample proves to be equally well-behaved with our tests.  The run of the Hubble parameter with systemic velocity seen in Figure~\ref{fig:spit} oscillates around a constant value.  Distance modulus offsets are roughly constant in comparisons between ``spitzer" values and alternatively our {\it riz} and {\it W}1 band values as seen in Figure~\ref{fig:delmodspt}. The distributions with velocity in these plots reveal a dual-selection property of the ``spitzer" sample.  The enhanced number within $3000$~\kms\ results from an earnestness to be complete locally with coverage extended as far as possible to low galactic latitudes. The double $3\sigma$ rejection test found fault with two of our {\it riz} measurements and two of those from ``spitzer" among 984 comparisons, as well as three and four of 1666 cases with our {\it W}1 measurements.  Again, rejections are well below 1\%.

The situation is less satisfactory with the fourth external sample, ``2mtf".  That study derived distances separately in $J$, $H$, and $K$ bands; our comparisons are with the $K$-band material. Values of the Hubble parameter are plotted against systemic velocity in Figure~\ref{fig:2mtf} where distances have been shifted from consistency with the published fiducial $H_0=100$~\kmsMpc\ to fiducial $H_0=75$.  There is a drift in averaged values of about 8 Hubble units ($\sim10\%$) over the range $2000-10,000$~\kms.  The drift is manifested in values of distance modulus differences between ``2mtf" and alternatively our {\it riz} and {\it W}1 measurements as seen in Figure~\ref{fig:delmod2mtf}. 

We propose an adjustment to ``2mtf" moduli.  The straight red line in Fig.~\ref{fig:2mtf} obeys the formula
$${\rm log}H=1.875-5.76~10^{-4}(V_{LS}-6239)$$
where the line crosses fiducial $H_0=75$ at 6239~\kms.  This equation can be reformulated as an adjustment to ``2mtf" distance moduli.  The result of the adjustment in comparison with our distance moduli is seen in Figure~\ref{fig:delmod2mtfadj}. The previously observed trends have been satisfactorily removed over the range $2000-10,000$~\kms. Very large positive offsets are still seen at velocities less than $2000$~\kms. Also, the results from the double $3\sigma$ tests are less favorable.  With 832 galaxies in common with our {\it riz} photometry sample, five of our measurements fail while 27 ``2mtf" fail.  With 1122 galaxies overlapping our {\it W}1 sample, five of our measurements fail while 32 ``2mtf" fail.  Our failures remain below 1\% while the ``2mtf" failure rate is $\sim3\%$.  The ``2mtf" sample has attractive all-sky coverage features.  The sample is useful for inclusion in the {\it Cosmicflows-4} compilation with adjustments for the bias with redshift and the absolute scaling.  We are reticent to use contributions at velocities below $2000$~\kms.

Scatter in the differential measurements ranges from $\pm0.25$ mag between our WISE {\it W}1 and ``spitzer" moduli (photometry at $3.4\mu m$ and $3.6\mu m$ respectively) to $\pm0.35$ mag between our SDSS{\it riz} and the 2MASS $K$ moduli.  This scatter arises from alternate observations and analysis procedures of the same targets.

\section{\hnut~ from Field Galaxies} \label{sec:hnut}

Hubble parameter values for individual galaxies can be constructed using the distance moduli and velocities listed in Table \ref{tab:distance_catal}.  As discussed in \S\ref{sec:literature_comp}, it is expected that averaged $H_0$ values should be roughly constant with redshifts beyond the domain of substantial peculiar velocity perturbations.

\begin{figure}[t]
\centering
\includegraphics[width=\linewidth]{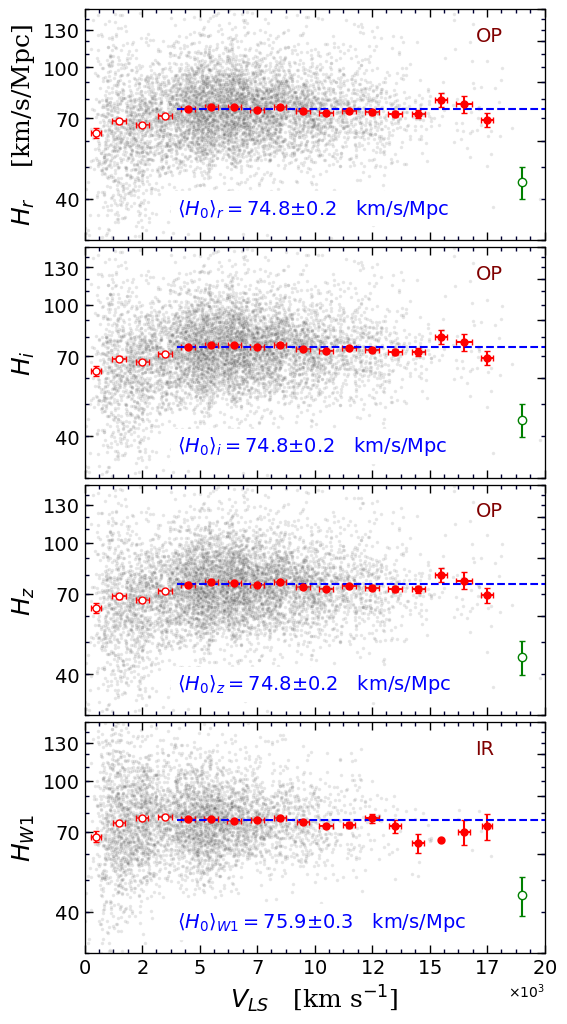}
\caption{ Hubble parameter as a function of radial velocity in various passbands. Blue horizontal line lies at the log average of the Hubble parameter of galaxies beyond 4000~\kms. Galaxies are shown by gray points. Red points display the average of data points within velocity intervals of 1500~\kms, with open point representing the the average velocities at intervals less than 4000~\kms. Error bars on the red points show the 1$\sigma$ uncertainty of the average Hubble parameter within the corresponding bins. Green error bar in the bottom right of each panel illustrates the typical uncertainty of an individual Hubble parameter.  ``OP' and ``IR" labels have the same meanings as in Figure~\ref{fig:DM_compare}.
}
\label{fig:Hubble_params}
\end{figure}

In Figure \ref{fig:Hubble_params} there are plots of Hubble parameters at the {\it r}, {\it i}, {\it z} and {\it W}1 bands as a function of radial velocity in the Local Sheet reference frame. 
The uncertainties of the measured distance moduli are described by normal distributions. Accordingly, we average Hubble parameters in log space to calculate the Hubble constant. In our averaging process, we exclude galaxies with radial velocities less than 4000 \kms, the domain where peculiar velocities can be a significant fraction of the Hubble expansion rate and cause large and potentially systematic scatter. It is shown in Figure \ref{fig:Hubble_params} that we find a value of the Hubble constant at {\it i} band of $74.8\pm0.2$ \kmsMpc\ and similar values at other optical bands. The Hubble constant determined at {\it W}1 band is $75.9\pm0.3$ \kmsMpc, about $1.5\%$ larger. 

\begin{figure}[t]
\centering
\includegraphics[width=\linewidth]{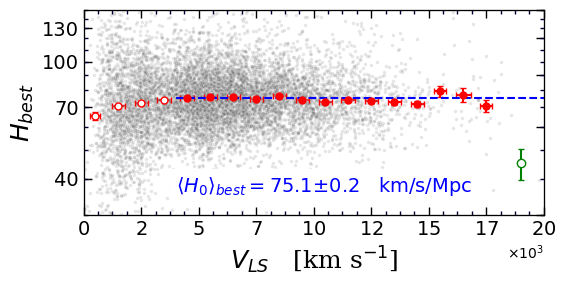}
\caption{ 
Similar to Figure \ref{fig:Hubble_params} but for the Hubble parameter calculated from the best distance moduli reported in column (5) of Table \ref{tab:distance_catal},
}
\label{fig:Hubble_params_best}
\end{figure}

The optical and infrared measurements are combined in Figure~\ref{fig:Hubble_params_best}.  The optical contributions are more firmly grounded, with superior control of color terms, but pertain to a restricted part of the sky.  The infrared contributions are not as robust, but they do uniformly cover the full sky. 
Accepting the best measured distance modulus for each galaxy given in column (5) of Table~\ref{tab:distance_catal} leads to the final result: $H_0 = 75.1\pm0.2$~\kmsMpc.

\begin{figure}[t]
\centering
\includegraphics[width=\linewidth]{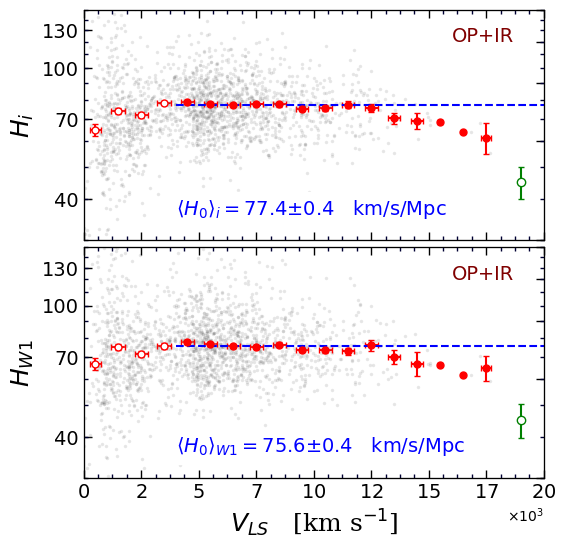}
\caption{ 
Similar to the bottom panel of Figure \ref{fig:Hubble_params} for the subsample of galaxies with both optical and infrared photometry coverage. 
}
\label{fig:Hubble_params_OPIR}
\end{figure}

The quoted errors are statistical, and are small because our samples are large.  Systematic errors totally dominate.  Comparison of the bottom panels of Figures~\ref{fig:Hubble_params} and \ref{fig:DM_compare} provides an instructive lesson.  In the latter plot, over a wide velocity range, the {\it W}1-band moduli are on average larger than those measured at the {\it i} band, implying that the inferred Hubble constant at the {\it W}1 band should be smaller than that at the {\it i} band. However, in Figure~\ref{fig:Hubble_params} we find a value for the Hubble constant at the {\it W}1 band that is {\it larger} than the optical values.  The explanation lies in the fact that the comparison with the {\it i} band in Figure~\ref{fig:DM_compare} involves {\it only} galaxies jointly observed at both optical and infrared bands, while the result shown in Figure~\ref{fig:Hubble_params} involves {\it all} galaxies with infrared photometry. 4The point is reinforced with a comparison between Figures~\ref{fig:Hubble_params} and  
\ref{fig:Hubble_params_OPIR}. Relative differences in average $H_0$ values between optical and infrared determinations vary with the details of the samples.

In the case of overlap with optical photometry, the sky coverage is that of SDSS, whereas in the case of the entire WISE sample, the sky coverage is much more extensive (see Figure~\ref{fig:aitoff_equatorial}). 
Moreover, as discussed in the appendices, there are statistical color differences between subsamples with joint SDSS and WISE photometry (OP+IR) and those with SDSS photometry without WISE photometry (OP$-$IR). Consequently, there are variations in $H_0$ at the level of $\pm1.4\%$.

\section{Potential Systematics} \label{sec:systematics}

Contemplation of possible systematics is sobering.  Their potential effect is felt at two levels: those relating to relative distances and those relating to absolute distances.  Measurements of peculiar velocities are insensitive to the absolute scaling, because $V_{pec} = V_{obs} - H_0 d$ and $H_0 \propto 1/d$.  Consequently, it is appropriate to compartmentalize between these two regimes.

The greater concern of the {\it Cosmicflows} program is the mapping of deviations from Hubble expansion.  It is worrying that, in our present study, our optical photometry from SDSS covers only a part of the sky.  Our infrared photometry from WISE covers the entire sky, but could there be systematic differences?  The alternate TFR studies discussed in \S \ref{sec:literature_comp} provide bridges.  Each of these samples covered quasi-full sky domains.  We can compare overlaps separately with our SDSS and WISE samples and look for differences.  Recall, we are looking for {\it relative} differences in distances (in moduli), because absolute distances are not our concern at this point.

Drawing on \S \ref{sec:literature_comp}, we give attention to $\langle DM_{cf4} - DM_{alt}\rangle$ where $DM_{cf4}$ is either based on the optical $\langle riz \rangle$ modulus or the infrared {\it W}1 modulus and $DM_{alt}$ is based in turn on the ``cf2+sfi", ``spitzer", or ``2mtf" samples.  Since we are interested in relative, not absolute, differences, we next consider $\Delta DM = \langle DM_{cf4}^{\langle riz \rangle} - DM_{cf4}^{W1} \rangle$ for each of the external samples.  Results in the three cases are $\Delta DM_{cf2+sfi} = -0.018\pm0.009$, $\Delta DM_{spitzer} = +0.007\pm0.011$, and $\Delta DM_{2mtf} = +0.040\pm0.016$.  The cumulative average is $\Delta DM = -0.000\pm0.006$.  There is no evidence of a problem either with respect to any of the three individual external samples or with the ensemble.  Our rms uncertainties are larger in the parts of the sky covered only by WISE but no offset is manifested at the level of 1\% in distances.

Recall that, with the cluster calibration foundation reported in K20, the {\it W}1 calibration gave averaged distances to clusters ($V_{LS}>4000$~\kms) 0.9\% shorter ($H_0$ 0.9\% greater) than the average of $r,i,z$ values. 

Tests of the constancy of Hubble parameter values illustrated in Figures~\ref{fig:Hubble_params} and \ref{fig:Hubble_params_OPIR}
manifest fluctuations in binned averages below 2\% over the velocity domain 4000-15,000~\kms\ (averaged velocity deviations below 80-300~\kms).

Hints of systematics are greater in absolute distances.  For example, there is the controversy over the tip of the red giant branch and Cepheid scales \citep{2019ApJ...882...34F,2019ApJ...876...85R} with differences at the level of $\pm2.9\%$. In K20, we found systematic differences in individual tip of the red giant branch and Cepheid distances at the level of $\pm3.5\%$.  Also in K20, color differences between galaxies that define the slope of the ITFR and the galaxies with independently established distances that set the zero-point, differences with only $1\sigma$ significance, still called for adjustments between optical and infrared bands at the level of 3\%.  Indeed, differences between the average colors of galaxies in the calibrating clusters can have distance effects at the level of 4\% between clusters. 

Then we must note the differences between the successive absolute calibrations by our own team over the progression {\it Cosmicflows-2,3,4}.  Differences from average values are at the level of $\pm3.3\%$ (2 and 4) and $\pm2.3\%$ (3 and 4), with a drift toward smaller distances and larger $H_0$.

In summary, 1$\sigma$ systematic uncertainties in the absolute zero point could sum to as high as 4\% or $\Delta H0 = 3$~\kmsMpc.  There are fewer signs of potential relative systematics, but we are loath to suggest that they lie less than 2\% or $\Delta H_0 = 1.5$~\kmsMpc.  We can make qualified estimates of known unknowns, but there are still the unknown unknowns.

\section{Summary} \label{sec:summary}

This era of contiguous wide-field and even all-sky surveys is creating opportunities for vastly expanded samples for cosmological studies.  The current program to acquire galaxy distances from the correlation between the rotation rates and luminosities of spirals benefits from the kinematic information provided by the full coverage of the sky outside the galactic plane in the declination range of the Arecibo Telescope (ALFALFA), the photometric information at optical bands over a closely overlapping part of the sky (SDSS), and photometric information at infrared bands over the entire sky (WISE).

The ALFALFA survey \citep{2018ApJ...861...49H} provided a particularly valuable impetus.  {\it Cosmicflows-3}, the last release of our program \citep{2016AJ....152...50T}, was heavily weighted toward the south celestial hemisphere by the numerically dominant 6dFGSv contribution \citep{2014MNRAS.445.2677S}.  Once this new sample of distances is integrated with other available material, there will be much more satisfactory coverage of the full unobscured sky extending to $\sim 0.05c$.  

Targets enter our sample in two ways.  An ALFALFA component enters by virtue of sufficient \hi flux.  Galaxies with sufficient $S/N$ are evaluated by our standard program criteria (morphology, inclination, concern for confusion or disturbance).  We personally evaluated the \hi profiles of cases that were available at the time of the 40\% release of ALFALFA and entered the results of our analysis in the {\it All Digital \hi} catalog in the {\it Extragalactic Distance Database}.  The 100\% release has arrived too recently for our full attention but the substantial overlap in analysis results between the ALFALFA team and ourselves with the 40\% release permits a reliable merging of line width information. The second path to inclusion in this study is initiated by optical images.  Potential targets filtered by systemic velocity, magnitude, axial ratio, and morphological type are further evaluated for appropriateness and subsequently observed with a radio telescope appropriate to their declination.  \hi information is stored in the {\it All Digital \hi} catalog which at this date has entries for almost 19,000 galaxies. 

Upon evaluation, we settled on 10,737 galaxies with appropriate \hi information as candidates for photometry.  Of these, we have optical photometry in five bands for 70\% that lie within the SDSS footprint. We have WISE infrared photometry in two bands for 51\% of our sample, of which 21\% are in common with our SDSS sample.  There is WISE coverage of our entire sample, but we only carried out photometry on a fraction of the objects with SDSS coverage, given the superior results with the optical material.

By employing the luminosity$-$line width relations calibrated in K20, we measure the distances of our sample galaxies at SDSS {\it riz} and WISE infrared {\it W}1 passbands.  The domain of application of the relations in K20 are restricted to systems brighter than $M_i=-17$ or $M_{W1}=-16.1$, causing the rejection of cases determined to be fainter.  Furthermore, extremely deviant cases evaluated by their implicit Hubble parameters or through comparisons with other available distance information were rejected.  We are left providing distances for 9792 galaxies.

In principle, only a line width and a photometric magnitude are needed to derive a distance to a galaxy (supplemented by inclination and reddening terms).  We have an abundance of other information in hand (colors, \hi fluxes, surface brightnesses, error constraints).  Factors that cause dispersion or bias can be investigated with these extra parameters.  We have investigated concerns including Malmquist bias and color dependencies.  In the substantial majority of cases with available optical photometry, we can make adjustments for bias and subordinate parameters that result in coherent results with minimal dispersion.  The situation is less robust if only infrared information is available, because the cross-correlation with optical bands turns out to be so useful.  We make due with adjustments coupled to the First Component parameter described in K19 and K20 that only needs \hi and infrared information.  We remain concerned about potential systematics between the parts of the sky covered by SDSS and parts only covered by WISE.

Our final derived distances are in close agreement at the $r, i, z$ bands thanks to the tight parameter coupling through our adjustments.  Considering only galaxies beyond the domain of substantial velocity anomalies, those with velocities greater than 4000~\kms, we determine $H_0^{\langle riz \rangle} = 74.8\pm0.2$~\kmsMpc\ for cases with optical photometry.  If only infrared photometry is available and with the same velocity restriction, we find $H_0^{W1} = 75.9\pm0.3$ ~\kmsMpc. We conclude from this study that $H_0 = 75.1\pm0.2$~\kmsMpc.  Errors are statistical and systematic errors are larger.  In \S \ref{sec:systematics}, rough estimates of relative and absolute systematic uncertainties are $\pm1.5$ and $\pm3$~\kmsMpc\ respectively.

The next step in this program is to produce a new master list of galaxy distances: {\it Cosmicflows-4}. This projected compilation will merge the luminosity$-$line width results of this paper with those discussed in \S \ref{sec:literature_comp} and with distances from other methodologies.  The ensemble will be heterogeneous, but with the great virtue that multiplicity brings to the assessment of systematics.

\acknowledgments

We are pleased to acknowledge the citizen participation to scientific research of undergraduate students at University of Hawaii, members of the amateurs astronomy clubs in France Plan\'etarium de Vaulx-en-Velin, Association Clair d'\'etoiles et Brin d'jardin, Soci\'et\'e astronomique de Lyon, Club d'astronomie Lyon Amp\`ere, Club d'astronomie des monts du lyonnais, and Club d'astronomie de Dijon, as well as friends who helped us with measuring inclinations of spiral galaxies in our sample.

Support for E.K. and R.B.T. was provided by the NASA Astrophysics Data Analysis Program through grant number 88NSSC18K0424. A.D. acknowledges financial support from the Project IDEXLYON at the University of Lyon under the Investments for the Future Program (ANR-16-IDEX-0005).
H.C. acknowledges support from Institut Universitaire de France. This project is partly financially supported by Region Rhone-Alpes-Auvergne

This research has made use of the NASA/IPAC Extragalactic Database\footnote{\url{ttp://nedwww.ipac.caltech.edu/}} which is operated by the Jet Propulsion Laboratory, California Institute of Technology, under contract with the National Aeronautics and Space
Administration. This research made use of Montage, funded by the National Aeronautics and Space Administration’s Earth Science Technology Office, Computational Technologies Project, under Cooperative Agreement Number NCC5-626 between NASA and the California Institute of Technology. The code is maintained by the NASA/IPAC Infrared Science Archive.

The Pan-STARRS1 Surveys (PS1) and the PS1 public science archive have been made possible through contributions by the Institute for Astronomy, the University of Hawaii, the Pan-STARRS Project Office, the Max Planck Society and its participating institutes, the Max Planck Institute for Astronomy, Heidelberg and the Max Planck Institute for Extraterrestrial Physics, Garching, The Johns Hopkins University, Durham University, the University of Edinburgh, the Queen's University Belfast, the Harvard-Smithsonian Center for Astrophysics, the Las Cumbres Observatory Global Telescope Network Incorporated, the National Central University of Taiwan, the Space Telescope Science Institute, the National Aeronautics and Space Administration under grant No. NNX08AR22G issued through the Planetary Science Division of the NASA Science Mission Directorate, the National Science Foundation grant No. AST-1238877, the University of Maryland, Eotvos Lorand University (ELTE), Los Alamos National Laboratory, and the Gordon and Betty Moore Foundation.

\appendix
\begin{figure}
\centering
\includegraphics[width=0.92\linewidth]{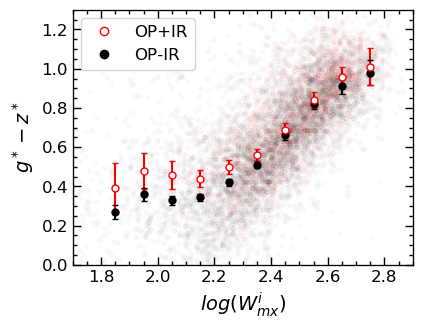}  \\
\includegraphics[width=0.92\linewidth]{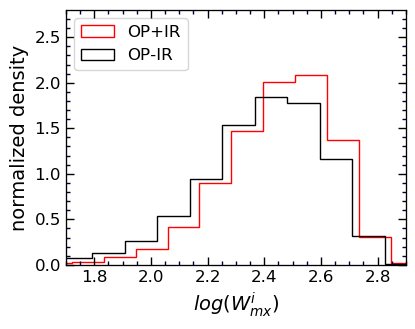} \\
\includegraphics[width=0.92\linewidth]{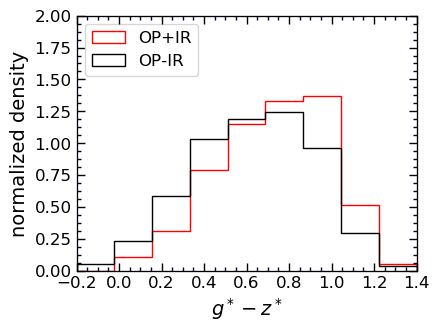} 
\caption{
{\bf Top:} $g^*-z^*$ color versus the \hi line width for two subsamples. The OP$-$IR sample consists of galaxies with available optical SDSS photometry and missing infrared WISE photometry. The OP+IR sample includes 2244 spirals with photometry coverage with both SDSS and WISE. Large points are positioned at the median value of galaxy color indices within line width bins of constant size. Error bars exhibit the 1$\sigma$ scatter of galaxy colors within each bin. 
{\bf Middle:} normalized line width distributions.
{\bf Bottom:} color distribution of both samples.
\label{fig:gz_logW}}
\end{figure}

\section{Comparing the optical and infrared subsamples}  \label{sec:OPIR-OP-IR}

The top panel of Figure \ref{fig:gz_logW} plots the $g^*-z^*$ colors of our spirals versus line width for two subsamples with the SDSS coverage, OP$-$IR and OP+IR. The OP+IR galaxies have both SDSS and WISE photometry data that have been analyzed while the OP$-$IR galaxies have analyzed SDSS data but were not given attention in our WISE photometry program.
 Evidently, galaxies of our OP+IR sample are on average redder at all line widths. 

The middle and bottom panels of Figure \ref{fig:gz_logW} compare the line width and color distribution of both subsamples and reveal a significant difference. A Kolmogorov–Smirnov test implies a $p$-vale of $2\times 10^{-34}$, meaning that it is very unlikely that both subsamples are drawn from the same distribution. Our WISE photometry program mainly prioritizes galaxies with missing SDSS photometry, regardless of their size/luminosity. However, in the presence of the SDSS photometry, larger spirals have the highest priority of inclusion in our infrared photometry program which explains why our OP+IR sample is biased toward redder galaxies. Larger galaxies tend to be redder, as they are hosts of older stellar populations.

\begin{figure}[t]
\centering
\includegraphics[width=0.92\linewidth]{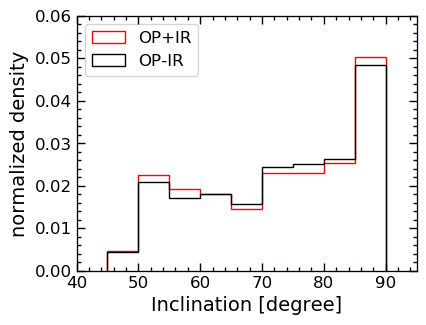}
\caption{
The normalized Inclination distribution of the spirals in our sample. OP+IR and OP$-$IR are the same as in Figure \ref{fig:gz_logW}.
\label{fig:histograms_logW_inc}}
\end{figure}

To complete our assessments, in Fig.~\ref{fig:histograms_logW_inc} we plot the distribution of galaxies in both OP+IR and OP$-$IR samples in terms of inclination. Both histograms (black and red) look similar and there is no evidence of inclination-dependent sample selection bias.

\section{Prediction of the dust attenuation}  \label{dust:prediction}

Attenuation of galaxy magnitudes due to line-of-sight path lengths through inclined disks is negligible in the WISE {\it W1} ($3.4 \mu m$) and especially so in the {\it W2} ($4.6 \mu m$) bands, but can be substantial at optical bands. The attenuation at SDSS {\it ugriz} and WISE {\it W1} bands as a function of inclination is described by Eqs.~\ref{eq:dust} and \ref{Eq:Fli}. The dust attenuation factor $\gamma_{\lambda}$ was empirically determined by a principal component analysis \citep{2019ApJ...884...82K} with the leading component being the linear combination of the three properties specified in Equation~\ref{Eq:Mu50_i}: log$W^i_{mx}$ as a proxy for absolute magnitude, $\langle \mu_j \rangle^{(i)}_e$ as a measure of surface brightness, and $C_{21W2}$ to monitor the relative importances of \hi content and old stars.  The complex forms of $\gamma_{\lambda}$, peaking at luminous, metal- and gas-rich systems and falling off toward alternatively metal poor dwarfs and gas-poor, dominantly old population giants, can be seen in Fig.~10 of \citet{2019ApJ...884...82K}. 

This formalism to calculate the dust obscuration requires the availability of infrared photometry information to quantify the $C_{21W2}$ term.  
Out of 10,737 accepted spirals in our program, only 21\% have full optical and infrared photometry coverage.  We have 7501 galaxies in the sector of the sky covered by the Sloan survey for which we have {\it ugriz} photometry.  Over the remainder of the sky, we have 3234 galaxies, all with WISE {\it W1} and {\it W2} photometry.  In principal, WISE photometry is equally available for all the targets in the SDSS coverage area; however, at this time, WISE photometry has been carried out for only 2243 of these cases.  In this appendix, methods are described that give predictions of {\it W2} magnitudes with sufficient accuracy to act as proxies for the calculation of attenuation factors. 

\begin{figure*}[t]
\centering
\includegraphics[width=0.9\linewidth]{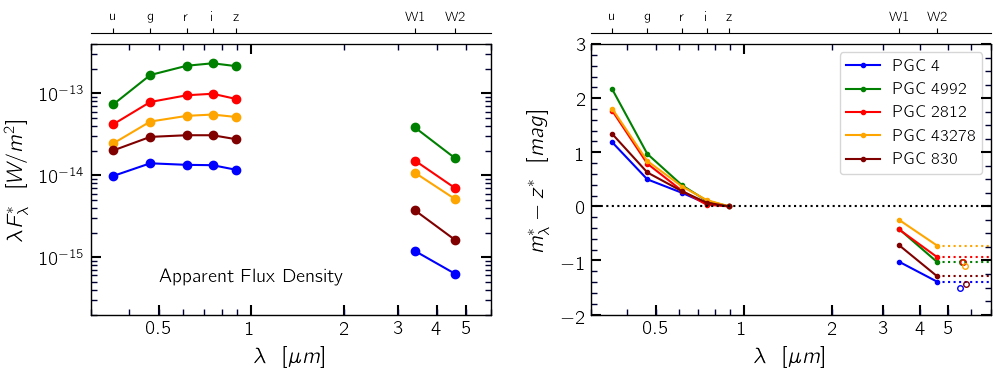}
\caption{
{\bf Left:} apparent spectral energy distribution of five galaxies with available optical/infrared photometry. 
{\bf Right:} $m_\lambda^*-z^*$ color indices for the same galaxies plotted in the left panel. Dotted horizontal lines are drawn at the level of $W2^*-z^*$. Open circles are the values predicted via the algorithm discussed in \ref{sec:pred_algo}. Names and positions of each passband are displayed on top of each panel. 
\label{fig:sed}}
\end{figure*} 

\begin{figure}[t]
\centering
\includegraphics[width=0.8\linewidth]{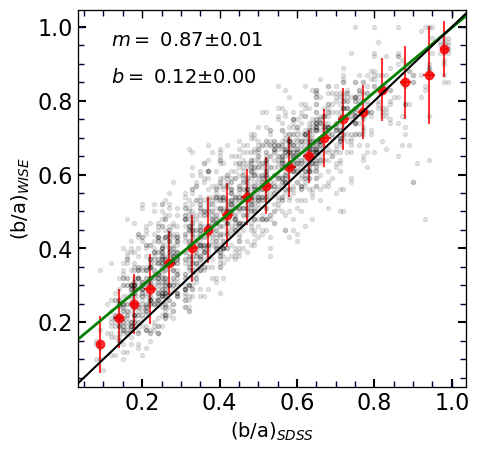}
\caption{
WISE vs. SDSS axial ratios of the elliptical apertures used for photometry. Each galaxy is represented by a gray point. Red points display the median of scattered points within horizontal bins with a size of $0.05$ and their error bars show the 1$\sigma$ scatter. Green solid line is the best-fitted straight line to the gray points, and has the form $(b/a)_{WISE}=0.87(b/a)_{SDSS}+0.12$. If there were equality between the parameters, data would scatter about the black line. 
\label{fig:WISE_SDSS_ba}}
\end{figure} 

In concept, given the luminosity of a galaxy at optical wave bands along with some other information about its physical properties, such as intrinsic size and/or morphology, one can predict the galaxy luminosity at longer wave bands. 
A spectral energy distribution (SED) can be fit over the observed magnitudes using a set of template SEDs that represent the morphology, size, and physical properties of the sample galaxies. Based on the fitted SED, one is able to estimate the luminosity of the galaxy at missing passbands. The fitting of SEDs and the building of such models is beyond the scope of this research program. Our objective is only to monitor dust obscuration levels, not to determine the infrared luminosities of galaxies. Therefore, we emphasize that we do NOT take the SED fitting approach in this study, but rather perform an empirical analysis to build a simpler model in order to predict the missing values at infrared bands---and ultimately, the dust extinction. 

We have orchestrated a random forest algorithm together with a set of distance-independent observables in order to predict the missing infrared information. 
Our prediction algorithm is trained using $\sim$2200 spirals with the full optical/infrared magnitude coverage. The trained algorithm is capable of predicting {\it W2} magnitudes with an rms uncertainty of $\sim 0.2$ mag. Based on the predicted {\it W2} magnitudes, the $1\sigma$ uncertainty of the predicted $\gamma_{\lambda}$ is $\sim 0.04$ mag for the optical bands and is smaller at longer wavelengths. Next, $\gamma_{\lambda}$ is multiplied by $\mathcal{F}_\lambda$ to obtain the dust attenuation $A^{\lambda}_i$. Here, $\mathcal{F}_\lambda$ is a monotonic increasing function of inclination that is maximal for fully edge-on galaxies between $1.5$ and $1.75$ for the optical wave bands and $0.75$ for {\it W1} band. The overall uncertainty on our predicted $A^{\lambda}_i$ values is always no worse than $\sim 0.07$ mag. 

As examples, Fig. \ref{fig:sed} displays the apparent flux density for five spirals with distinct apparent luminosities that have photometry information at optical/infrared passbands. The filled circles display the positions of the actual photometry measurements. We are trying to estimate the best values for infrared magnitudes, {\it W1} and {\it W2}, given the optical luminosities at {\it u, g, r, i}, and {\it z} bands, as well as other observables that probe the physical properties of spirals. We need to focus on distance-invariant parameters, because these inputs will later be used to determine distances. In the right panel of Fig.~\ref{fig:sed}, the colors of the same galaxies are plotted versus wavelength, normalized to the {\it z} band. Now, the goal is to predict the values on the right side of the diagram, based on the information on the left side and other extra pieces of distance-invariant parameters available for each galaxy. In \ref{sec:mag:adjustment} we presented the distance invariant parameters  $\log (W^i_{mx})$, $C_{21W2}$, and $\langle \mu_j \rangle^{(i)}_e$, the constituents of the main principal component, $P_1$, correlated with dust attenuation. The missing infrared information contributes to the calculations of the parameters $C_{21W2}$ and $\langle \mu_j \rangle^{(i)}_e$. 

\begin{figure}[t]
\centering
\includegraphics[width=0.8\linewidth]{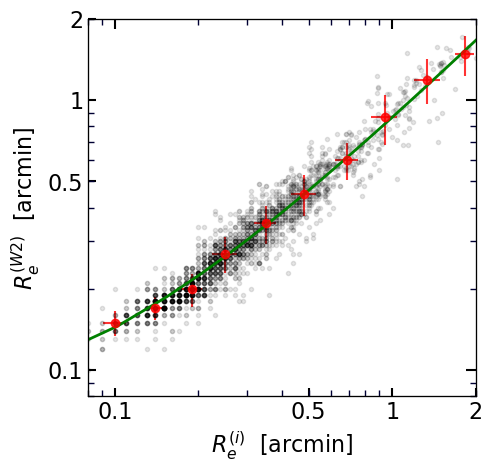} \\
\includegraphics[width=0.8\linewidth]{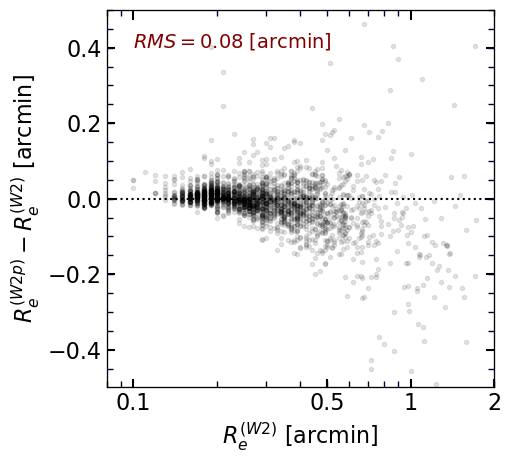}
\caption{ {\bf Top:} {\it W2} half-light radii vs. SDSS {\it i}-band effective radius. Each black point represents a galaxy. Green curve illustrates the relation between effective radius measured at optical {\it i} band and infrared {\it W2} band, given as $R_e^{(W2p)}=0.78 R_e^{(i)} + 0.07$, where $R_e^{(i)}$ is the {\it i}-band effective  radius.
{\bf Bottom:} deviation of the predicted half-light radius at {\it W2} band from the measured value. The rms scatter is no worse than $0.1$'. 
\label{fig:W1_griz_half_light}}
\end{figure} 

\begin{figure}[t]
\centering
\includegraphics[width=\linewidth]{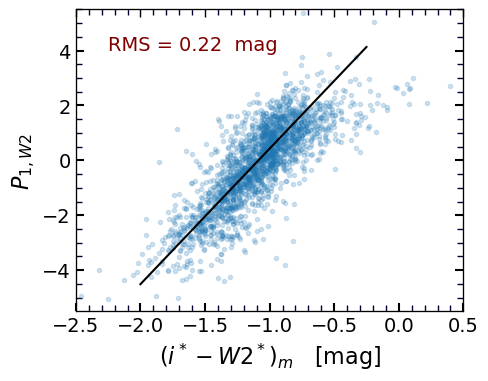}
\caption{
First principal component, $P_{1,W2}$ vs. extinction-corrected optical-infrared color. Each blue point represents a galaxy. Solid black line is the fitted fiducial line that minimizes the rms of residuals along the color axis.
\label{fig:P1_col}}
\end{figure} 

\begin{figure*}[t]
\centering
\includegraphics[width=0.75\linewidth]{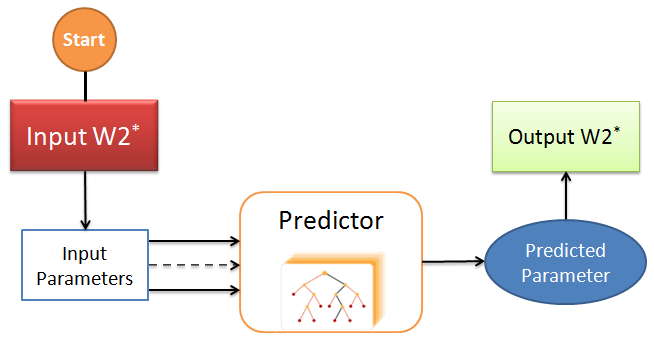}
\caption{
Schematic diagrams for the algorithm that predict missing {\it W2} magnitudes.
\label{fig:rf_schem}}
\end{figure*} 

\begin{figure*}[t]
\centering
\includegraphics[width=\linewidth]{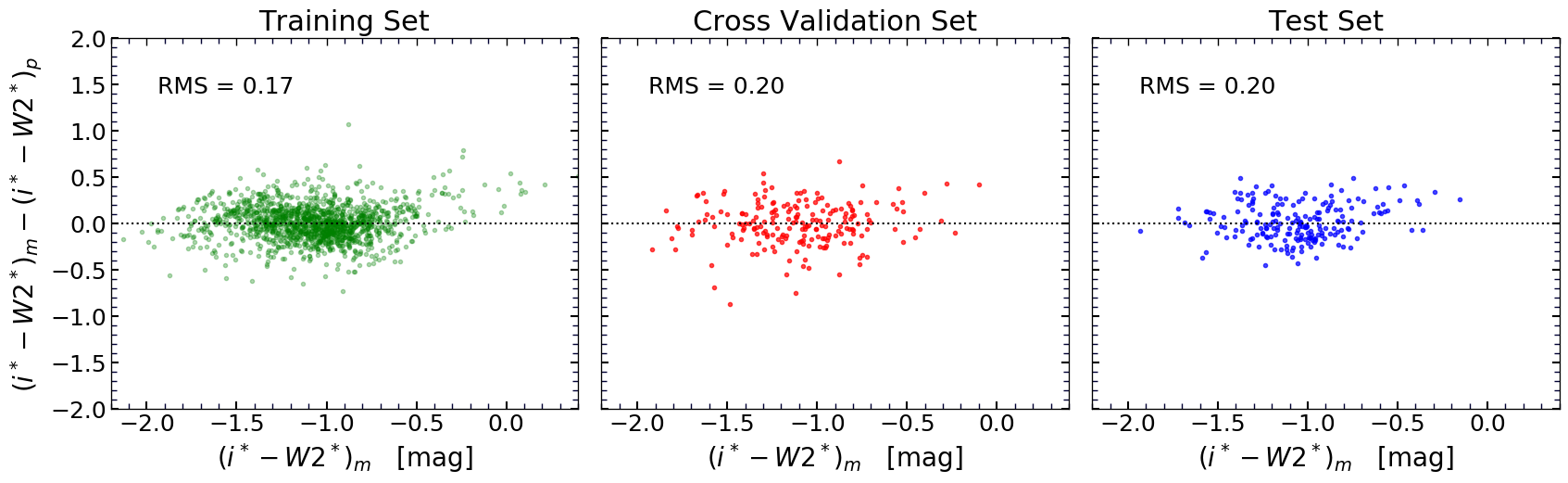}
\caption{Color differences and scatters from the random forest training, cross-validation, and test sets. In this particular random forest, the output parameter is the $i^*-W2^*$ color and inputs are $g^*-r^*$, $r^*-i^*$, $i^*-z^*$, and $P_{1,W2}$.
\label{fig:RF_sets}}
\end{figure*}

\begin{figure*}
\centering
\includegraphics[width=0.45\linewidth]{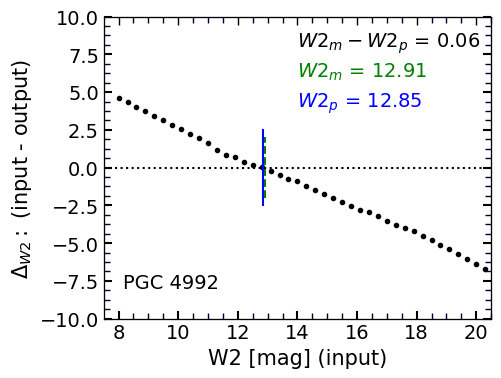} 
\includegraphics[width=0.45\linewidth]{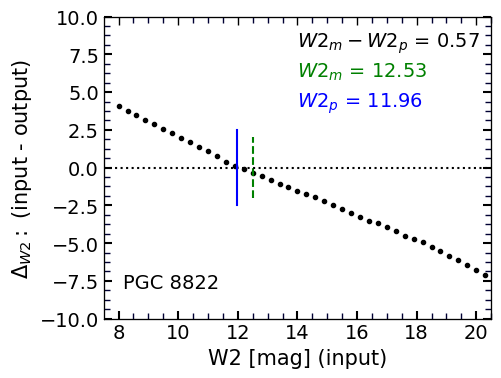}
\caption{
Differences between input and output $W2^*$ values for PGC 4992 (left) and PGC 8822 (right). Vertical solid blue lines mark where $\Delta_{W2}=0$ mag specifying $W2^*_p$, the predictive best input $W2^*$ value. Vertical green dashed lines mark the actual measured values $W2^*_m$.
\label{fig:predict_w2}}
\end{figure*} 

\subsection{SDSS vs. WISE axial ratios and effective radii}

One of the parameters that is involved in the calculation of dust extinction is $\langle \mu_2 \rangle^{(i)}_e$, the average surface brightness of a galaxy in the {\it W2} band within its effective radius (the radius that encloses half of the total light of that galaxy), corrected for the geometric effect of inclination. Here, $\langle \mu_2 \rangle^{(i)}_e$ is given by
\begin{equation}
\langle \mu_j \rangle^{(i)}_e=\langle \mu_j \rangle_e+0.5 \log_{10} (a/b), 
\end{equation}
where $j=W2$ and $a$ and $b$ are the semimajor and semiminor axes of the photometry aperture. The apparent effective surface brightness is calculated from 
\begin{equation}
\langle \mu_2 \rangle_e = W2 +2.5 {\rm log_{10}}(2\pi R^2_e) ~,
\end{equation} 
where $R_e$ is the effective radius of the galaxy and {\it W2} is its apparent magnitude. 
The value of $\langle \mu_2 \rangle^{(i)}_e$ depends on the {\it W2}-band magnitude, the axial ratio, and the half-light radius of the aperture used with the WISE images in the process of photometry. 

 The axial ratios of the apertures used for the photometry of SDSS and WISE images are plotted against each other in Fig.~\ref{fig:WISE_SDSS_ba}. The axial ratios of WISE apertures are systematically greater than the SDSS aperture axial ratios. In general, the morphology and visible size of a galaxy are different at optical and infrared wavelengths. Galaxy bulges are much more prominent at longer wavelengths. In addition, the point spread function (PSF) of WISE images is large compared to that of SDSS. Therefore, capturing all the galaxy light requires choosing larger axial ratios ($b/a$) when dealing with WISE images, where $a$ and $b$ are semi-major and semi-minor axes of the elliptical photometry apertures.  In more face-on spirals ($b/a$ approaching unity), the effects of bulges and PSF are less pronounced. 

Galaxy effective radius is defined as the radius within which half of its total light is emitted. As an example, Fig.~\ref{fig:W1_griz_half_light} plots the effective radius at the WISE {\it W2} band against SDSS {\it i} band. 
We use linear models to describe the relation between the effective radii at optical {\it g}, {\it r}, {\it i}, and {\it z} bands and that at the {\it W2} band. These relations are given as
\begin{align}
  R_e^{(W2p)}&=0.71 R_e^{(g)} + 0.08 \\
  R_e^{(W2p)}&=0.76 R_e^{(r)} + 0.07 \\
  R_e^{(W2p)}&=0.78 R_e^{(i)} + 0.07 \\
  R_e^{(W2p)}&=0.82 R_e^{(z)} + 0.07. 
\label{eq:Re2o} 
\end{align}

Bulges of spiral galaxies are more luminous at longer wavelengths because they are formed by red and dead stars. At longer wavelengths, we expect half-light radii to be smaller due to the centralized flux contributions of bulges. In Equation~\ref{eq:Re2o}, this effect gives rise to larger coefficients at redward wave bands. 

\begin{table*}[t]
\begin{center}
\caption{Different Combinations of Input/Output Observables Used to Train the Random Forest Algorithm. }
\begin{tabular}{rccccccccc|cc} \hline
    No. & ${\rm log}W^i_{mx}$ & $\langle \mu_2 \rangle^{(i)}$ & $C_{21W2}$ & $P_1$ &  $g^*-r^*$ & $g^*-i^*$ & $r^*-i^*$ & $r^*-z^*$ & $i^*-z^*$ & Output  & rms  \\
    & \% & \% & \% & \% & \% & \% & \% & \% & \% & mag &  \\
    & (1) & (2) & (3) & (4) & (5) & (6) & (7) & (8) & (9) & (10) & (11)   \\ 
    \hline \hline
    1 &   &  &  &  67.1  &  28.3 &     &  1.9 &       & 2.8   & $g^*-W2^*$ & 0.18  \\
    2 &   &  &  &  74.1  &  18.8 &     & 2.4  &  4.7  &       & $r^*-W2^*$ & 0.19  \\
    3 &   &  &  &  74.0  &  19.0 &     & 2.3  &  4.7  &       & $i^*-W2^*$ & 0.19  \\
    4 &   &  &  &  54.4  &  10.4 &     & 3.1  &  32.0 &       & $z^*-W2^*$ & 0.20  \\
    5 &   &  &  &  91.5  &       &     & 3.8  &  4.7  &       & $r^*-W2^*$ & 0.19  \\
    6 &   &  &  &  88.2  &       & 5.5 &      &       & 6.3   & $i^*-W2^*$ & 0.20  \\
    7 &   &  &  &  90.9  &  3.9  &     &      &  5.2  &       & $r^*-W2^*$ & 0.19  \\
    8 &   &  &  &  92.5  &  4.0  &     &  3.4 &       &       & $r^*-W2^*$ & 0.19  \\
    \hline 
    9 & 18.2  & 63.6 & 13.6 &        &  4.6  &      &      &       &       & $i^*-W2^*$ & 0.19  \\
   10 & 2.7   & 11.9 &       &       &  68.5 &      & 17.0 &       &       & $C_{21W2}$ & 0.26  \\
   11 & 7.5   &      & 59.3  &       &  28.6 &      & 4.7  &       &       & $\langle \mu_2 \rangle^{(i)}$ & 0.30  \\
    \hline \hline
\end{tabular}
\label{tab:features}
\end{center}
\begin{quote}
{\bf Note.} Input parameters are denoted by their corresponding importance percentages. Column (11) tabulates the rms of $\Delta W_2=W2_m-W2_p$, where $W2_m$ is measured {\it W2} and $W2_p$ is the prediction of the algorithm displayed schematically in Fig. \ref{fig:rf_schem}. In column (10), all color indices are in magnitude and $\langle \mu_2 \rangle^{(i)}$ is in mag arcsec$^{-2}$.
\end{quote}
\end{table*}

\subsection{Prediction Algorithm} \label{sec:pred_algo}

Our goal here is to predict missing infrared {\it W1} and {\it W2} magnitudes. To begin, we need to establish a fiducial relation between distance-independent observables for a set of galaxies with both optical and infrared measurements. 
In Fig.~\ref{fig:P1_col}, there is a plot of $P_{1,W2}$, the principal component parameter defined by Equation~\ref{Eq:P1}, versus $(i^*-W2^*)_m$, a color index corrected for host dust obscuration, for ~2200 galaxies with both SDSS and WISE photometry. A linear relation describes this correlation and can be used to solve for $W2^*$ given every other involved parameter. An optimal value for {\it W2} can be found through testing a range of plausible values and adopting the value that forces the galaxy to obey the fiducial correlation. 

Our algorithm, shown in Fig. \ref{fig:rf_schem}, starts with an input guess value, for $W2^*$. The guess value along with other known parameters, such as optical magnitudes (presented in the form of optical-infrared colors), first principal component, \hi line width, effective surface brightness at {\it W2}, etc., are then fed into the fiducial correlation, which outputs a parameter that is used to extract the $W2^*$ parameter. 
Only one value of $W2^*$ is consistent with the fiducial correlation (the input and output $W2^*$ parameters agree with each other), and it is adopted as the predicted value of $W2^*$.
Adopting the fiducial correlation illustrated in Fig.~\ref{fig:P1_col}, we can derive $P_{1,W2}$ over a range of $W2^*$ and use these values to calculate the corresponding $i^*-W2^*$ color indices. 
The output $W2^*$ is then derived, given the {\it i} magnitude is known.

In Appendices \ref{sec:RF} and \ref{sec:optimization}, we use the {\it random forest} concept to explore a more complicated predictor and to incorporate a large number of galaxy features.

\begin{figure}[t]
\centering
\includegraphics[width=\linewidth]{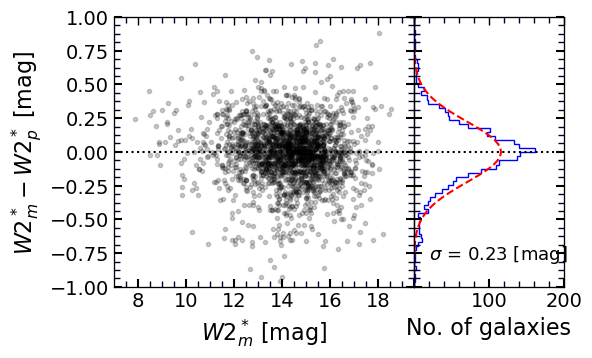} \\
\includegraphics[width=0.8\linewidth]{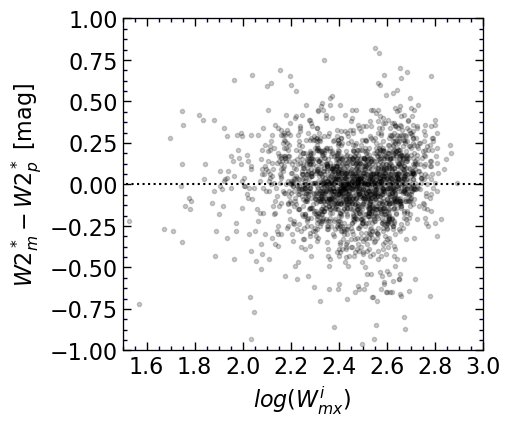}
\caption{
Measured minus predicted $W2^*$ values vs. the measured $W2^*$ (top) and \hi line width (bottom). The 1$\sigma$ standard deviation of the scatter in the prediction error ($W2^*_m-W2^*_p$) is 0.23 mag.  
\label{fig:pred_error}}
\end{figure} 
\subsection{Random Forests} \label{sec:RF}

{\it Random forest} is a supervised machine-learning technique that uses
training sets to capture the general trends in data and reach optimal performance in producing desired outputs.
Normally, 80\% of the data are used in the training process; these data are called the ``training set''. Another 10\% of the data are usually used to evaluate the algorithm performance and to optimize its structural hyperparameters in order to maximize its efficiency. This set is called the ``cross validation set''. The remaining 10\% of the data, called the ``test set", are held in reserve to evaluate the ultimate performance of the algorithm. These data are never used in the training and optimization process. 

The building blocks of random forests are decision trees. Each decision tree in the forest is trained separately and is different from the other trees in the forest. The training of each tree can be randomized based on the following factors: (1) the number of input parameters and the order in which they are used for splitting and branching, and (2) the training of each tree being based on a randomly chosen subsample of the training set. The outputs of all random decision trees in the forest are averaged and reported as the ultimate output of the forest. 

Decision trees are binary trees; their training is a recursive process and involves partitioning the training sample to maximize the information gain. 
At its root, each decision tree begins with the entire training sample. At each node, one of the input features is used to split the data set and pass it to the left and right children. The features and the splitting criteria are chosen to reach the largest possible information gain, given as 

\begin{equation}
\begin{split}
IG(D_p, f) & = I(D_p) \\
           & -\Big(\frac{N_{left}}{N_p}I(D_{left})+\frac{N_{right}}{N_p}I(D_{right}) \Big)~,
\end{split}
\end{equation}
where $f$ is the chosen feature that is used to perform the split, $D_p$ is the parent node dataset, $D_{left}$ and $D_{right}$ are the left and right children datasets, respectively. $N$ is the number of data points in each sample, and  $I$ is the ``Impurity Metric'', which is usually chosen to be the mean squared error (MSE). The MSE is the variance of the output feature, $y$, in the sample $D$

\begin{equation}
I(D) = MSE(D) = \frac{1}{N}\sum_{i=1}^{N}\big(y^{(i)} - \overline{y} \big)^2~,
\end{equation}
where $\overline{y}$ is the average of $y$ in the sample. 
The key to the machine-learning aspect of decision trees is that the machine decides how to split up the feature space at each node of the tree in order to minimize the cost criterion.
The splitting at all nodes continues until reaching one of the training criteria. These criteria could be the maximum depth of the tree, the minimum number of data points at the leaf nodes, and/or a threshold on the increase of the information gain as the training advances. Intuitively, maximizing the information gain is equivalent to minimizing the total variance of the successive nodes. At the end of the training process, the leaf nodes contain data points whose target features, $y$ are almost analogous. MSE values closer to zero are better, as they indicate lower deviations of the predicted values from their true values.

\begin{figure}[t]
\centering
\includegraphics[width=0.9\linewidth]{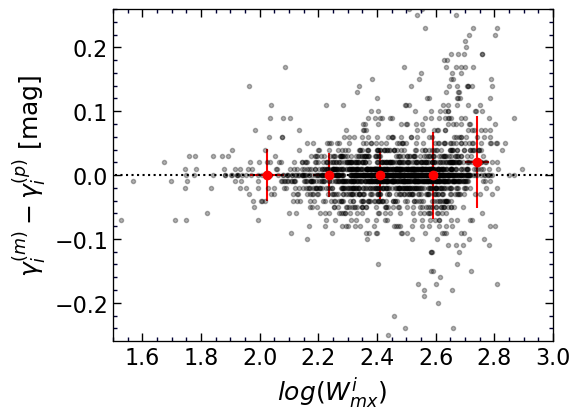}
\caption{
Differences between predicted and measured dust extinction factors versus \hi line width. Each black point represents a galaxy. Red points are the average of black points within bins of $0.2$ along the line width axis. Error bars show the 1$\sigma$ scatter of points.
\label{fig:gamma_pred1}}
\end{figure}

\begin{figure*}[t]
\centering
\includegraphics[width=0.9\linewidth]{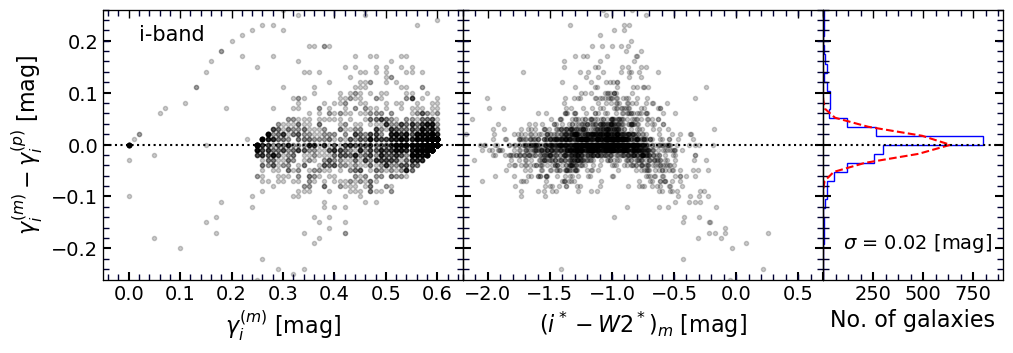}
\caption{
Differences between predicted and measured dust extinction factors at {\it i} band versus measured $\gamma^{(m)}_i$ and color $(i^*-W2^*)_m$. The 1$\sigma$ scatter of the measurement$-$prediction discrepancies, $\gamma{(m)}_i-\gamma{(p)}_i$, is 0.02 mag. 
\label{fig:gamma_pred2}}
\end{figure*} 

\subsection{Optimizations and Predictions} \label{sec:optimization}
We use {\tt RandomForestRegressor} from the Python package {\tt scikit-learn}\footnote{\url{https://scikit-learn.org/stable/modules/generated/sklearn.ensemble.RandomForestRegressor.html}} to train and evaluate our random forest models. 
A decision tree has a few parameters over which the user has control, called hyperparameters. Hyperparameters can be any internal parameters that control the training process, e.g. the number of levels in the tree (how deep the tree is) and how many leaves are allowed on each branch are two hyperparameters we set manually. The ``cross-validation set" is used to optimize these parameters. Our optimized values for the maximum number of levels in trees and the minimum number of leaf galaxies are 14 and 9, respectively.

Figure \ref{fig:RF_sets} plots the differences between the measured and predicted $i^*-W2^*$ values for three different sets, where the random forest was trained using the main principal component, $P_{1,W2}$, and $g^*-r^*$, $r^*-i^*$, and $i^*-z^*$ colors. The output parameter is chosen to be the $i^*-W2^*$ color index. Each of the cross-validation and test sets contains 200 galaxies. The training set consists of 1800 galaxies. All of these sets are chosen randomly.
The RMS scatter of differences between predicted and measured $i^*-W2^*$ colors is $\sim$0.20 mag. The RMS scatter for the training set is slightly better, which indicates the possibility of overtraining the random forest.

In this study, input features and the output feature can be any of the distance-independent parameters introduced earlier. We attempt to understand which features are important in our best-fit model and which features are not as important. The importance of each feature is a function of (1) the number of nodes in the tree that use that particular feature and (2) the level of improvement in the `gain' parameter. 
Those features that are used more frequently and contribute more in achieving more `gain' are considered to be more important. 

With the various parameters available, we attempt to find the combination of parameters that is optimal while including as much nonredundant information as possible. Some sets can achieve good results with fewer parameters, but they may force the algorithm to rely on a single parameter. The {\it r} and {\it i} bands have the smallest uncertainties, so the $r^*-W2^*$ and $i^*-W2^*$ outputs were expected to perform best.
Table \ref{tab:features} shows a portion of the results. 
Each row of this table represents a random forest, the input features of which are indicated by the value of their importances, denoted as percentages. Column (10) lists the output parameter in each case. Column (11) tabulates the RMS scatter of the differences between measured and predicted {\it W2} magnitudes in each case. Please refer to Appendix \ref{sec:pred_algo} for more on how to derive {\it W2} magnitudes. In practice, in the ``Predictor" block of our algorithm (see Fig. \ref{fig:rf_schem}), we use a trained random forest as an encoder that holds all the fiducial information determined to be useful in our problem.    

Table \ref{tab:features} only lists the most promising combinations, and it tries to visit a variety of scenarios where one or a couple of magnitudes are missing due to poor image quality or other photometry problems. We find that the best performance is achieved when the output parameter is one of optical-infrared ($m^*_{\lambda}-W2^*$) colors. The input parameters are combinations of optical-optical colors and the first principal component, $P_{1,W2}$. For a given galaxy, we use all or a few of the random forest predictors and use their average predictions. 

We explained our prediction algorithm in Appendix \ref{sec:pred_algo}. Instead of the simple linear fiducial correlation presented in Fig. \ref{fig:P1_col}, we use different random forest encoders presented in Table \ref{tab:features}.  We run the algorithm for a range of input $W2^*$ values and end up with an output $W2^*$ parameter. Fig. \ref{fig:predict_w2} plots the difference between input and output $W2^*$ parameters, $\Delta_{W2}=W2_{in}-W2_{out}$, as a function of the input parameter, for two galaxies taken from the test sample. The predicted value, $W2^*_p$, is at $\Delta_{W2}=0$, where the input and output of the random forest model are in agreement with each other. 

\subsection{Evaluating the predicted parameters} \label{sec:predict_eval1}

The performance of our prediction method can be evaluated by comparing the predicted and measured parameters of the control sample (training and cross-validation sets).
Fig. \ref{fig:pred_error} plots the differences between predicted and measured $W2^*$ values. 
In the top panel, the discrepancies are plotted against the measured values, $W2^*_m$. The 1$\sigma$ deviation of errors in the predicted values is $0.23$ mag. In the bottom panel, the deviations are plotted against the \hi line widths that probe the absolute luminosity (and hence the size) of galaxies. No meaningful systematic is evident in either plot.

In the final part of the process, we use predicted {\it W2} magnitudes, axial ratios, and effective radii to calculate the dust attenuation coefficient, $\gamma_\lambda$ (Equation~\ref{eq:dust}). Uncertainties in predicted $\gamma_\lambda$ decrease with wavelength, ranging from 0.05 mag at {\it g} band to 0.01 mag at {\it z} band. As examples, Figures \ref{fig:gamma_pred1} and \ref{fig:gamma_pred2} plot the discrepancies between our predicted values of $\gamma_\lambda$ and the measured values versus various parameters. We observe no significant systematic correlations between discrepancies and the parameters used in these plots. 

\begin{figure}[t]
\centering
\includegraphics[width=0.95\linewidth]{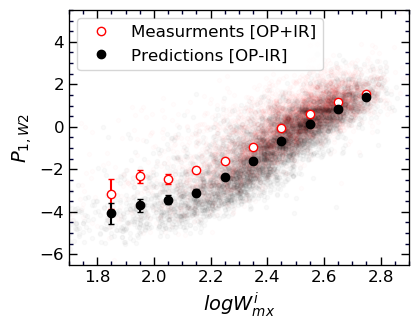}
\caption{
Main principal component $P_{1,W2}$ (see Equation \ref{Eq:P1} for the definition) versus line width.
Red points represent the training sample (OP+IR) that consists of 2244 galaxies. Black points are spirals with a lack of WISE infrared photoemtry (OP$-$IR), where their $P_{1,W2}$ are predicted following the discussed algorithm in Appendix \ref{sec:pred_algo}. Large points represent the average of data points within the bins of constant size, and their error bars show the 1$\sigma$ scatter of data points.
\label{fig:P1_gamma_logW}}
\end{figure} 

\subsection{Evaluating the predicted parameters} \label{sec:predict_eval2}

In the previous section, we evaluated the performance of our method solely based on the OP+IR galaxies, meaning that the training and control galaxies have both optical and infrared data. Our tests show that our algorithm is capable of producing reasonable predictions that are in reasonable agreement with the measurements. However, we appreciate that the trained model is practically applied on spirals with missing infrared data. 
As discussed in Appendix \ref{sec:OPIR-OP-IR}, the OP$-$IR sample has different statistical characteristics than the OP+IR sample. 

Figure \ref{fig:P1_gamma_logW} plots the main principal component, $P_{1,W2}$, versus line width, with the red and black points representing the OP+IR and OP$-$IR galaxies, respectively. 
The $P_{1,W2}$ values of the red points are calculated from measurement, whereas those of the black points are predictions of our algorithm. 
All predicted $P_{1,W2}$ parameters seem to be smaller than the measured values. However, we attribute the offset to the average color difference of the two samples. 
Here, $P_{1,W2}$ is the linear combination of line width, the $C_{21W2}$ color, and the average surface brightness at the {\it W2} band (see Equation \ref{Eq:P1}). 
As illustrated in the top panel of Figure \ref{fig:gz_logW}, the OP+IR galaxies are slightly redder on average at a given line width.
At a constant line width, redder spirals are formed by older stellar populations. They have a smaller ratio of \hi to stellar mass, and hence their $C_{21W2}$ color index is larger (redder). Moreover, they are brighter at longer passbands, implying their $\langle \mu_2 \rangle^{(i)}_e$ to be smaller.
These, together with the form of Equation \ref{Eq:P1} explain the differences between the average $P_{1,W2}$ values of OP+IR and OP$-$IR galaxies. In fact, the similarity between Figure \ref{fig:P1_gamma_logW} and the top panel of Figure \ref{fig:gz_logW} indirectly indicates that our predictive algorithm has successfully attained the essential aspects of our physical arrangement. 

\section{On the calculations of uncertainties}  \label{sec:uncertainties}

In this work, multiple sources of uncertainties contribute to the final reported moduli uncertainties in Table \ref{tab:distance_catal}:

\begin{itemize}
    \item {\it The uncertainty of the measured observable parameters that are directly used to calculate the galaxy distance modulus.} These parameters are as follows: (1) apparent magnitudes of the galaxy at optical and infrared wave bands. (2) \hi flux and line width, (3) galaxy spatial inclination, and (4) other geometrical features of the elliptical aperture that fits the galaxy.
    
    The uncertainty of our measured magnitudes are no worse than $0.05$ mag at all wave bands except for the {\it u} band, where we adopt the conservative error of $0.1$ mag.
    To calculate the error of the surface brightness from Equations \ref{Eq:Mu50_i} and \ref{Eq:Mu50_e}, in addition to the error of the measured apparent magnitudes, we considered an error of $1$ pixel ($0.4$" for SDSS and $1$" for WISE) for the galaxy's projected dimensions. We do not consider any error on the radial velocities of the galaxies because this parameter is never used in the determination of distances in this study.
    The uncertainties of all other parameters are reported in columns (3-37) of Table \ref{tab_data}. 
    
    \item {\it The uncertainty in the fitted parameters of the applied relations.} 
    The main source of uncertainty is associated with the calibration of TFRs. The TFRs' parameters and their uncertainties are listed in Table \ref{tab:revised_ITFR}. We refer readers to K20 for a detailed discussion. The main concern regarding the TFRs is the vertical scatter of galaxies along the magnitude axis, which is larger than the statistical uncertainties we report for the measured distance moduli. Part of the scatter can be explained by the statistical uncertainties of the measured parameters that are involved in establishing TFRs (magnitudes and \hi line widths), while there might be other contributions by some unknown physical processes that are not captured by the observables at our disposal. We list the RMS scatter about the TFR in Table \ref{tab:distance_catal}. The covariances of the parameters of the fitted adjusting/regularizing relations in \S \ref{sec:col_dep_sys} and \S \ref{sec:H0_regularization} are taken into account when they are employed to revise distance moduli.
    
    \item {\it The uncertainty of the predicted quantities, for cases where a parameter is not directly available.} In case of the predicted half-light radii and axial ratios, we combine the uncertainties of the dependent variable and uncertainties of the parameters of the fitted relations. The predicted dust attenuation depends on the estimated apparent magnitudes at the $W2^*_p$ band. The RMS scatter of differences between measured and predicted values is $\sim$0.2 mag. The parameter $W^*_p$ is solely used to calculate $\gamma_{\lambda}$, which is connected to $A^{(p)}_{\lambda}$ through Equation \ref{eq:dust}. It is more reasonable to consider the scatter of differences between measured and predicted attenuation factors, $\Delta \gamma_{\lambda} = \gamma^{(m)}_{\lambda}-\gamma^{(p)}_{\lambda}$. Similarly to what is presented in Figure \ref{fig:P1_gamma_logW}, we determine respective scatters of $\sim$0.03 mag for $\lambda = u,g,r$ and $\sim$0.02 mag for $\lambda=i,z$. Accordingly, we add these additional statistical scatters in quadrature to the error budget of the attenuation factors.
    
\end{itemize}

All the uncertainties in our calculations are accounted for following the Gaussian formalism of error propagation.

\clearpage
\bibliography{paper}

\end{document}